\documentclass[epj,nopacs,final]{svjour}
\usepackage{graphics}
\usepackage{latexsym}
\usepackage{subfigure}
\usepackage{epsfig,color,rotating,amsmath,delarray,array}
\usepackage{float,amssymb}
\usepackage{epstopdf}
\usepackage[percent]{overpic}
\usepackage{lineno}

\definecolor{mygreen}{rgb}{0.0,0.5,0.0}
\definecolor{myblue}{rgb}{0.0,0.0,0.55}
\definecolor{myblueh}{rgb}{0.0,0.0,0.65}
\definecolor{myred}{rgb}{0.6,0.0,0.}
\definecolor{myyellow}{rgb}{0.6,0.6,0.}
\definecolor{myyellow2}{rgb}{0.8,0.5,0.1}
\definecolor{myyellow2b}{rgb}{0.9,0.9,0.1}
\definecolor{mycyan}{rgb}{0.0,0.6,0.6}
\definecolor{mylila}{rgb}{0.6,0.,0.6}
\definecolor{mygray}{rgb}{0.37,0.37,0.37}

\begin{document}

\title{Measurement of the Helicity Asymmetry \boldmath$E$ for the reaction \boldmath $ \gamma p\to \pi^0 p$}
\titlerunning{Helicity Asymmetry $E$}
\authorrunning{The CBELSA/TAPS Collaboration}
\author{The CBELSA/TAPS Collaboration 
\medskip \\  
 M.~Gottschall$\,^{1}$,
 F.~Afzal$\,^{1}$,
 A.V.~Anisovich$\,^{1,3}$,
 D.~Bayadilov$\,^{1,3}$,
 R.~Beck$\,^{1}$,
 M.~Bichow$\,^{6}$,
  \mbox{K.-Th.~Brinkmann$\,^{4}$,}
 V.~Crede$\,^{7}$,
 M.~Dieterle$\,^{5}$,  
 F.~Dietz$\,^{4}$,
 H.~Dutz$\,^{2}$,
 H.~Eberhardt$\,^{2}$,
 D.~Elsner$\,^{2}$,
 R.~Ewald$\,^{2}$,
 K.~Fornet-Ponse$\,^{2}$,
 St.~Friedrich$\,^{4}$,
 F.~Frommberger$\,^{2}$,
 A.~Gridnev$\,^{3}$,
 M.~Gr\"uner$\,^{1}$,
 E.~Gutz$\,^{4}$,
 Ch.~Hammann$\,^{1}$,
 J.~Hannappel$\,^{2}$,
 J.~Hartmann$\,^{1}$,
 W.~Hillert$\,^{2}$,
 Ph.~Hoffmeister$\,^{1}$,
 Ch.~Honisch$\,^{1}$,
 T.~Jude$\,^{2}$,      
 S.~Kammer$\,^{2}$,
 H.~Kalinowsky$\,^{1}$,
 I.~Keshelashvili$\,^{5}$,
 P.~Klassen$\,^{1}$,   
 F.~Klein$\,^{2}$,
 E.~Klempt$\,^{1}$,
 K.~Koop$\,^{1}$,
 B.~Krusche$\,^{5}$,
 M.~Kube$\,^{1}$,
 M.~Lang$\,^{1}$,
 I.~Lopatin$\,^{3}$,
 P.~Mahlberg$\,^{1}$,  
 K.~Makonyi$\,^{4}$,
 V.~Metag$\,^{4}$,
 W.~Meyer$\,^{6}$,
 J.~M\"uller$\,^{1}$,
 J.~M\"ullers$\,^{1}$, 
 M.~Nanova$\,^{4}$,
 V.~Nikonov$\,^{1,3}$,
 R.~Novotny$\,^{4}$,
 D.~Piontek$\,^{1}$,
 G.~Reicherz$\,^{6}$,
 T.~Rostomyan$\,^{1}$,
 A.~Sarantsev$\,^{1,3}$,
 Ch.~Schmidt$\,^{1}$,
 H.~Schmieden$\,^{2}$,
 T.~Seifen$\,^{1}$,
 V.~Sokhoyan$\,^{1}$,
 K. Spieker$\,^{1}$,
 A.~Thiel$\,^{1}$,
 U.~Thoma$\,^{1}$,
 M.~Urban$\,^{1}$,
 H.~van~Pee$\,^{1}$,
 D.~Walther$\,^{1}$,
 Ch.~Wendel$\,^{1}$,
 D.~Werthm\"uller$\,^{5}$, 
 U.~Wiedner$\,^{6}$,
 A.~Wilson$\,^{1,7}$, 
 A.~Winnebeck$\,^{1}$,~and
 L.~Witthauer,~and        
Y.~Wunderlich.}           

\institute{
$^1\,$Helmholtz--Institut f\"ur Strahlen-- und Kernphysik, Universit\"at Bonn, Germany\\
$^2\,$Physikalisches Institut, Universit\"at Bonn, Germany\\
$^3\,$Petersburg Nuclear Physics Institute, Gatchina, Russia\\
$^4\,$Physikalisches Institut, Universit\"at Gie{\ss}en, Germany\\
$^5\,$Physikalisches Institut, Universit\"at Basel, Switzerland\\
$^6\,$Institut f\"ur Experimentalphysik I, Ruhr--Universit\"at Bochum, Germany\\
$^7\,$Department of Physics, Florida State University, Tallahassee, FL 32306, USA\\[1mm]
}
\date{Received: \today / Revised version:}

\abstract{A measurement of the double-polarization observable $E$ for the reaction $\gamma p\to \pi^0 p$ is
reported. The data were taken with the CBELSA/TAPS experiment at the ELSA facility in Bonn using the Bonn frozen-spin
butanol (C$_4$H$_9$OH) target, which provided longitudinally-polarized protons.
Circularly-polarized photons were produced via bremsstrahlung of longitudinally-polarized electrons.
The data cover the photon energy range from $E_\gamma =600$~MeV to $E_\gamma =2310$~MeV and nearly the complete
angular range. The results are compared to and have been included in recent partial wave analyses. 
  }
\maketitle
%

\section{Introduction}

Nucleons, protons and neutrons, are no elementary particles, they contain three light valence
quarks. As three-body system, the nucleon is expected to exhibit a large number of excitation
modes (see e.g. \cite{Capstick:1986bm,Loring:2001kx,Giannini:2015zia}), and the prediction of
hybrid baryons increases the number of expected resonances even further
\cite{Capstick:2002wm,Dudek:2012ag}. Lattice QCD seems to confirm the large number of predicted
states, at least when a quark mass corresponding to $m_\pi=396$\,MeV is used~\cite{Edwards:2011jj}.
When the pion mass is varied between 255 and 596\,MeV, the masses of baryons with
$J^P=1/2^\pm,3/2^\pm$ are still in good agreement with the observed states \cite{Engel:2013ig}. To
understand the problem of the {\it missing resonances}, alternative approaches have been suggested:
Possibly, i) the three-quark dynamics is frozen to a quark-diquark system \cite{Anselmino:1992vg},
ii) excited baryons could be generated dynamically by the interaction of mesons and octet or
decuplet (ground-state) baryons
\cite{Kaiser:1995cy,Meissner:1999vr,Kolomeitsev:2003kt,Sarkar:2009kx,Oset:2009vf,Bruns:2010sv}, or
iii) the number of excited baryons might be reduced in models based on
AdS/QCD~\cite{Forkel:2008un,deTeramond:2014asa,Klempt:2010du}. However, i) predicts fewer
resonances than observed, ii) and iii) do not give a prediction which resonances should be observed
and which ones not. Recent surveys of the field can be found in
\cite{Klempt:2009pi,Klempt:2012fy,Crede:2013kia,Eichmann:2016yit}.

\begin{figure*}[ht]
 \centering
 \includegraphics[height=0.37\textwidth]{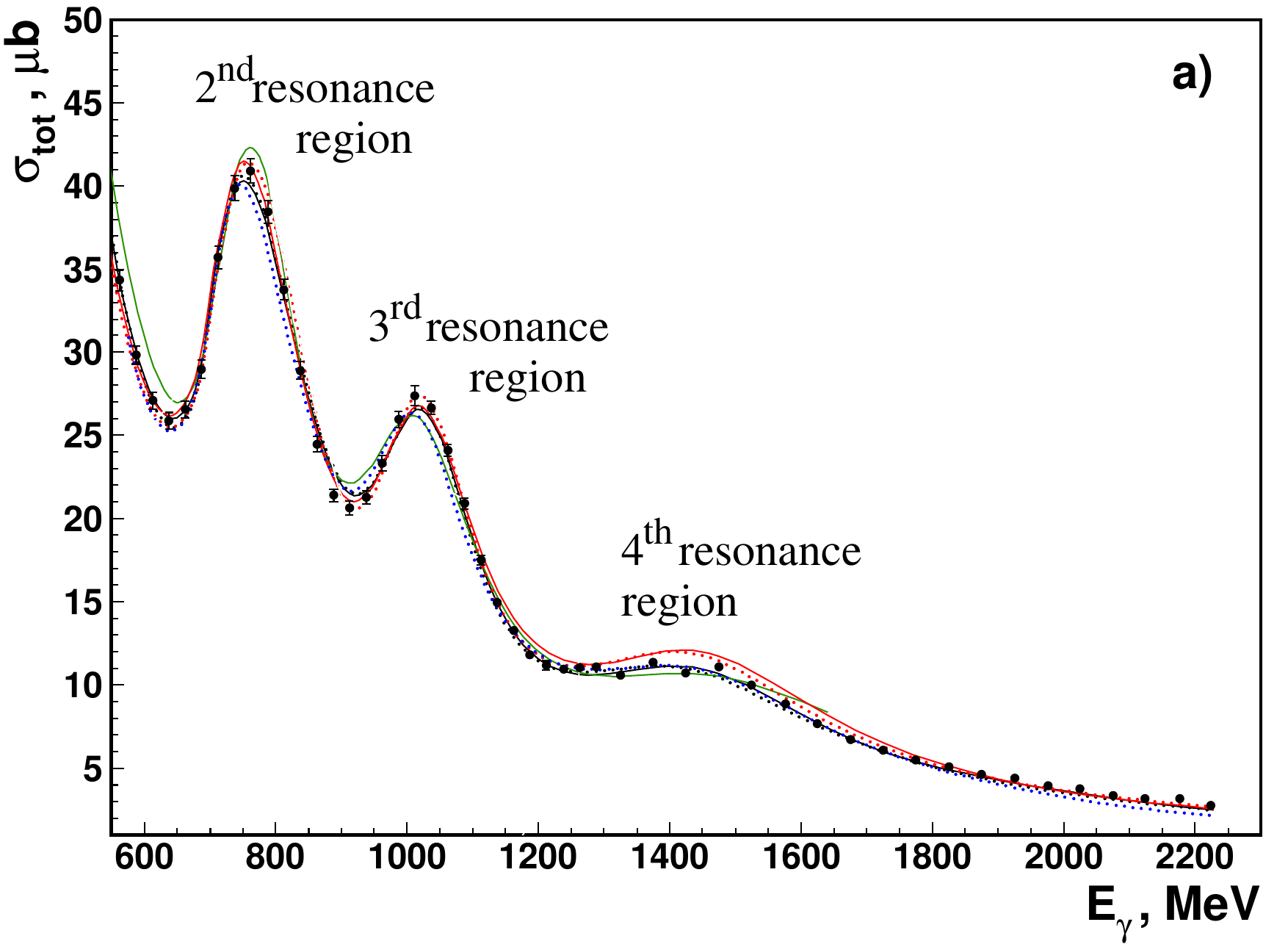} \qquad
 \includegraphics[height=0.37\textwidth]{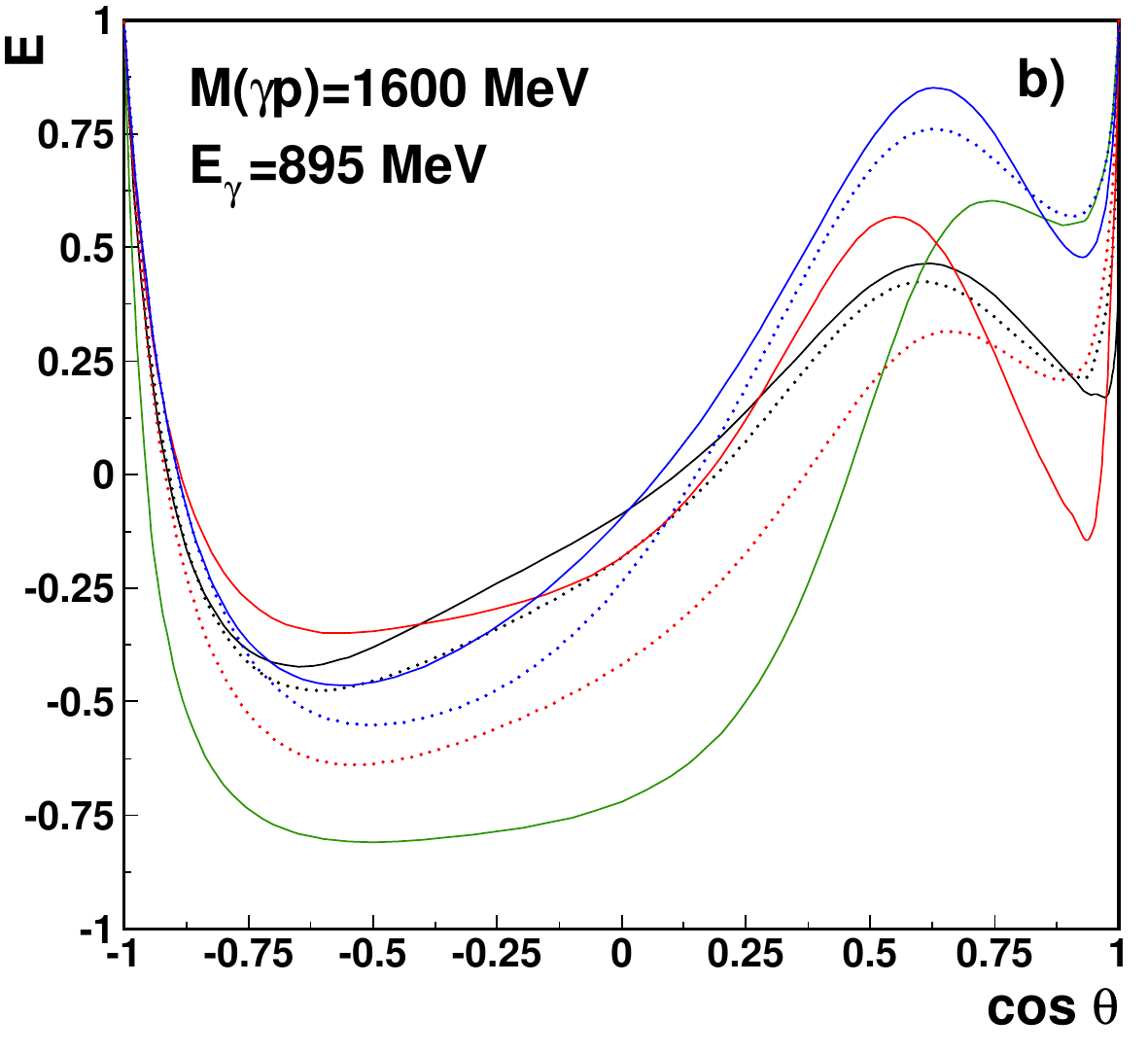}
 \caption{\label{unpolWQ} Left: Total $\gamma p \to p\pi^0$-cross section.
Right: Predictions for the double-polarization variable $E$.
The different PWA solutions for which the predictions are shown, did not yet include the new
polarization data discussed in this paper: BnGa 2011-02 (black solid)~\cite{Anisovich:2011fc},  BnGa 2011-01 (black dotted)~\cite{Anisovich:2011fc},
SAID CM12 (red solid)~\cite{Workman:2012jf}, SAID SN11 (red dotted)~\cite{Workman:2011vb}, J\"uBo 2015-B (blue solid)~\cite{Ronchen:2015vfa}, J\"uBo 2013-01 (blue dotted)~\cite{Ronchen:2014cna}, MAID 2007 (green solid)~\cite{Drechsel:2007if}. }
\end{figure*}

Alternatively, the number of baryon resonances predicted in quark models could be correct, but a
sizable fraction of them may have escaped discovery in appropriate experiments. The traditional
path to the excitation spectrum of nucleons was the study of $\pi N$ scattering. For a long time,
our knowledge on the excitation spectrum was based on the two classical analyses at
Karlsruhe-Helsinki (KH) \cite{Hohler:1979yr} and Carnegie Mellon (CM) \cite{Cutkosky:1980rh}, and
the more recent analysis at George Washington University (GW) \cite{Arndt:2006bf} which included
high-precision data from meson factories and measurements of the phase-sensitive polarization state
of the recoiling nucleon. Only recently, the experimental efforts at ELSA, GRAAL, JLab, MAMI,
and Spring-8 have provided high-statistics data on photoproduction of mesons off nucleons, and
several new nucleon resonances were discovered (see e.g.~\cite{Anisovich:2011fc}). 
But still, the number of predicted states exceeds the number of observed states. There is 
hope that further data and further analyses will
finally clarify the {\it missing resonance} problem.

Polarization experiments are key to improving the data\-base of photoproduction reactions. For the
two best studied photoproduction processes, $\gamma p\to \pi^0 p$ and  $\gamma p\to \pi^+ n$, data were accumulated for
several observables. Here, we quote a few recent measurements; references to data reported before
the year 2000 can be found elsewhere~\cite{GWU}. The recent data cover the differential cross
section $d\sigma/d\Omega$
\cite{Ahrens:2004pf,Bartholomy:2004uz,Bartalini:2005wx,Dugger:2007bt,Dugger:2009pn,Crede:2011dc,Adlarson:2015byy},
the beam asymmetry $\Sigma$ \cite{Bartalini:2005wx,Sparks:2010vb,Dugger:2013crn}, the target
asymmetry $T$ \cite{Hartmann:2014mya,Hartmann:2015kpa,Annand:2016ppc}, and the recoil polarization
$P$ \cite{Hartmann:2014mya,Hartmann:2015kpa,Luo:2011uy}. The single-polarization observables
$\Sigma, T$ and $P$ require the use of linearly-polarized photons, of transversely-polarized
protons, or the analysis of the polarization of the outgoing protons, respectively. The observable $P$ can,
however, also be determined from a measurement using linearly-polarized photons and
transversely-polarized protons~\cite{Hartmann:2014mya,Hartmann:2015kpa}.

More information can be gained when two of the three polarizations are controlled experimentally.
Photons can be polarized linearly or circularly, and target protons can be polarized along the
photon-beam axis or in a transverse direction to measure the quantities  $G$ \cite{Ahrens:2005zq,Thiel:2012yj,Thiel:2016chx},
$E$ \cite{Ahrens:2002,Ahrens:2004,Gottschall:2013uha,Strauch:2015zob,Dieterle:2017myg}, $H$
\cite{Hartmann:2014mya,Hartmann:2015kpa}, and $F$~\cite{Annand:2016ppc}. The correlation between
the photon polarization and the recoil polarization yields the observables
$O_{x^\prime},O_{z^\prime}, C_{x^\prime}, C_{z^\prime}$. The  correlation between the target
polarization and the recoil polarization is governed by $T_{x^\prime},T_{z^\prime}$, $L_{x^\prime},
L_{z^\prime}$. For $C_{x^\prime}$ and $C_{z^\prime}$, data have been published for a few specific
energies and angles~\cite{Luo:2011uy,Wijesooriya:2002uc,Sikora:2013vfa}. This is also the case for
$O_{x^\prime}$ and $O_{z^\prime}$, where only a few older data points
exist~\cite{Avakyan:1991pj,Bratashevsky:1980dk}. To our knowledge, no direct measurements of the
target-recoil polarization for single-$\pi^0$ photoproduction have been undertaken.
Following theory, not all 16 observables need to be determined to obtain a unique solution for the
photoproduction amplitude. The scattering process is governed by four complex amplitudes, the so-called
Chew-Goldberger-Low-Nambu (CGLN) amplitudes \cite{Chew:1957tf}, and a minimum of seven carefully chosen
measurements should be sufficient for their determination (up to an unknown phase).
Due to the nonlinear relation between amplitudes and observables, an eighth measurement is required to
resolve discrete ambiguities \cite{Barker:1975bp,Chiang:1996em,Keaton:1995pw}, and all measurements have to be precise 
and cover the full solid angle \cite{Sandorfi:2010uv}. 
This approach leads - if successful - to the reconstruction of the four complex
amplitudes for every bin in photon energy and angle. For every
kinematic bin, one phase is arbitrary.
The CGLN amplitudes can then be expanded into a series of electric and magnetic multipoles which drive the
excitation of particular resonances. In practice, this expansion needs to be truncated using a finite
number of multipoles only.

A more direct approach constructs the multipoles directly by fitting
the data. In this approach, the minimum number of known observables
can even be smaller than eight, depending on the data quality and the
number of contributing multipoles. Even a small number of observables can be sufficient to arrive at a unique solution \cite{Omelaenko1981,Wunderlich:2014xya}. This is especially true at low energies: Below the 2$\pi$ threshold, only $S$ and $P$ waves contribute to $\pi^0$ photoproduction and the strong-interaction phase is fixed due to the Watson theorem \cite{Watson:1954uc} . A measurement of differential cross sections $d\sigma/d\Omega$ and the photon beam asymmetry $\Sigma$ is thus sufficient to determine the contributing multipoles in the first resonance region where the $\Delta(1232)$ resonance dominates \cite{Beck:1997ew,Blanpied:1997zz}. With the number of known observables increasing, $S$, $P$, and $D$ waves can be determined in an energy-independent fit to the data covering the second resonance region \cite{Hartmann:2014mya}. Here, higher partial waves have been fixed to their contributions determined from an energy-dependent fit.

The higher the incident photon energy, the more resonances contribute. This in turn makes the measurement
of additional polarization observables necessary to finally pin down the contributing resonances.
Fig.~\ref{unpolWQ} (left) shows fits to the $\gamma p\to \pi^0 p$ total cross section. The data
reveal two clear peaks and a smaller enhancement which are assigned to the 2$^{\rm nd}$, 3$^{\rm
rd}$, and 4$^{\rm th}$ resonance region where several resonances contribute with masses around
1500~MeV, 1650~MeV and 1900~MeV, respectively. The MAID, SAID, J\"uBo and BnGa partial wave
analyses all reproduce the data well. However, even at rather low energies,  the predictions for
the double-polarization observable $E$ spread over a wide range, see Fig.~\ref{unpolWQ} (right).
Obviously, the amplitudes included in the different PWA solutions shown in the figure are by far
not identical. This finally leads to different resonances and resonance properties which are
extracted based on the different PWAs.

In this paper, we report on a measurement of the double polarization observable $E$ for the reaction $\gamma p\to \pi^0p$ in the
photon energy range from $E_\gamma =600$~MeV to $E_\gamma =2310$~MeV. Selected data have been
presented in a Letter \cite{Gottschall:2013uha}. Here, we give experimental details 
and report the  full data set on $E$.

The paper is organized as follows: This Introduction is followed by Section~\ref{Edef} which
defines the double-polarization variable $E$ and describes the experimental method.
Section~\ref{CBTAPS} introduces the experimental set up. Decisive ingredients for polarization
measurements are of course the beam and target polarization, which are reviewed in
Section~\ref{Pol}. It is followed by Section~\ref{Calib} on the calibration procedures and the
selection of the events. In Section~\ref{Analysis}, the analysis method used to determine the
helicity asymmetry $E$ is explained and the results on $E$ are presented and discussed. The paper
ends with a short summary.

\begin{figure*}[ht]
 \begin{center}
  \includegraphics[width=0.75\textwidth]{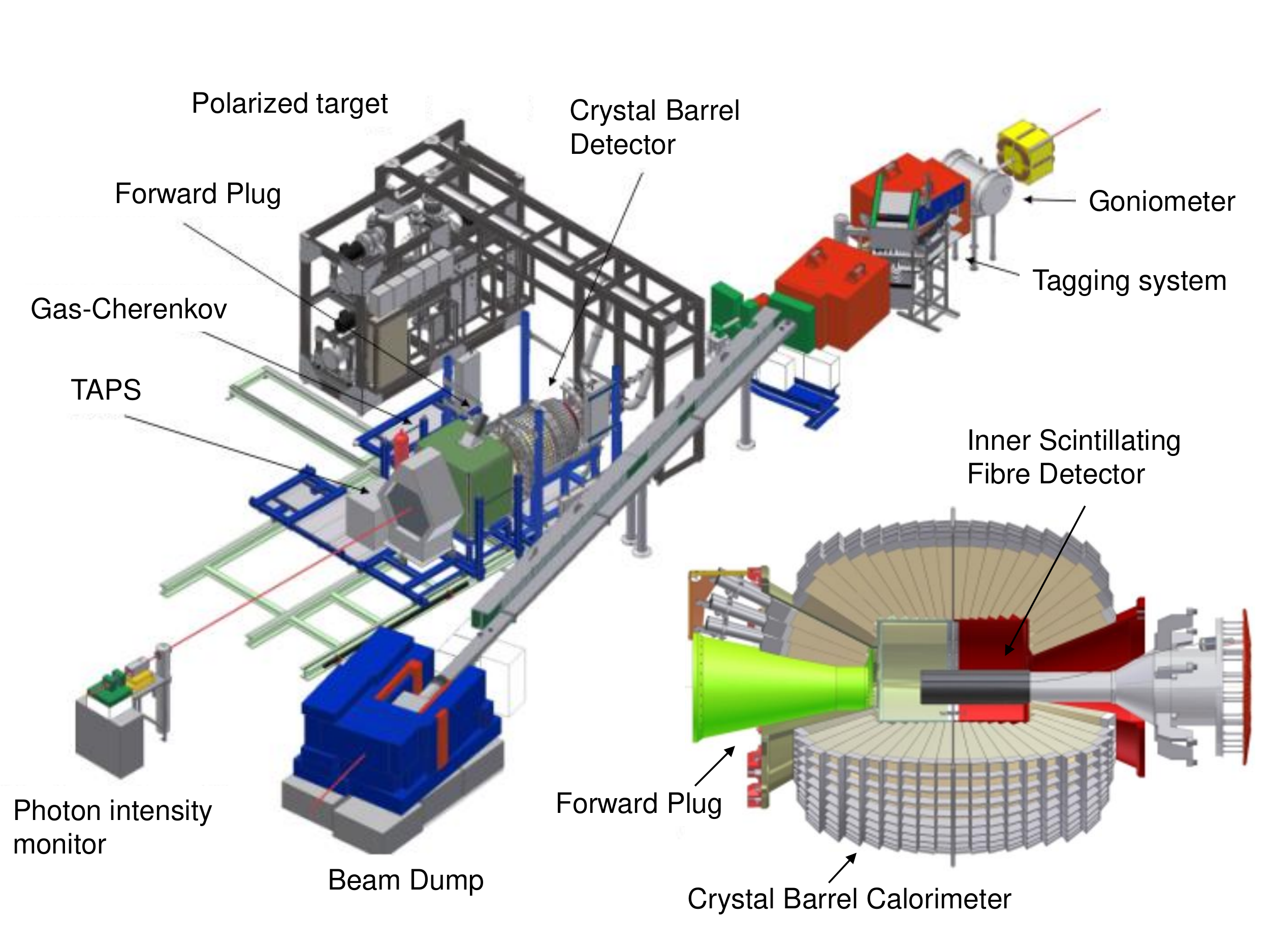}
  \caption{\label{FigExp} Experimental Setup of the CBELSA/TAPS Experiment.
}
 \end{center}
\end{figure*}

\section{\label{Edef}\boldmath The double-polarization observable $E$ for the
reaction $\gamma p \to \pi^0 p$} The double-polarization observable $E$ is defined as the normalized
difference of the cross sections for antiparallel and aligned spins of beam photons and target
protons, $\sigma_{1/2}$ and $\sigma_{3/2}$, respectively:
\begin{equation}
 E=\frac{\sigma_{1/2}-\sigma_{3/2}}{\sigma_{1/2}+\sigma_{3/2}}=\frac{\sigma_{1/2}-\sigma_{3/2}}{2\sigma_0}\,,
\end{equation}
where the unpolarized cross section $\sigma_0$ is written as $2\sigma_0=\sigma_{1/2}+\sigma_{3/2}$.

To measure the  observable $E$, a circularly-polarized photon beam and a longitudinally-polarized
target was needed. Circularly-polarized photons were produced using lon\-gitudinally-polarized
electrons (see Section~\ref{Pol-gamma}) which then transferred part of their polarization to the
bremsstrahlung photon. Their polarization was monitored by M\o{}ller
polarimetry~\cite{Kammer:2009}. Polarized protons were obtained by using the Bonn frozen-spin
polarized butanol (C$_4$H$_9$OH) target (see Section~\ref{Pol-proton}).

The observable $E$ can be expressed in terms of count rates in the following way:
\begin{equation}
\label{E_formula}
E= \frac{N_{1/2}-N_{3/2}}{N_{1/2}+N_{3/2}}\frac{1}{P_TP_z}\frac{1}{d}\,.
\end{equation}
$N_{3/2}$ and $N_{1/2}$ are the count rates for the beam-target setting 
with the spins of the photon and proton parallel and
antiparallel, $P_T$ is the target polarization, $P_z$ the beam polarization and $d$ the dilution
factor. The dilution factor takes into account that the butanol contains not only polarizable
protons in hydrogen, but also nucleons bound in carbon and oxygen nuclei, which are not
polarizable. Thus, the fraction of free polarizable protons contributing to the reaction had to be
determined and is given by the dilution factor $d$.

 The measurement of $E$ required the identification of the reaction
\begin{equation}
\label{reaction}
\gamma p \to  \pi^0 p \to \gamma\gamma p 
\end{equation}
and a separate measurement the number of events in the two spin configurations. In addition, the
dilution factor was needed.

After a calibration procedure (see Sections~\ref{sec:calib_time}-\ref{sec:calib_energy}) for the
detector components, the energy of the bremsstrahlung beam-photon was deduced from a measurement of
the energy loss of the corresponding electron (see Section~\ref{sec:beam-photon}). The four-vectors
of photons produced in the reaction were reconstructed from the Crystal Barrel and TAPS
electromagnetic calorimeters (see Section~\ref{sec:reconstruction}).  In the 2$\gamma$ invariant
mass spectrum, the $\pi^0$ was clearly observed. 
The proton was identified in
the forward direction by plastic scintillators in front of the crystals ($<30^\circ$) or at larger
angles by a scintillating fiber detector surrounding the target (see Section~\ref{matching}). Its
direction was determined either from its respective signal in one of the calorimeters or from the
signal of the scintillating fiber detector. The kinematics of the event was then over-constrained
by energy and momentum conservation, and the events of interest were identified nearly background
free (see Section~\ref{Selec}).

\section{\label{CBTAPS}Experimental Setup}
\subsection{Overview}
An overview of the CBELSA/TAPS experiment is shown in Fig.~\ref{FigExp}. The experiment was located
at the electron accelerator ELSA~\cite{Hillert:2006yb} in Bonn, Germany. ELSA provided either
unpolarized or longitudinally-polarized electrons with energies up to 3.2~GeV (see
Section~\ref{Pol-gamma}). The electrons hit a radiator target inside a goniometer, where photons
were produced via bremsstrahlung. For the data presented here, a M{\o}ller target was used as the
bremsstrahlung target which allowed the monitoring of the electron  polarization via M\o{}ller
scattering in parallel to the data taking.
The M\o{}ller target was a 20$\mu$m 
thin foil with a magnetization of 8.2\%, enclosing an angle of $\pm$20$^\circ$ with the electron beam
(see Section~\ref{Pol-gamma}).
After the bremsstrahlung process, the electron was deflected in a magnetic field and
momentum-analyzed in a ladder of organic scintillator bars and fibers (see
Section~\ref{tagging_system}). The photon energy was then deduced from the energy difference of the
incoming and the deflected electron  after the bremsstrahlung process.

The bremsstrahlung photons impinged on the polarized target (see Section~\ref{Pol-proton}). The
target was surrounded by a three-layer scintillation fiber detector used for the identification of
charged particles and by the Crystal Barrel electromagnetic calorimeter consisting of 1230
CsI(Tl)-crystals. In the forward direction, below polar angles of 30$^\circ$, two further
calorimeters, the Forward Plug (90 CsI(Tl) crystals) and the forward TAPS-wall (216 BaF$_2$
crystals), provided  calorimetric information. Plastic scintillators in front of these crystals
allowed for the identification of charged particles.

The signals from the forward calorimeters, the fiber detector and the tagging system were used in
the first-level trigger. A second-level trigger identified the number of clusters in the Crystal
Barrel calorimeter. In between the Crystal Barrel and the forward TAPS detector, a CO$_2$ Cherenkov
counter vetoed signals of electron or positron hits due to the electromagnetic background produced
in the target.

\subsection{\label{tagging_system}The Tagging System}
The Tagging System~\cite{Fornet-Ponse:2009} consisted of 96~partly-overlapping scintillation bars with a width
between $1.4\,$cm and $5\,$cm. These allowed us to tag electrons in the energy range between
2.1\,\% and 82.5\,\% of the electron beam energy $E_{0}$. To improve the energy
resolution, the scintillation bars were complemented by 480~cylindrical scintillating fibers
with 0.2~cm in diameter covering the electron energy range from 16.6\,\% to 88.1\,\% of $E_{0}$.
Combined, this resulted in a tagging range for photons ($E_{\gamma}=E_0-E_e$ with $E_e$: measured electron energy)
from 11.9\,\% to 97.9\,\% with an energy resolution varying from
$\sim$0.5\,\% $E_{0}$ at low photon energies to 0.1\,\% $E_{0}$ at high
photon energies. While the high photon energy regime is only covered by bars, the very low 
photon energy regime is only covered by fibers.
The bars and the fibers were
read out by photomultipliers reaching a time resolution for the bars (fibers) 
of about 0.5\,ns (1.2\,ns) 
FWHM.
The signals from the tagger bars were included in the first-level trigger of the experiment.

\subsection{Collimator and the Beam Dump}
After the goniometer in the photon beamline, six collimators each with an inner (outer) diameter of 4 (20)~mm
and a length of 4~cm reduced the beam halo. Behind the collimators, a permanent magnet deflected
low energetic charged particles coming along the beamline, which then hit a lead wall.
Electrons from the primary beam not producing bremsstrahlung were deflected into a beam dump.
The latter consisted of 70\,t of steel and 470\,kg iron. It was placed far behind the calorimeters to avoid
background in the detector systems coming from the beam dump.

\subsection{Targets}
The main data used for this work were taken with the Bonn frozen spin target (C$_4$H$_9$OH)
of 2~cm length described in Section~\ref{Pol-proton}.
In a butanol target, polarizable free protons as well as nucleons bound in the
unpolarized carbon and oxygen nuclei contribute to the count rate.
To determine the contribution from the bound nucleons, a carbon foam target was used
inside the cryostat of the polarized target. The target nose was also filled with liquid helium
to keep the conditions comparable to the butanol measurements.
The foam target had approximately the same density as the carbon and oxygen component
of the butanol target. For further tests, data were available which were taken with a 
liquid hydrogen (LH2) target - as used, e.g., in
\cite{vanPee:2007tw} and \cite{Gutz:2014wit}.

\subsection{The Inner Scintillating Fiber Detector}\label{sec:chapi}
The target was surrounded by the Inner Scintillating Fiber Detector~\cite{Suft:2005cq}. It was 400\,mm long,
had an inner (outer) diameter of 116\,mm (131\,mm), and covered the polar angle range of
$21^\circ < \theta < 167^\circ$. The detector consisted of 513 scintillating fibers with a diameter of
2\,mm in 3 layers. The outer layer (191 fibers) was positioned parallel to the beam axis, the middle layer
(165 fibers) was oriented at an angle of $+25.7^\circ$, and the innermost (157 fibers) at an angle of
$-24.5^\circ$ with respect to the beam axis. The angles resulted from the requirement for the bent
fibers to go exactly halfway around the detector. Charged particles were identified and their impact
points reconstructed by coincident hits in two or three layers. The readout was organized via 16-channel
photomultipliers connected to the fibers via light guides. The inner detector was included in the first-level
trigger. A valid inner detector trigger signal was defined by two out of three layers firing.

\subsection{The Calorimeters}
The electromagnetic calorimeters of the detector system covered the full azimuthal range and
polar angles from $1^{\circ}$ in the forward direction to $156^{\circ}$ in the backward direction.
The combined solid-angle coverage was about 95\,\% of 4$\pi$.

\paragraph{The Crystal Barrel Calorimeter and the Forward Plug }
\label{det_cb_fp}
The Crystal Barrel calorimeter consisted of 1230 CsI(Tl) crystals, each crystal pointing to the
target center. Its 21~rings were arranged in a barrel shape around the production target, 10~rings
covered the forward hemisphere, 11~rings the backward hemisphere. Every ring consisted of 60
crystals, each covering an azimuthal- and polar-angle range of $6^{\circ}$ except for the last ring in the
backward direction which consisted of 30 crystals only, each covering an azimuthal-angle range of $12^{\circ}$. The crystals were read out via a photo-diode mounted at the edge of a wavelength shifter
collecting the light emitted by the crystal at its backside~\cite{CB}. This component of the
calorimeter covered the polar angles of 30\,-\,156$^{\circ}$.

A fast cluster encoder (FACE)~\cite{Diss_Flemming} based on a cellular
logic defined the number of contiguous clusters of crystal hits with
energies above $\sim$15~MeV,
typically within less than 10\,$\mu$s. This information was used as a second-level trigger. For rejected events,
a fast reset was generated which cleared the readout electronics in less than 15\,$\mu$s.

In the forward direction, the Crystal Barrel calorimeter was complemented by the Forward Plug (FP),
which consisted of additional 90 CsI(Tl) crystals arranged in three rings covering the polar angle
from $11.2^{\circ}$ to $27.5^{\circ}$. Due to an additional holding structure, there was an area
between the two detectors, where the acceptance was slightly reduced. The Forward Plug was read out
by photomultipliers providing a fast enough timing signal to be used in the first-level trigger. A
cluster finder provided the number of clusters in the FP~\cite{Diss_Funke}. It was based on a
comparison of the crystal hit pattern in the FP with a Field Programmable Gate Array (FPGA)
internal lookup table and delivered the number of clusters in less than 200\,ns. It could thus be
incorporated in the first level-trigger. Typically the cluster finder considered crystals with more
than 25~MeV as hit crystals. For the identification of charged particles, 180 overlapping
scintillator plates were mounted in two layers in front of these crystals. The azimuthal
granularity of each scintillator plate in angle was $12^\circ$. Each scintillator in the first
layer, close to the crystals, covered one of the 90 CsI(Tl) crystals. The second layer was turned by
$6^\circ$ relative to the first one, covering now one half of two crystals. This doubled the $\phi$
angular resolution for the charged particles.

The Crystal Barrel calorimeter (including the Forward Plug) was optimized for photon detection.
The energy resolution was empirically determined to be~\cite{CB}
\begin{equation}
\frac{\sigma_E}{E}\;\approx\;\frac{2.4\,\%}{\sqrt[4]{E~[{\rm GeV}]}}\,.
\end{equation}
Since photons produce a shower in the calorimeter, an angular resolution better than the crystal
granularity can be reached. As an example, using photons with energies above 200~MeV~(500~MeV) which
were not hitting the detector boundaries, an angular resolution of better than
$1.75^{\circ}$~($1.4^{\circ}$) was reached by determining of the shower center. The angular
resolution improved for higher-energy photons. The given resolution takes into account the
extension of the butanol target. For a point-like target, the respective resolutions would improve
to $1.5^{\circ}$~($1.1^{\circ}$)~\cite{Diss_Jonas}.

\paragraph{The TAPS detector }
At a distance of 2.10\,m behind the target center, the TAPS detector with 216 BaF$_2$
crystals~\cite{TAPS} closed the hole in the forward direction. It had a high granularity and
covered polar angles from $12^{\circ}$ down to $1^{\circ}$ . This calorimeter reached an energy
resolution~\cite{Gabler:1994ay} of
\begin{equation}
\frac{\sigma_E}{E}\;=\;\frac{0.59\,\%}{\sqrt{E~[{\rm GeV}]}}+1.9\,\%\,
\end{equation}
for photons,
and a time resolution of $\sigma\approx 370$\,ps measured relative to the tagging system. 
The 5-mm-thick scintillator plates in front of
the crystals discriminated charged against uncharged particles. The fast photomultiplier readout of the
TAPS BaF$_2$-crystals allowed the inclusion of its signals into the first-level trigger. TAPS was
divided into four trigger sectors (each 25\,\% of TAPS) and provided two types of first-level
trigger signals. The TAPS1-trigger was provided by at least one trigger sector showing a trigger
signal above 80~MeV. The TAPS2-trigger indicated that at least two trigger sectors had fired with
an energy deposit above 80~MeV; in this case, signals in the two innermost rings were excluded
since they suffered from a large $e^+e^-$~background. In the case of the TAPS2 trigger, no further
hits in the other calorimeters were required, in contrast to the TAPS1 trigger.

\subsection{The Cherenkov detector}
A CO$_2$ gas Cherenkov detector was used to identify electromagnetic background. It consisted of 
a parabolic \mbox{mirror}, which focussed the Cherenkov light onto a single photomultiplier.  With an 
energy threshold of $E_{\rm thres}=17.4$~MeV for electrons and $E_{\rm thres}=4.76$~GeV for
charged pions, it was well suited to suppress pair production and Compton events at the trigger
level without affecting charged hadrons.

\subsection{The gamma intensity monitor and flux monitor}
At the end of the beamline, directly in front of the photon beam dump, the gamma intensity monitor
(GIM) consisted of 16~PbF$_2$ crystals. Photons impinging on the GIM induced an electromagnetic
shower whose Cherenkov light signals were read out by photomultipliers. For photon rates
$\gg$\,1\,MHz, the GIM efficiency decreased due to deadtime effects. Therefore, a second detector
was placed in front of the GIM. This flux monitor consisted of a 100-$\mu$m-thick Pb foil used as a
thin conversion target followed by two organic scintillators to detect the $e^+e^-$-pairs. A third
scintillator placed in front of the Pb foil was used for the suppression of charged-particle
background. Calibrated relative to the GIM at low photon rates, the flux monitor counted only a
well known fraction of the total photon flux. It was also used to determine the rate-dependent GIM
detection efficiency.

\subsection{The trigger}
Trigger conditions were imposed at two levels. The first-level trigger required a valid hit in the
tagger not vetoed by the Cherenkov counter. This condition was combined with specific first- and
second-level configurations for at least two detected particles in the calorimeters to enhance the
fraction of useful data on disk. At least two hits (above the respective trigger threshold) in any
combination of the Forward Detector and TAPS hits triggered the readout. If there was only one hit
in these two detector components (or no hit, combined with a hit in the inner detector), the
second-level trigger FACE had to identify at least one cluster (or at least two clusters) in the
Crystal Barrel Detector. These trigger conditions covered all possible combinations  for having at
least two clusters somewhere in the calorimeters.

\section{\label{Pol}\boldmath The polarized beam and target }
\subsection{\label{Pol-gamma}\boldmath Photon polarization}
\paragraph{\boldmath The source of polarized electrons: }
Polarized electrons were produced via photoemission from a GaAs/GaAsP strained-layer superlattice
photocathode. By optical pumping using circularly-polarized laser light, electrons in a specific
spin state were transferred from the valence to the conduction band. Due to the energy gap between
the conduction band and the vacuum (electron affinity (EA) of typically 5.2~eV), these electrons
could not escape to the vacuum without further measures. In order to lower the EA, a 5-nm-thick
surface layer of the crystal was heavily p-doped ($5 \times 10^{19}/{\rm cm}^3$). A further
reduction of the EA to negative values was achieved by depositing a monolayer of cesium and oxygen.
This Cs-O layer caused the vacuum level to fall below the conduction band, and the electrons could
tunnel through the remaining thin potential barrier. Using a flashlamp-pumped Ti:Sa laser,
microsecond-long electron pulses with a polarization above 80\,\% and a pulse current of 120\,mA
were generated. The electrons underwent a first acceleration in the source and were transferred
first to a linear accelerator (Linac), guided through magnetic deflection and focusing fields. An
additional electrostatic deflector turned the electron polarization into transversal direction. The
electrons were pre-accelerated in the Linac and the following Booster Synchrotron, injected into
the ELSA ring, post-accelerated to their final energy, and then slowly extracted for the experiment.
The polarization of the laser light, and hence the polarization of the emitted electrons, was
changed for every filling-accelerating-extracting cycle. Maintaining the electron polarization
during the acceleration process was a difficult task. Depolarizing resonances were compensated by
carefully centering the beam in the quadrupoles, by applying harmonic corrections of
resonance-driving magnetic fields, and by fast changes of the accelerator optics (tune jumping) at
specific energies. After the slow extraction, the spin was rotated in a solenoid and two bending
magnets into the longitudinal direction. At an electron energy of 2.335~GeV, a longitudinal
electron polarization at the radiator target of 64\,\% (51\,\%) was achieved in the 2009 (2007)
beam time.

\paragraph{\boldmath Circularly-polarized photons: }
The longitudinally-polari\-zed electrons impinged on a target and produced bremsstrahlung. The 
polarization of the electron beam was partly transferred to the photon beam yielding a photon polarization
\cite{Olsen:1959zz} of
\begin{equation}
\label{eqn_olsen}
P_{\odot}=\frac{4x-x^2}{4-4x+3x^2}P_{\rm
e^-}\quad\mbox{with~}x=\frac{E_{\gamma}}{E_{e^{-}}}\,.
\end{equation}
The polarization had its maximum at the highest tagged-photon energies ($64\,\%$ at 2286~MeV for
the 2009 beam time) and then decreased toward lower energies ($19\,\%$ at 600~MeV).

\paragraph{The M\o ller Polarimeter: }
\label{moeller}
The M\o ller target was mounted inside of the vacuum tank of the goniometer. The polarimeter served
two purposes: It produced the brems\-strahlung and acted as a device to measure the longitudinal
polarization of the beam electrons at the radiator position by exploiting the spin dependence of
the electron-electron scattering process. The M\o ller target consisted of a $20\,\mu$m vacoflux
foil, an alloy of 49\,\% Fe, 49\,\% Co and 2\,\% V. It was oriented at an angle of $\pm 20^\circ$
relative to the photon beam and could be flipped from the $+ 20^\circ$ position to the $- 20^\circ$
position. A solenoidal magnetic field of $0.08$\,T saturated the magnetization of the alloy. 
An electron polarization of the M\o ller target of 
($8.163 \pm 0.067$)\,\% was reached~\cite{dipl_holger_eberhardt,Kammer:2009}. 

The M\o ller detector consisted of two lead-glass counters, which were placed in the forward
direction above and below the plane of the bremsstrahlung electrons. Electrons from M\o ller
scattering were separated from bremsstrahlung electrons by asking for two coincident hits;
accidental hits in both lead-glass counters were measured in parallel and subtracted. 
The foil was
flipped frequently between its $\pm 20^\circ$ positions to eliminate potential transverse
components of the beam polarization. The relative systematic uncertainty of the polarization
measurement was determined to be 3.3\,\%
including the systematic uncertainty of target foil
polarization, the simulated effective asymmetry coefficient of the  M\o{}ller detector setup as
well as of the analysis-related systematic uncertainties such as background contributions in the
M\o{}ller measurement.

\subsection{\label{Pol-proton}\boldmath Proton polarization}
The Bonn frozen-spin target~\cite{Bradtke:1999zg} used beads of butanol (C$_4$H$_9$OH) as the
target material and provided polarized protons. The protons of the hydrogen atoms within the
butanol - carbon and oxygen are spinless - were polarized using the method of dynamic nuclear
polarization (DNP). In the DNP process, the polarization of the free electrons in the target
material was achieved by doping the material with paramagnetic radicals, and then transferring 
the polarization 
to the nucleons using microwave irradiation. This was done in an external magnetic field of 2.5~T
at a temperature of $\sim$300~mK. Depending on the microwave frequency, inducing hyperfine
transitions, the spin could be aligned parallel or antiparallel to the magnetic field. For the
measurement with the Crystal Barrel detector, the polarizing magnet needed to be removed. During
the data-taking, the polarization was preserved by a very thin superconducting coil of 500\,$\mu$m
in thickness which provided a magnetic field of 0.6~T. At the same time, the temperature of the
target was lowered to less than 70~mK. This allowed continuous measurements of
several days with relaxation times for the polarization of $\sim$ 500\,h. During the measurements,
a mean polarization of +65\,\%\,/$-$\,71\,\% (2007) and +70\,\%\,/$-$\,74\,\% (2009) was reached.
Not considering the polarization direction, a mean polarization of 71\,\% with a systematic error
of 2\,\% was reached for the whole beam time.

\begin{figure}
\centering
\includegraphics[width=0.458\textwidth]{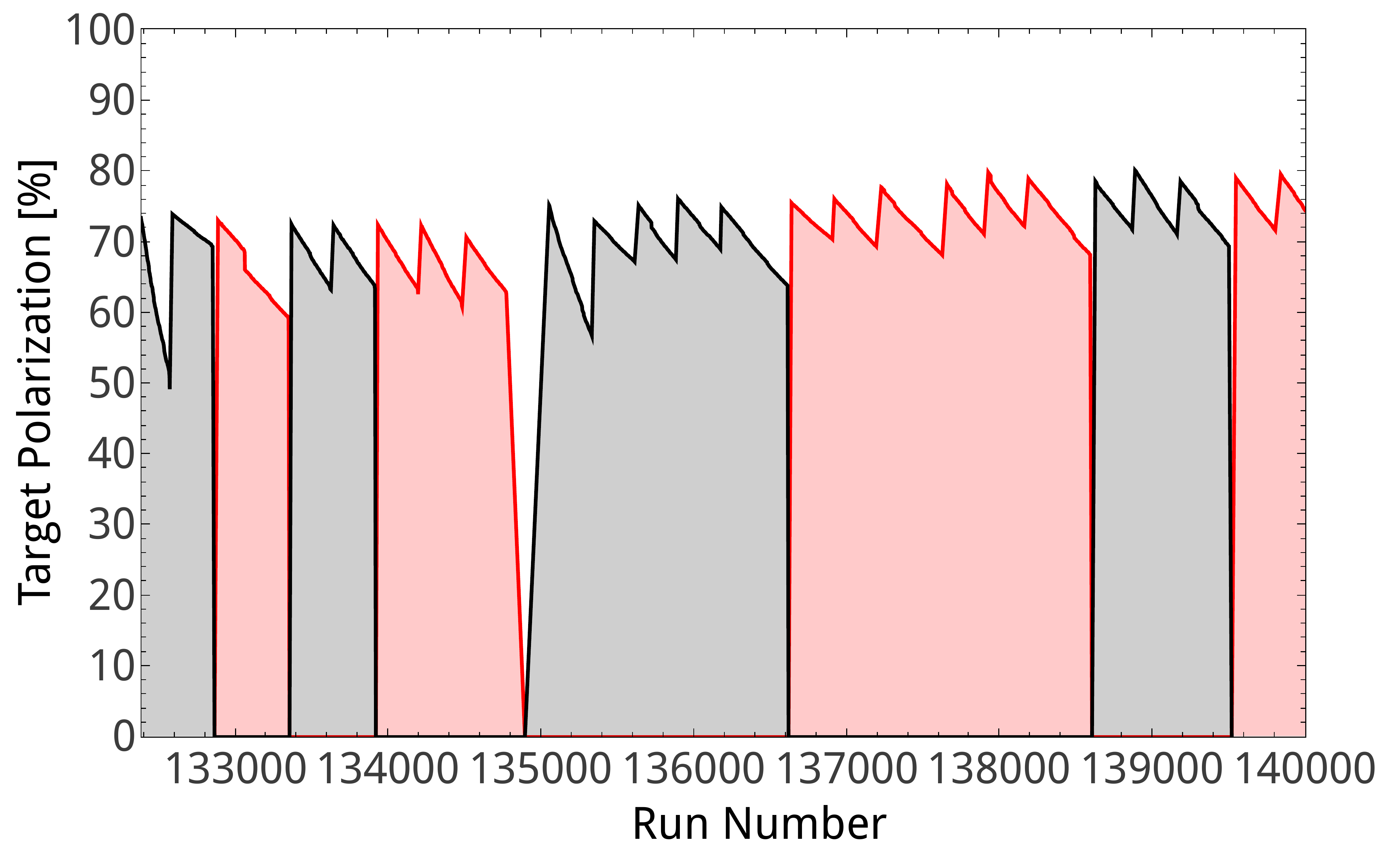}
\caption{The target polarization for the two beam times (Sept.~2009 and Nov.~2009) with positive (red) and negative (black)
polarization.}
\label{pic:targetpol}
\end{figure}

The direction of the polarization was regularly changed (Fig.~\ref{pic:targetpol}) either by flipping the
direction of the external magnetic field or by adjusting the microwave frequency.
Both methods were used to study systematic effects. The butanol target material was placed inside a
target container of 2~cm in length and 2~cm in diameter, which was immersed in liquid~He.

The target material could be replaced by a carbon foam target inside the cryostat which had about
the same target area density as the carbon and oxygen components of the butanol target.
Measurements with the carbon foam target allowed us to determine the so-called dilution factor
which determined the fraction of the polarizable protons in the butanol target for
each energy and angular bin.

\section{\label{Calib}\boldmath Calibration and Event reconstruction}
For the final analysis, precise knowledge of the four-vectors of the different
particles belonging to a certain event was needed.
In a first step, the digital raw information stored in
the analog-to-digital (ADCs) and time-to-digital converters (TDCs) of the detector systems
had to be translated into energy, position and timing information of the final-state particles.
In the following, the time and then the energy calibration of the different detector systems
will be discussed before sections on particle reconstruction follow.

\subsection{\label{sec:calib_time}Time Calibration}
The time calibration ensured that hits occurring at the same time in the experiment were also
reconstructed at the same event time (the time of flight for photons was set to zero for all
detector systems). The time calibration proceeded in several steps. First, all values of the
time-to-digital converters (TDCs) were transformed to relative times by a TDC-specific factor and
the prompt peak in the TDC spectrum, defining the time at which the trigger occurred, was moved to
$t=0$. In the Crystal Barrel / TAPS setup, the time provided by the tagger (96 scintillator bars) always
defined the trigger time. Due to varying signal running times (cable lengths, electronics) of the
different tagger channels, the trigger time was smeared out. Therefore, the above step provided only a
first rough calibration.

\begin{figure}[htp]
 \centering
\begin{tabular}{cc}
\hspace*{-0.35cm}
\includegraphics[width=0.2475\textwidth]{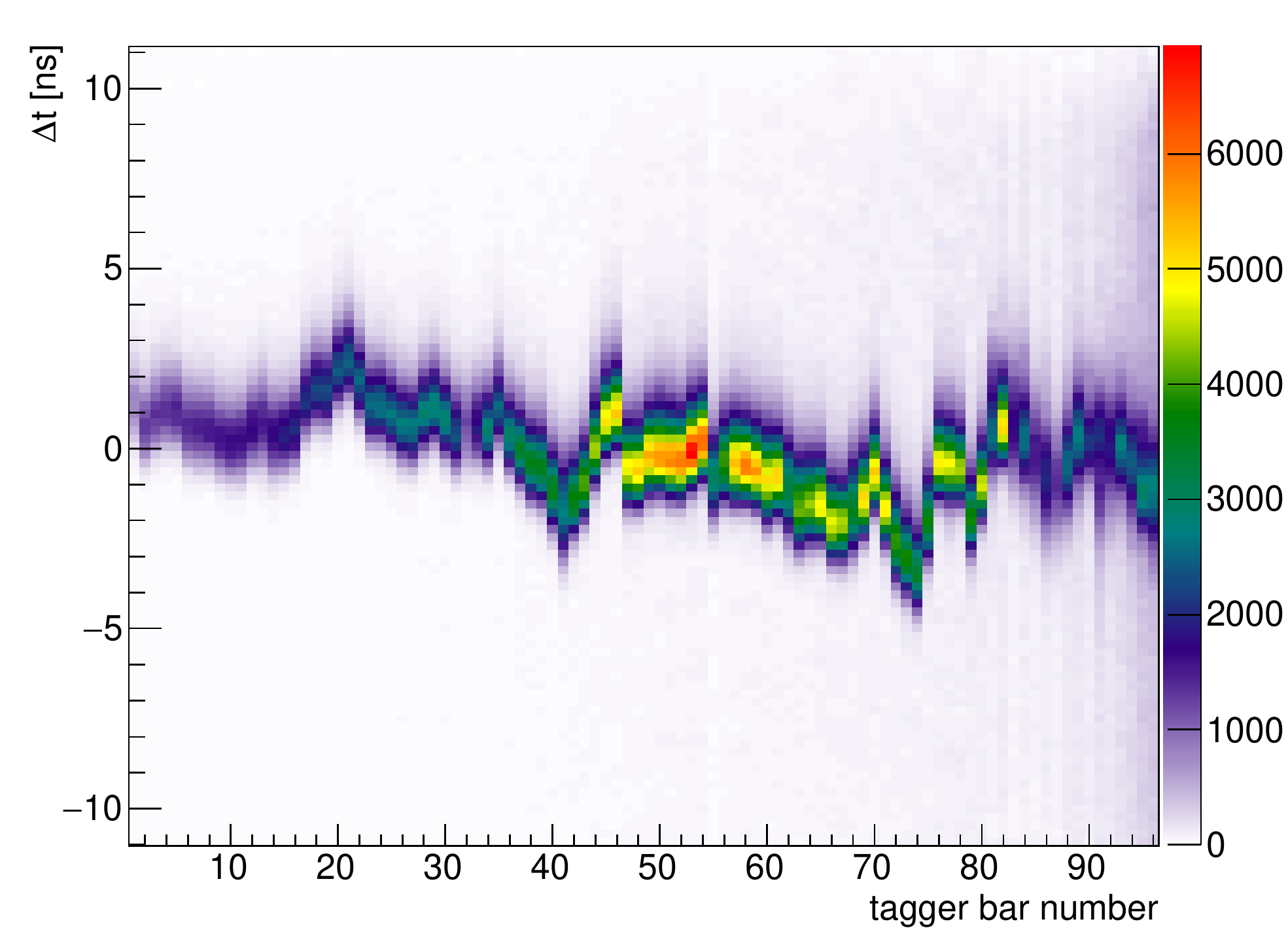} & \hspace*{-0.5cm}
\includegraphics[width=0.2475\textwidth]{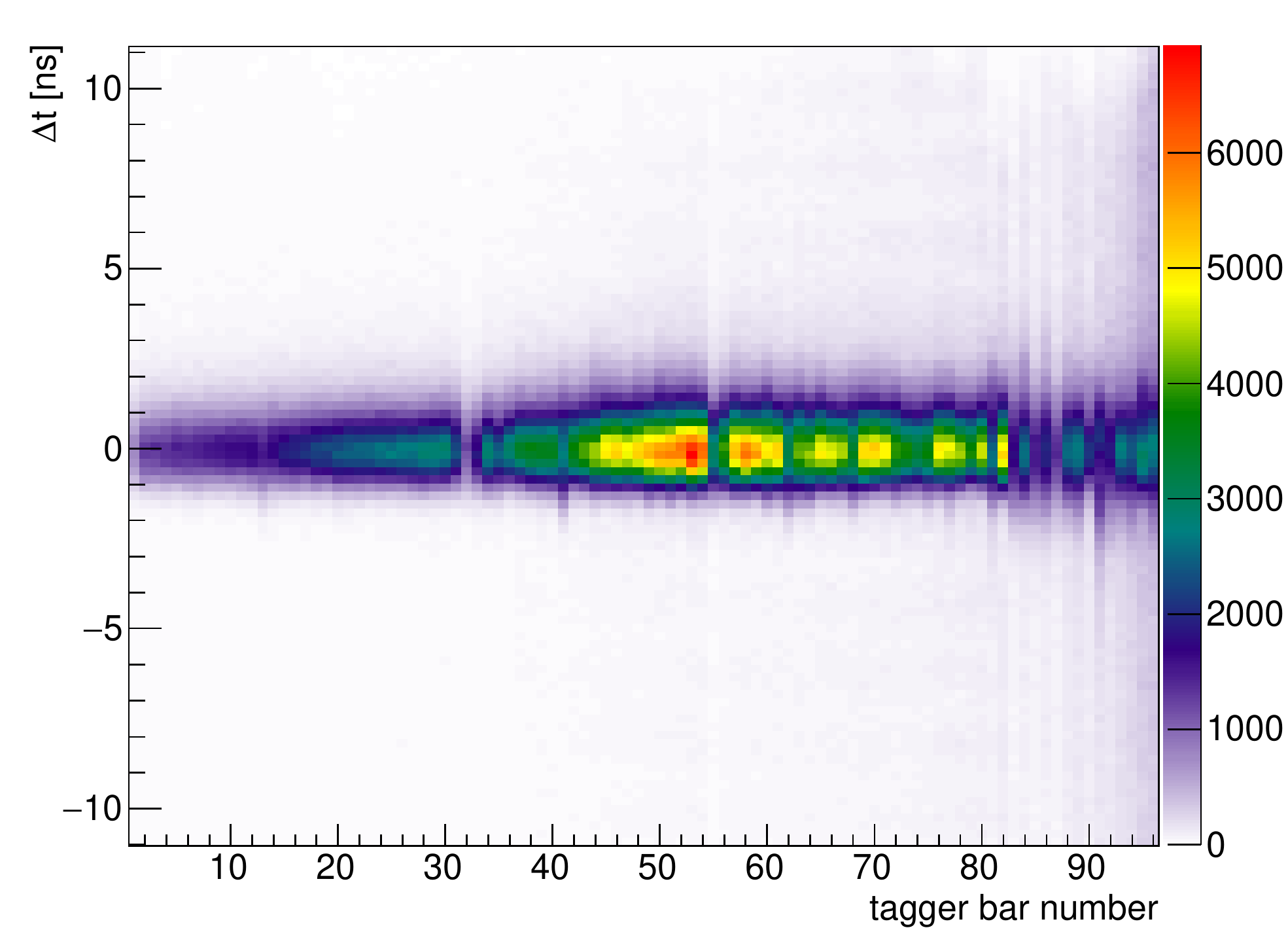}
\end{tabular}
 \caption{Time difference of the tagger bars relative to the Cherenkov detector
before (left) and after (right) the second calibration step~\cite{Hartmann:2008}.}
 \label{fig:finecalibration}
\end{figure}
Time differences of various detector systems 
on the other hand are independent
of the trigger time: $(t_1-t_{\rm trigger})-(t_2-t_{\rm trigger})=t_1-t_2$. Therefore, in a
second step, the tagger bars and fibers were calibrated relative
to the single-channel Cherenkov detector. The improvement of the tagger calibration
is shown in Fig.\ref{fig:finecalibration}.

The achievable time resolution was limited by the time-resolution of the Cherenkov detector
(FWHM$\,\gtrsim\,$1~ns). Especially for the tagger bars and the BaF$_2$ crystals of the TAPS
detector, a significantly better time resolution could be achieved. Therefore, they were
cross-calibrated in an iterative procedure. Subsequently, all other detectors were calibrated using
the calibrated tagger bars as a reference detector. In the last step, the time walk of the
leading-edge discriminators of the Forward Plug (FP) was removed by an  empirical energy-dependent
correction function. In contrast to the FP, TAPS used constant-fraction discriminators, which made
such a correction unnecessary.

\begin{figure}[h!]
 \centering
 \includegraphics[width=0.44\textwidth]{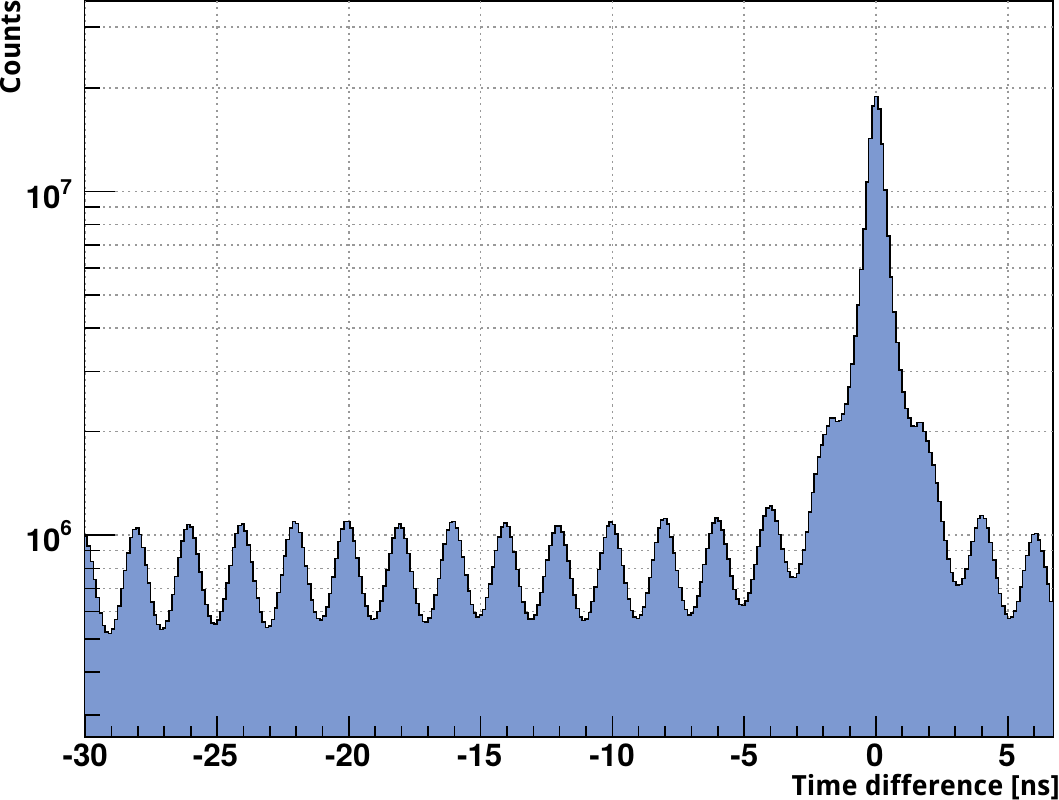}
 \caption{The time difference between two tagger bars after the calibration for all possible combinations.
The periodic structure arises due to the 500~MHz bunch structure of the ELSA accelerator~\cite{Hartmann:2008}.}
 \label{fig:timediff}
\end{figure}

Measured relative to the tagger bars, the following resolution values were reached:  
FWHM = $0.635 \pm 0.003$~ns (tagger bars), 
$1.694\pm0.06$~ns (tagger fibers),
$0.872 \pm 0.006$~ns (TAPS BaF$_2$ crystals), 
$3.06\pm0.05$~ns (TAPS plastic scintillators) 
$1.861 \pm 0.016$~ns (Forward Plug crystals), 
$4.434 \pm 0.013$~ns (plastic scintillators of the Forward Plug), 
and $2.093\pm0.013$~ns (inner detector fibers)~\cite{Hartmann:2008}. The quality of the final
calibration can also be seen in Fig.~\ref{fig:timediff}, where the 500~MHz bunch structure of the
ELSA accelerator is clearly observed.

\subsection{Energy calibration of the Tagging System}
 \label{sec:calib_tagger}
During the regular data taking, the photon energy was calculated using the known ELSA electron beam
energy, $E_0$, and the energy of the bremsstrahlung electron, $E_{e}$. The latter was determined
by measuring the deflection angle of the electron in the magnetic field, $B_0$, using the
scintillating bars and fibers of the tagging system (see Section~\ref{tagging_system}). For the
calibration, electrons with known beam energies $E^{\rm beam}_{\rm calib}$ -- smaller than during
the regular data taking -- were directly injected into the tagging system and their deflection
angle was measured by the hit in a certain scintillator. By varying the magnetic field strength
during the calibration measurements, the corresponding hit scintillator bars/fibers ($S_i$) were
identified and using
$$\frac{E_{e}}{B_0}\,(S_i) \,=\, \frac{E^{\rm beam}_{\rm calib}}{B^{\rm calib}_i}\,(S_i)\,,$$
the electron energy for a certain scintillator $i$ at the nominal field $B_0$ during the data taking 
could be determined. 
The resulting $B(S)$ correspondence was fit with a polynomial. Respective measurements were done
for several different electron energies to study the systematics.
The $E_{\rm e}(S)$~dependence was extended to lower $E_{\rm e}(S)$ via simulation for the area
not reachable via the direct calibration procedure.
The energy resolution of the tagging system ranged from $\Delta E_{\rm e} \approx 0.5\,\% \,E_0$
at the highest electron (lowest photon) energies to $\Delta E_{\rm
  e} \approx 0.1\,\% \,E_0$ at the lowest
electron (highest photon) energies. The systematic uncertainty is of the order of half a  
fiber within the energy regime
of the hardware calibration procedure, resulting
in 4.2~MeV at 600~MeV and 0.9~MeV at 1750~MeV for an incoming electron beam energy of
2.335~GeV, for instance.

\subsection{\label{sec:calib_energy}Energy Calibration of the Calorimeters}
\paragraph{Crystal Barrel Calorimeter: }
12-bit dual-range fastbus \\charge-to-digital converters (QDCs) were used to read out the Crystal
Barrel calorimeter. To reach maximal resolution for low energetic photons, the signal was split
with a ratio of 1:8 (high-range : low-range signal). This allowed the use of the whole QDC range
for signals at low energies. At an energy of about 130~MeV for the Crystal Barrel and 260~MeV for
the Forward Plug, the QDC switched from the low into the high range. To ensure a stable
high-range/low-range factor and to determine this factor precisely,
a light pulser was used. During dedicated light-pulser runs, pulsed signals of known relative
intensity were sent via light guides to each of the crystal modules. The light guide emitted its
light into the wave-length shifter behind the crystal, which was used to collect the light emitted
by the crystal. These signals were recorded by the QDCs and were later used for the high/low-range
calibration.

The Crystal Barrel shaper modules, placed in the signal path before the QDCs, allowed the
adjustment of the pedestal via an offset current. The pedestals of all crystals were adjusted to
one fixed value. Pedestal runs, which were performed at the start of each data-taking run,
determined the individual pedestal value for each of the crystals. To suppress empty channels
containing noise, only crystals with an ADC-entry 10 channels above this pedestal were written to
tape during the data taking. For all other channels, the pedestal value was subtracted from the
measured ADC value as a first step in the analysis.

\begin{figure}[htp]
 \centering
\hspace*{-0.2cm}\includegraphics[width=0.49\textwidth,angle=0]{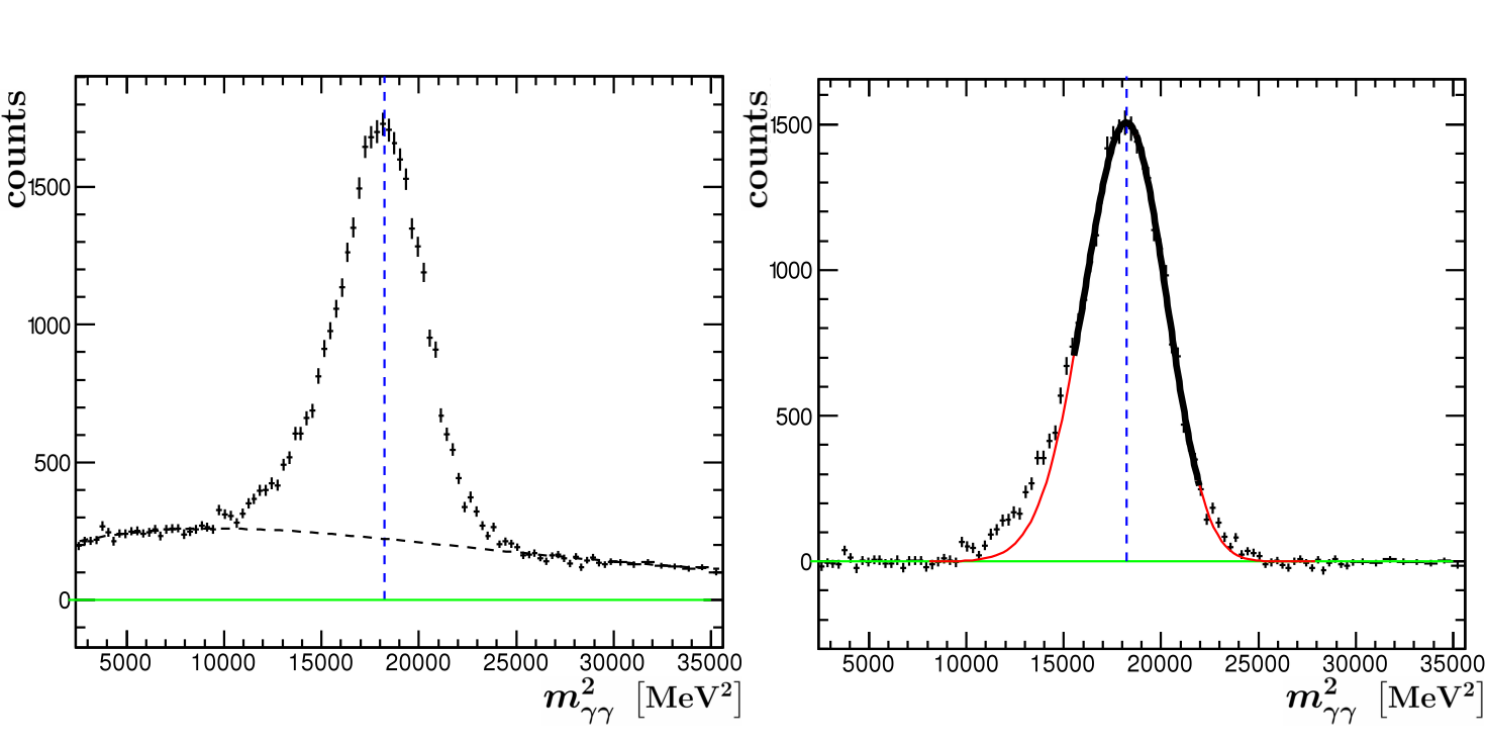}
\hspace*{-0.2cm} \includegraphics[width=0.495\textwidth]{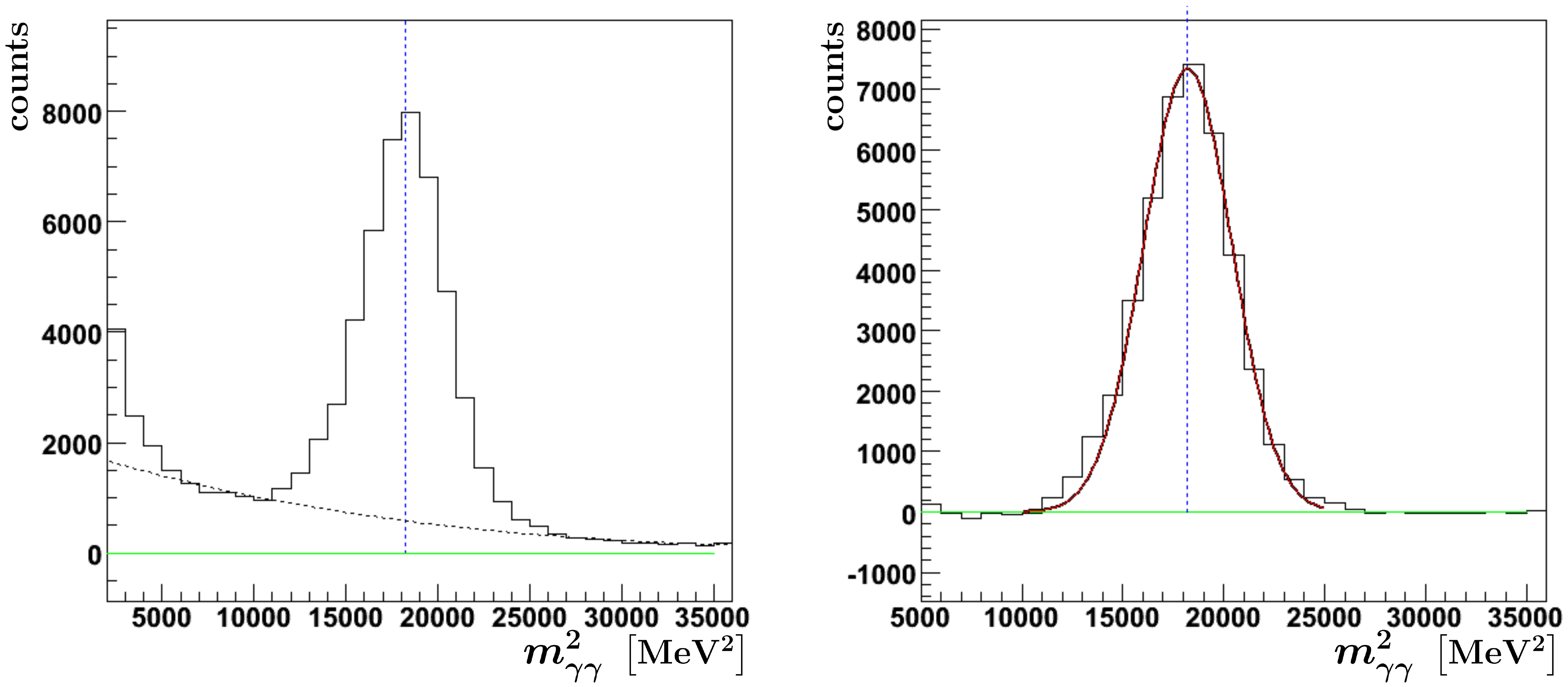}
 \caption{Examples of fits for selected crystals of the Crystal Barrel detector (top) and the TAPS detector (bottom).
Each squared invariant mass is fitted with a Chebyshev polynomial of fifth order outside of the
peak region to determine the background contribution (dashed line drawn over the entire mass
range). The background function is then subtracted from the spectrum and the remaining peak is
fitted by a Novosibirsk function\cite{Ikeda:1999aq} (red line, fit range: black line). For details
on the calibration method see~\cite{Diss_Jonas,Gottschall:2013}. }
 \label{fig:energy_calibration}
\end{figure}

As a starting value for the calibration of each crystal, an overall low-range calibration factor of
$c=0.033$~MeV per channel for the Crystal Barrel and of $c=0.061$~MeV per channel for the Forward
Plug (or simply the values from the preceding beam time) were used. For the detailed calibration of
the detector system, the well known $\pi^0$~mass was used ($\pi^0$~calibration), based on data
selected as follows: All events with two neutral and not more than one charged hit in the detector
system were initially selected using only events where the calorimeter hits were in the ADC low
range. For an optimal energy and angle determination, only signals due to non-overlapping photon
showers were used. The invariant mass of the two neutral hits showed a strong signal from the decay
$\pi^0\to \gamma \gamma$ (see Fig.~\ref{fig:energy_calibration}). Within the calibration procedure,
the squared invariant $\gamma\gamma$-mass was plotted separately for each crystal, which served as
the central crystal\,\footnote{The crystal with the highest energy deposit in the cluster.}. The
spectrum was then fitted outside the peak region with a Chebyshev polynomial of fifth order to
determine the background contribution. This background function was then subtracted from the
spectrum and the remaining peak was fitted with a Novosibirsk function~\cite{Ikeda:1999aq}. Based
on the fitted mass a gain correction factor was determined for the selected central crystal. The
procedure was repeated iteratively until the masses of the $\pi^0$~signal in the spectra of all
crystals were stable at the nominal pion mass (deviations of $\leq$\,100~keV were tolerated).
The high range was then calibrated by using the light-pulser system. Plotting the ADC channel
versus the transmission (intensity) $T$ of the light-pulser system allowed two straight line fits
($a_{\rm low}(T)$, $a_{\rm high}(T)$) and a conversion of the high range values into the low range
system by
\begin{equation}
a_{\rm low} \,=\, (a_{\rm high}\,-\,a_{{\rm high},\,x=0})\cdot g_{LP} \,+\, a_{{\rm low},\,x=0}\,,
\end{equation}
\begin{equation*}
g_{LP} \,=\, m_{\rm low}\,/\,m_{\rm high}\,,
\end{equation*}
where $a_{{\rm low},\,x=0}~ (a_{{\rm high},\,x=0})$  and $m_{\rm low}~(m_{\rm high})$
denote the $y$-intercept and the linear slope of the low (high) range line, respectively.

\paragraph{TAPS Calorimeter: }
Before each beam time, the TAPS detector was pre-calibrated to an energy deposit of \mbox{$38$~MeV}
using cosmic muons traversing 6~cm in a $\rm BaF_2$ crystal of the TAPS detector. The response of
each TAPS crystal to the cosmic radiation did not only provide a first calibration factor for each
crystal but also allowed a cross check of the constant fraction discriminator (CFD) threshold of
the respective crystal. Only energy deposits passing the CFD threshold of the crystal were written
to tape.

The positions of the pedestals for all crystals were determined regularly during the beam time
since they showed some variations, e.g.\ with temperature. This was done repeatedly during the runs
using a pulser-signal as trigger. Furthermore, based on the cosmic calibration and the pedestals
determined for each data run, the TAPS data were also energy-calibrated using the
$\pi^0$-calibration method described above. Due to the limited angular coverage of the TAPS
detector, this step was performed after the calibration of the Crystal Barrel calorimeter using
events with one photon in TAPS
and one photon in the Crystal Barrel calorimeter.

\subsection{\label{sec:beam-photon}Beam photon reconstruction}
The incoming electrons, which produced bremsstrahlung on the radiator target, were deflected by the
tagger dipole magnet. Their deflection angle was measured by the known position of scintillating
bars and fibers (see Section~\ref{tagging_system}), providing precise energy and time information
for all tagged beam photons. Each electron hitting the tagger hodoscope left signals in up to
three bars depending on the exact impact point of the electron. To reconstruct only one beam photon
for each electron, all hits in neighboring bars, which occurred within a time interval of not more
than $t_\text{diff} < 4$~ns, were combined into a single hit. Only hits affecting at least two
scintillators,
were retained to suppress background.
Separately, the same was done for the scintillating fibers. Here, a missing fiber between two hits was allowed
to account for inefficient fibers. Due to the worse time resolution, the time difference within a cluster
of fibers was required to be below $t_\text{diff} < 7$~ns.
For the high energetic electrons, the scintillating bars and fibers overlapped. In this region, a coincident
hit of the scintillating bars and the fibers was required with a mean time difference between the two clusters
of less than $t_\text{diff} < 4$~ns.

For the reconstructed beam photon, the mean time of the bar cluster or, in the overlap region, of
bars and fibers, weighted by the respective time resolution, was used. The exact deflection angle
was given by the mean position of the fibers or just by the scintillating bars, where no fibers
were available. The energy of the beam photon was then determined by the tagging polynomial, as
described in Section~\ref{sec:calib_tagger}.

\subsection{Reconstruction of final-state photons}\label{sec:reconstruction}
A photon, which interacted with the crystals of the calorimeters, created an electromagnetic shower which
spread across the neighboring crystals. The Moli\`{e}re radius was 3.57~cm and 3.10~cm for the
CsI(Tl) and BaF$_2$ crystals, respectively. The clusters of energy entries 
needed to be combined to finally determine the four-vector of the photon. A single-particle cluster
was called Particle Energy Deposit (PED).

\paragraph{Crystal Barrel Calorimeter: }
In a first reconstruction step, the cluster algorithm was used to identify contiguous groups of
crystals with energies above a single-crystal threshold of 1~MeV (well above noise level).
In addition, a central crystal and PED threshold of 20~MeV was used to reduce split-offs.
In the Crystal Barrel this ensured the reconstruction of clusters
above the FACE-trigger threshold of about 15~MeV.
Split-offs, which were identified as separate local minima, were produced by shower fluctuations and could
mimic separate PEDs.

The direction of the incident particle (photon) can be reconstructed by an energy-weighted sum of the $\theta$
and $\phi$ coordinates of the crystals within the PED:
\begin{equation}
\label{eqn-w1}
\theta_{PED}=\frac{\sum_{i} w_i\theta_i}{\sum_i  w_i}\,, \qquad \phi_{PED}=\frac{\sum_{i} w_i\phi_i}{\sum_i  w_i}\,,
\end{equation}
\begin{equation}
\label{eqn-w2}
w_i = \max \left( 0; W_0 + \ln \frac{E_i}{\sum_{i} E_i} \right)\,,
\end{equation}
where the logarithmic weighting accounted for the exponen\-tially-declining transversal shower
profile. This resulted in a larger weight for the outer crystals and in an improved angular
resolution. The cut-off value $W_0$ defined the lowest energy fraction for a crystal, which was
used for the angle reconstruction. This value was optimized using Monte Carlo simulations
($W_0=4.25$). The four-momentum of a photon was then determined by assuming that it originated from
the target center, and moved in the $\theta,\,\phi$-direction of the electromagnetic shower.

So far only clusters produced by a single particle, resulting in a single-PED cluster, were
discussed. However, energy deposits of different particles could also overlap, still producing a
single connected cluster of crystals with energy deposits above the single-crystal threshold.
Therefore, all clusters were scanned for local maxima above 20~MeV. If more than one maximum was
found in a cluster, the cluster was considered a multi-PED cluster. For the angle reconstruction in
these overlapping clusters, only the crystal with the local maximum in energy and its direct
neighbors were used in Eqn.~(\ref{eqn-w1},\ref{eqn-w2}). Furthermore, the crystal energies $E_i$ were corrected
with the calculated energy overlap expected.
Here, the contribution of the second PED, calculated in terms of
the distance of the crystal to the shower center of the second PED in units of
the Moli\`ere radius, $d_{i\leftrightarrow\text{max}}/R_M$, was subtracted:
\begin{equation}
 E_i^\text{corr} = E_i - E_\text{max}\cdot e^{-d_{i\leftrightarrow\text{max}}/R_M}\,.
\end{equation}
The ratio of the PED central energies determined how the total cluster energy
was distributed among the two overlapping PEDs. Multi-PED clusters with more
than two local maxima were very rare and neglected in the analysis.\\
In addition, a correction function for the PEDs was used: The segmentation of the detector, important
for the angular resolution, as well as other support structures
of the detector introduced also dead material into the detector. Part of the energy of the shower was therefore
not measured. This loss depended on the energy as well as on the direction of the particle.
To correct for these losses an energy correction function was determined from simulations and was
applied to each PED.

\paragraph{TAPS Calorimeter: }
The reconstruction of the TAPS calori\-meter was very similar to the one of the Crystal Barrel calorimeter,
with three distinct differences: (1) TAPS provided not only an energy but also a time signal for each crystal.
In the reconstruction, all crystals that were considered belonging to
the same cluster were required to fulfill $t_\mathrm{diff} < 5$~ns.

The TAPS detector was located in the forward direction and for this reason, its crystals
detected much more low-energetic background from the beam halo. (2) To prevent background events from
reaching the read-out chain, the thresholds of the TAPS constant-fraction discriminators were
set higher than the thresholds for the Crystal Barrel calorimeter. The two inner rings, which
were closest to the primary photon beam, had thresholds of 17~MeV, while the outer-ring thresholds
were set to 13~MeV. In addition, the central crystal of a cluster was required to have an energy of
more than 20~MeV, and the total energy of a cluster had to be above 25~MeV. (3) The crystal axes of
the TAPS detector were not oriented toward the target direction. Hence, an additional position-
and energy-dependent correction factor was necessary. These were part of the energy-correction
function developed for TAPS, again based on Monte-Carlo simulations.
%
%

\subsection{Identification of charged particles}\label{matching}
Three different detectors served to identify a charged particle: (1) The inner scintillating-fiber
detector covering polar angles of $167^\circ > \theta > 21^\circ$, (2) the plastic scintillators in
front of the forward detector for angles $27.5^\circ > \theta > 11.2^\circ$, and (3) the plastic
scintillators in front of the TAPS crystals, closing the polar-angle gap down to $1^\circ$.

\paragraph{Inner scintillating fiber Detector: }
The two inner layers were rotated with respect to the beam axis
(see Section~\ref{sec:chapi}) while the outer layer had straight fibers (parallel to the beamline).
Signals in two of the three layers were hence sufficient to define a hit and its position.
First, the hits were clustered layer-wise. Between two fibers, a single missing fiber was allowed; The time
difference of the signals could not exceed 14~ns. The hits in the respective layers were then
combined to two- or three-layer hits if the time differences were again within 14~ns. The time and
position of the hits were given by the mean time and the mean position of all involved fibers. The polar
and azimuthal angle of the resulting crossing point was then calculated and the four-vector
determined under the assumption that the particle originated from the target center.

A cluster in the Crystal Barrel calorimeter was identified as a charged-particle cluster if the angle
between the trajectory from the target center to the cluster and of the trajectory from the target center
to the inner detector hit was less than $12^\circ$ for $|\Delta\phi|$ and for $|\Delta\theta|$.
If a calorimeter time was available in the area of overlap between the inner scintillating fiber
detector and the Forward Plug, an additional time cut of $\Delta t \leq 15$~ns was applied.

\paragraph{Plastic Scintillators of the Forward Plug: }
Two layers of plastic scintillators were placed in front of the forward part of the Crystal Barrel
detector. Each plastic scintillator covered $\Delta\phi=12^\circ,\,\Delta\theta=6^\circ$. The
second layer was turned by $6^\circ$ with respect to the first layer and doubled the angular
resolution to $\Delta\phi=6^\circ$. Hits in both layers of scintillators within 20\,ns were
required to define a charged particle. The center of the overlapping scintillators was used to
determine the polar and azimuthal angle. The particles' four-vectors were calculated assuming that
they originated from the target center.

A forward-detector hit was marked as a charged particle, if the azimuthal angle difference between
the trajectory from the target center to the cluster and to the scintillator hit was less than
14$^{\circ}$, the polar-angle difference was less than 10$^{\circ}$,
and the time difference between the hits in both layers was less than 20~ns.

\paragraph{Plastic Scintillators of the TAPS Detector: }
A plastic scintillator was also located in front of each crystal of the TAPS detector. Hits in
these scintillators were clustered based on the time and the location of the detector. However,
since the plastic scintillators did not overlap, hits in multiple neighboring detectors were rather
rare.

Since the TAPS detector was a planar detector, its crystals were not oriented toward the target center,
and trajectories, first penetrating a neighboring scintillating plate, were possible.
Therefore, a particle in the calorimeter was identified as a charged particle if the distance between the two trajectories
of the calorimeter cluster and the reconstructed scintillator
at the surface of TAPS was less than 6.5~cm
and $|\Delta t| \leq 15$~ns.

\subsection{\label{Selec}\boldmath Event selection}
The reaction $\gamma p\to p\pi^0\to p\gamma\gamma$ was reconstructed by a series of kinematic cuts.
Events with exactly two neutral and one charged particle were used. The charged particle, always
considered to be a proton, was required in order to suppress background from reactions off neutrons. Two
classes of events were considered in the analysis, 3-PED events as well as 2-PED events with a
charged particle identified in the inner detector.
In the case of 3-PED events, all events with three clusters in the calorimeters were considered for
further analysis. Exactly one of them had to be identified as a charged particle (see
Section~\ref{matching}). To include also events with low-energy protons, which did not produce a
cluster in the Crystal Barrel calorimeter above the energy threshold of 20~MeV, two uncharged PEDs
in the calorimeters and a hit in the inner detector were considered as well. In the subsequent
analysis, the proton was treated as a missing particle for both, the 2-PED and the 3-PED events.
Its four-momentum was calculated from the known $\gamma p$ initial state and from the final-state
photons measured in the calorimeters.

\begin{figure}[ht]
\begin{center}
\includegraphics[width=0.425\textwidth]{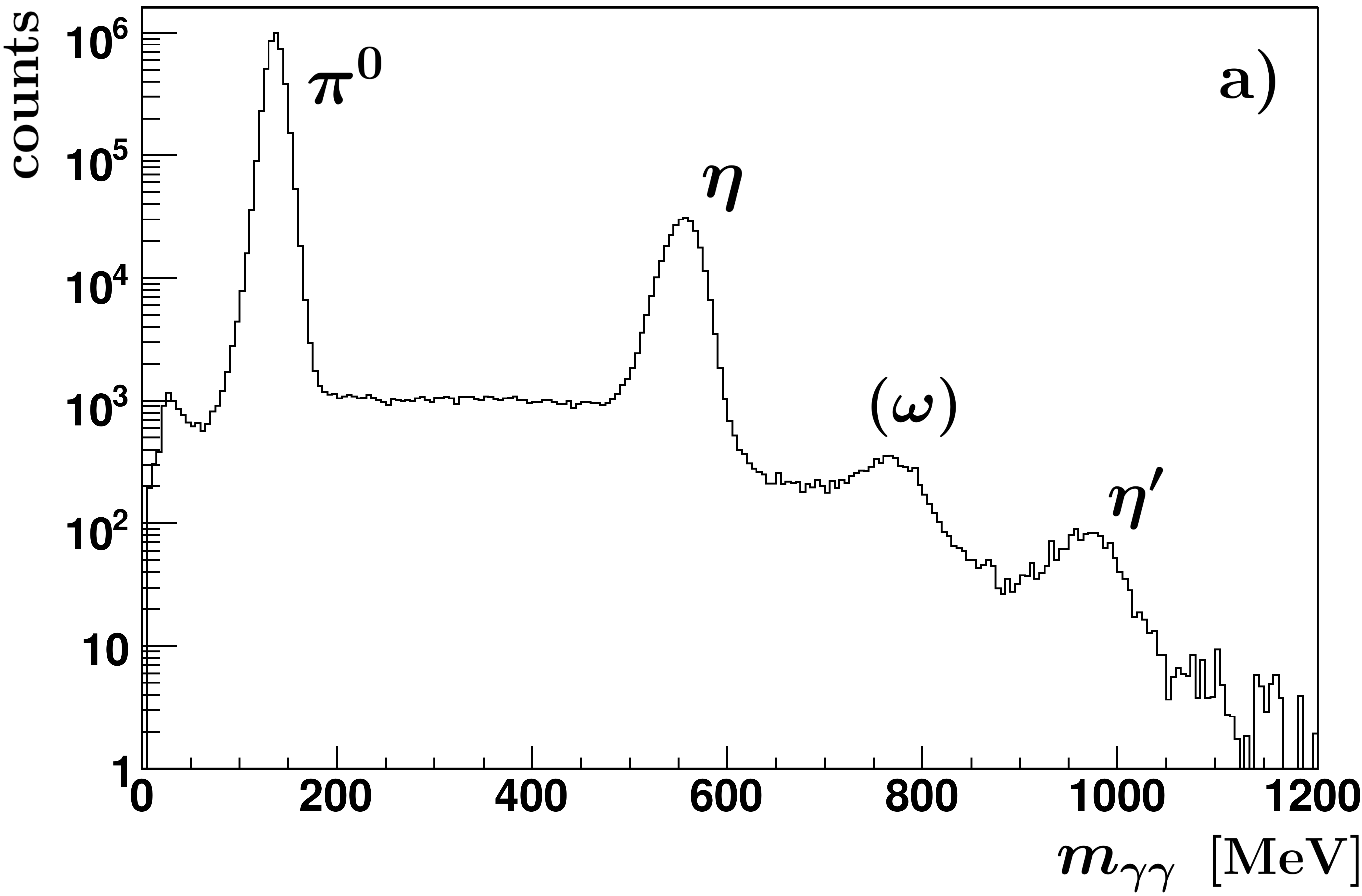}
\includegraphics[width=0.425\textwidth,angle=0]{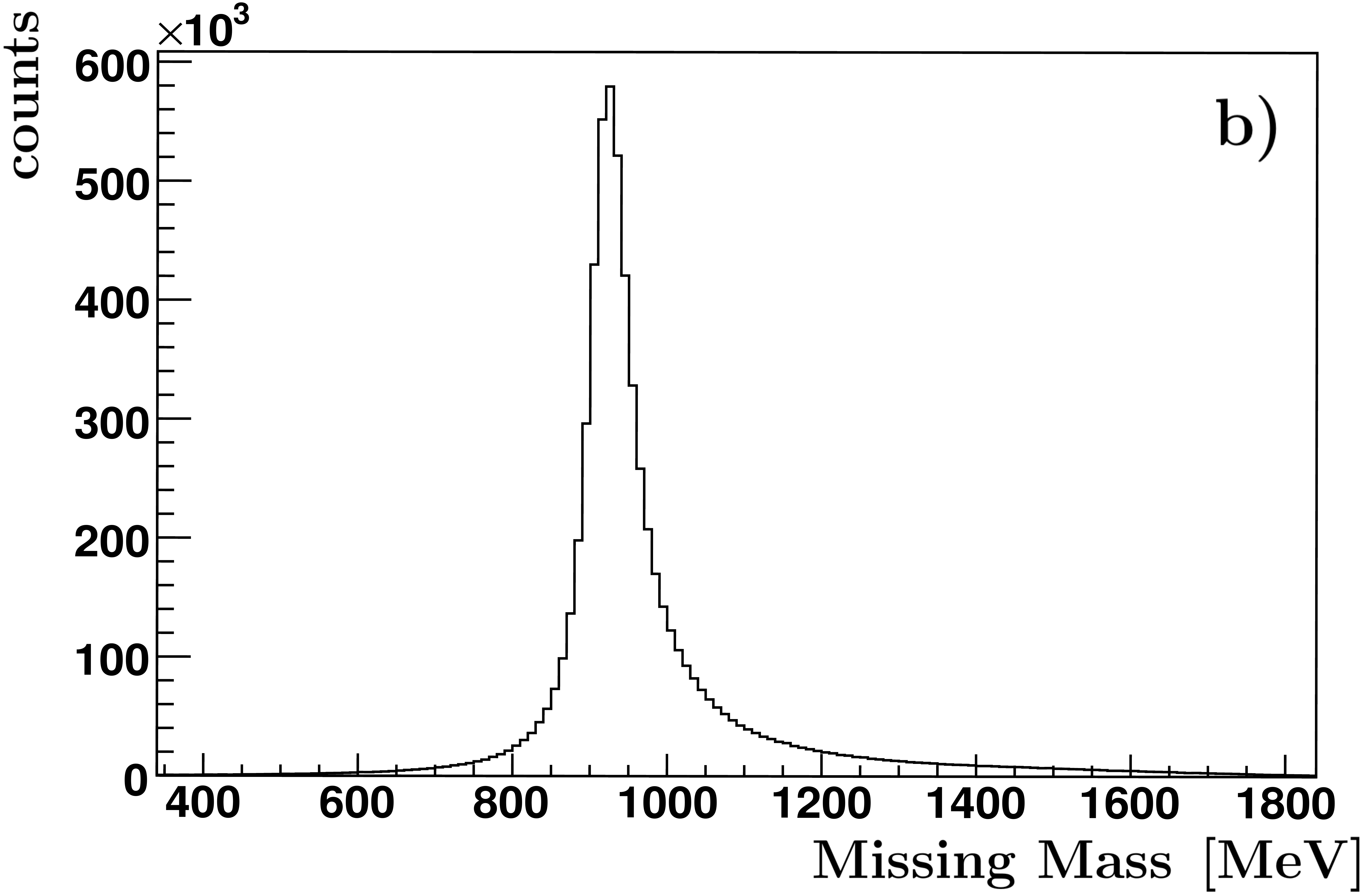}\\
\includegraphics[width=0.425\textwidth,angle=0]{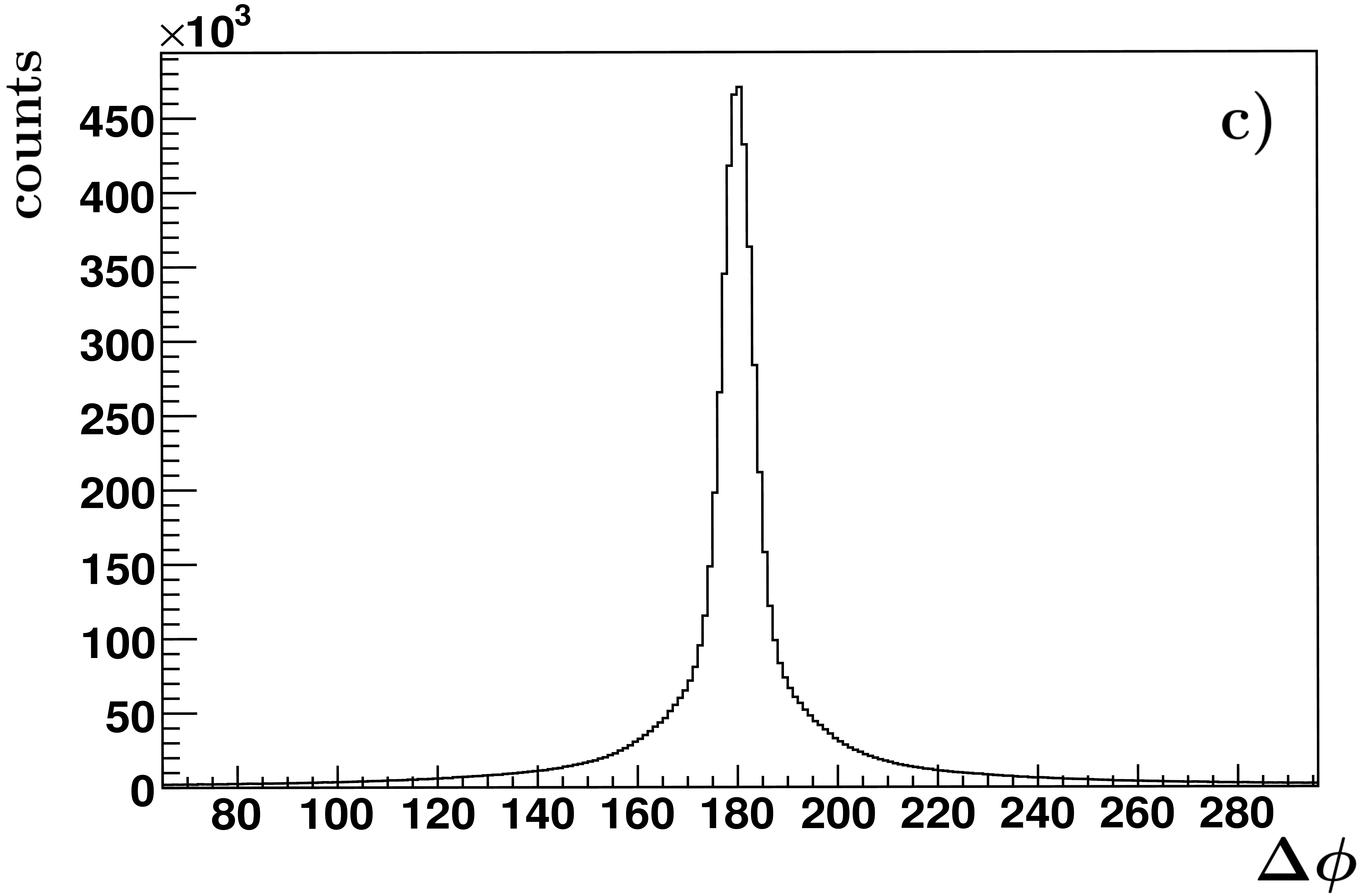}\\
\end{center}
 \caption{\label{pi0-MM-coplan} Invariant $\gamma\gamma$~mass (a), missing mass (b) and coplanarity (c) spectrum.
 All cuts except for the one on the respective spectrum have been applied and
  time background has been subtracted.
The $\omega$, eventhough decaying into $\pi^0\gamma\to 3\gamma$, is
visible in the invariant $\gamma\gamma$~mass
spectrum; here, either one low-energetic $\gamma$ remained undetected or two photons
merged into one cluster accidentally interpreted as one photon. 
}
\end{figure}

In a next step, a $\pm2\sigma$ cut on the invariant $\gamma\gamma$~mass around the nominal $\pi^0$
mass and on the missing mass around the nominal proton mass was applied. Since the width of the
missing mass distribution increased with energy, the cuts were adjusted for each incident photon
energy bin. Furthermore, the azimuthal angle between the the proton and the $\pi^0$, was required
to be 180$^{\circ}$ within an energy dependent $\pm2\sigma$ window (coplanarity).
To remove untagged events originating from photons below the tagging threshold (due to random
coincidences), the beam photon energy was calculated from the kinematics of the reaction using the
four-momentum of the $\pi^0$, the proton mass, and the condition that the total transverse momentum
in the event was zero. This value for the photon energy was then compared to the measured photon
energy in the tagging system. A cut on the calculated photon energy of $E_\gamma >550$~MeV was
applied, which was 50~MeV below the first investigated photon energy bin.

A time coincidence between the tagger hit and the reaction products was required for all shown
spectra. A subtraction of the random time-background was performed using side-band subtraction to remove
time-accidentals due to the high rate in the tagging system.

The resulting distributions for the butanol data are shown in Fig.~\ref{pi0-MM-coplan} after applying all
cuts with the exception of those for the respective distributions.
Figure~\ref{pi0-MM-coplan}a shows the invariant $\gamma\gamma$~mass distribution (log-scale). 
A clear peak with
little background is visible for the $\pi^0$. It is followed by peaks for the
$\eta\to\gamma\gamma$, $\omega\to\pi^0\gamma\to 3\gamma$ (with one undetected low-energetic
$\gamma$ or two merging photons in one cluster accidentally interpreted as one photon), and
$\eta^\prime\to\gamma\gamma$. The missing-mass distribution (Fig.~\ref{pi0-MM-coplan}b) shows a
clear peak at the proton mass.

Background is visible not only above but also below the proton mass peak, although accidental
time-background has been subtracted. The latter is not present for data taken with a liquid
hydrogen target since the missing mass X in a reaction $\gamma p \to \pi^0 X$ can only be larger
but not smaller than the mass of the proton. The low missing-mass events are due to the Fermi
motion of the nucleons bound in C or O. A similar effect is visible in the coplanarity spectrum
(Fig.\ref{pi0-MM-coplan}c), where the peak observed at 180$^\circ$ has a significantly larger width
than expected for a hydrogen target, two Gaussian distributions of different widths overlap. These
observations became important when the fraction of events produced off bound nucleons in the data
set was determined (see Fig.~\ref{dil}).

\paragraph{Data sample: }
After all cuts, about $1.2\cdot 10^6$ ($4.6\cdot 10^6$) events due to the reaction $\gamma p\to \pi^0 p$
from the butanol target were retained in the 2007 (2009) beam time. Additionally, $0.2\cdot 10^6$ events
taken with the carbon foam target survived after applying all cuts.
%
%

\section{\label{Analysis}Data Analysis and Results}
\begin{figure}[h!]
  \centering
  \includegraphics[width=0.4\textwidth, angle=0]{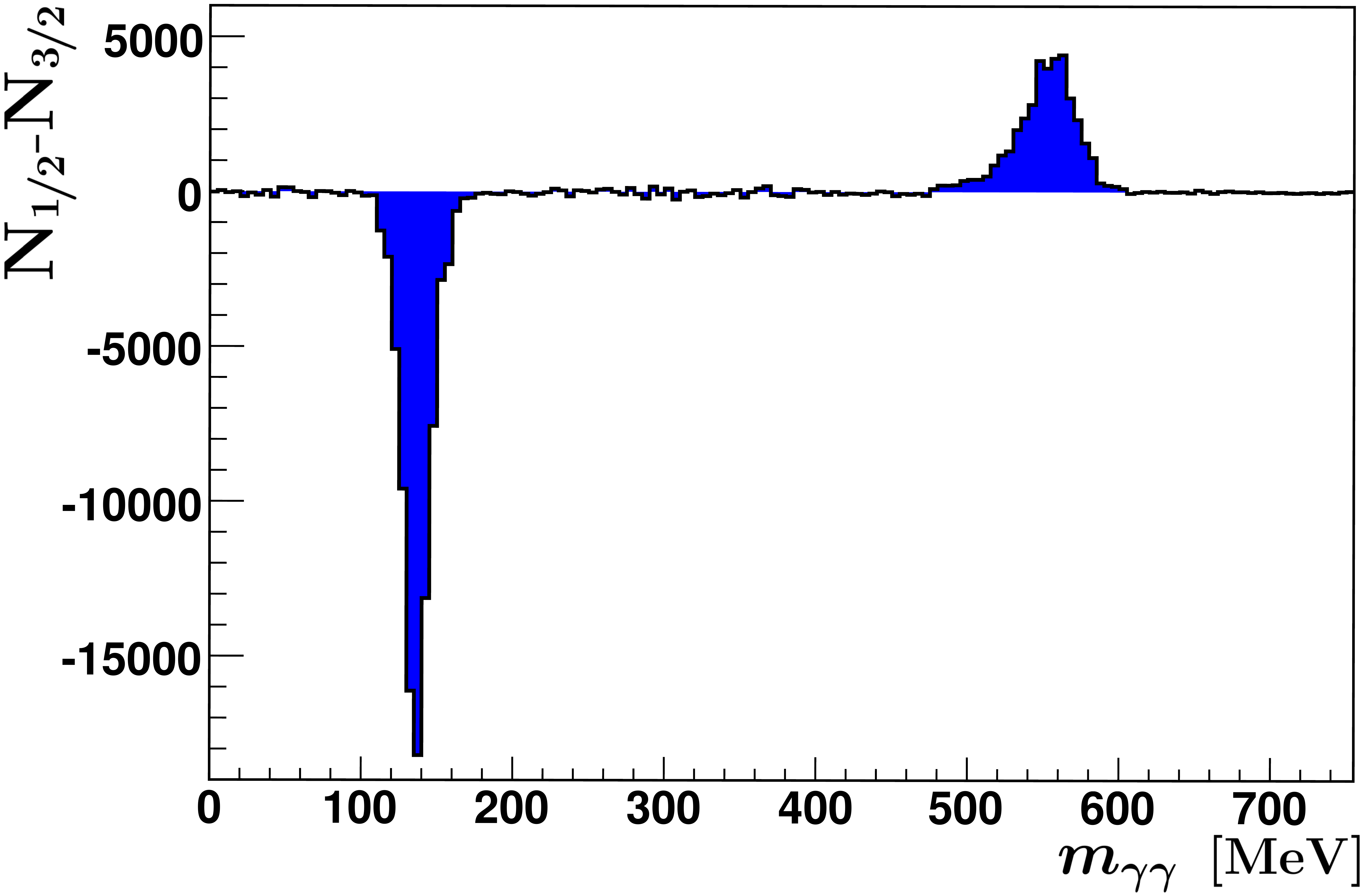}
  \caption{$N_{1/2}-N_{3/2}$ for the invariant mass spectrum. All events with energies above $E_\gamma=550$~MeV
are included.}
  \label{countrates}
\end{figure}

\paragraph{The count rates: }
The double-polarization observable $E$ was determined via formula (\ref{E_formula}).
Figure~\ref{countrates} shows the count rate difference $N_{1/2}-N_{3/2}$ of the invariant
$\gamma\gamma$~mass spectrum integrated over all energies and angles. A pronounced dip is visible
at the pion mass and a peak at the $\eta$ mass. In the photoproduction of neutral pions, the
$\sigma_{3/2}$ cross section obviously exceeds the $\sigma_{1/2}$ cross section in the investigated
energy range, while the reverse holds true in the photoproduction of $\eta$~mesons. This fact
illustrates the importance of investigating different reaction channels which highlight different
resonance contributions.

\begin{figure*}[t]
  \centering
\begin{tabular}{lll}
\hspace*{-0.34cm}
\hspace*{-0.04cm}
  \includegraphics[width=0.312\textwidth, angle=0]{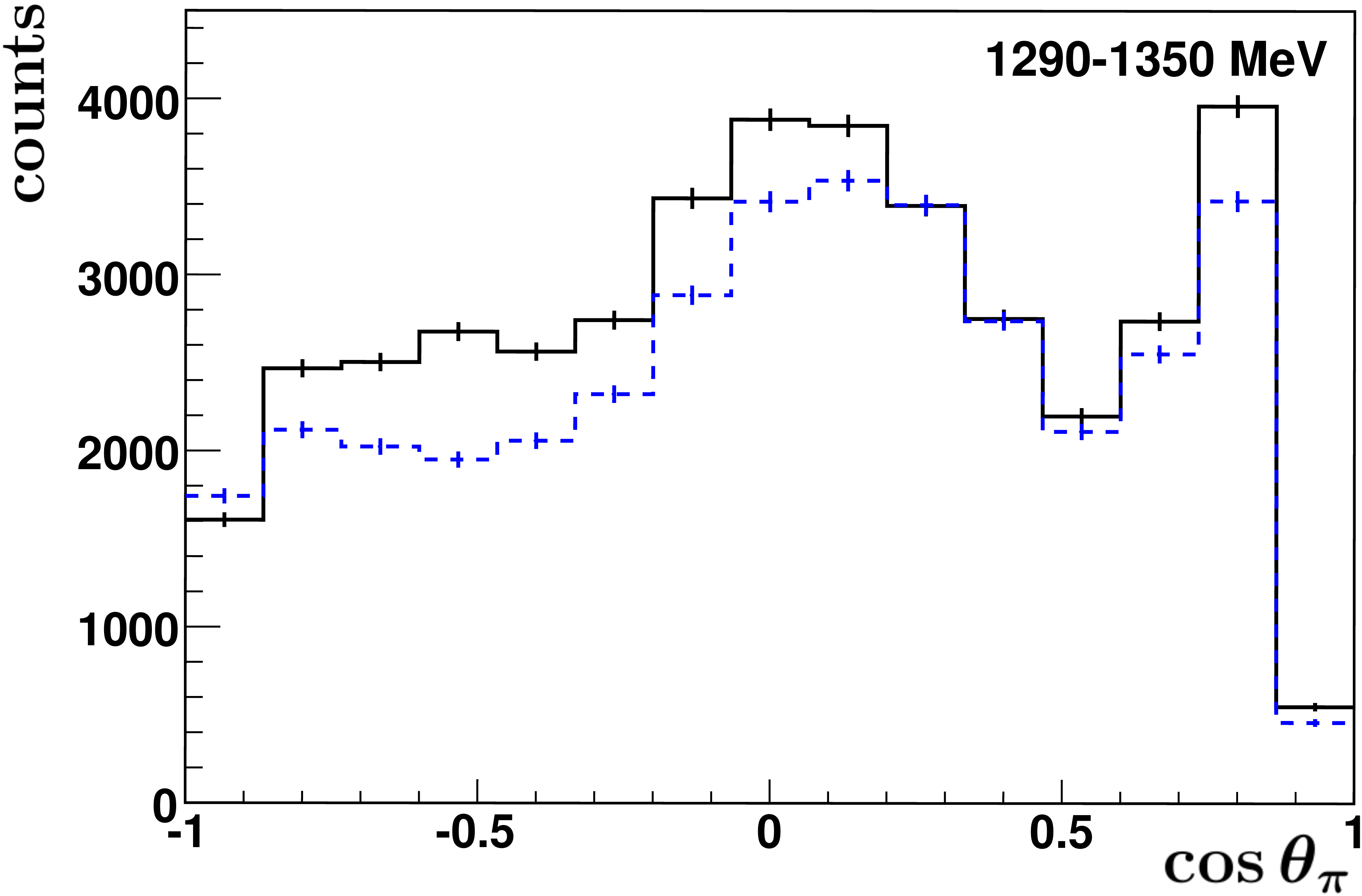}&\hspace*{-0.05cm}
\includegraphics[width=0.312\textwidth, angle=0]{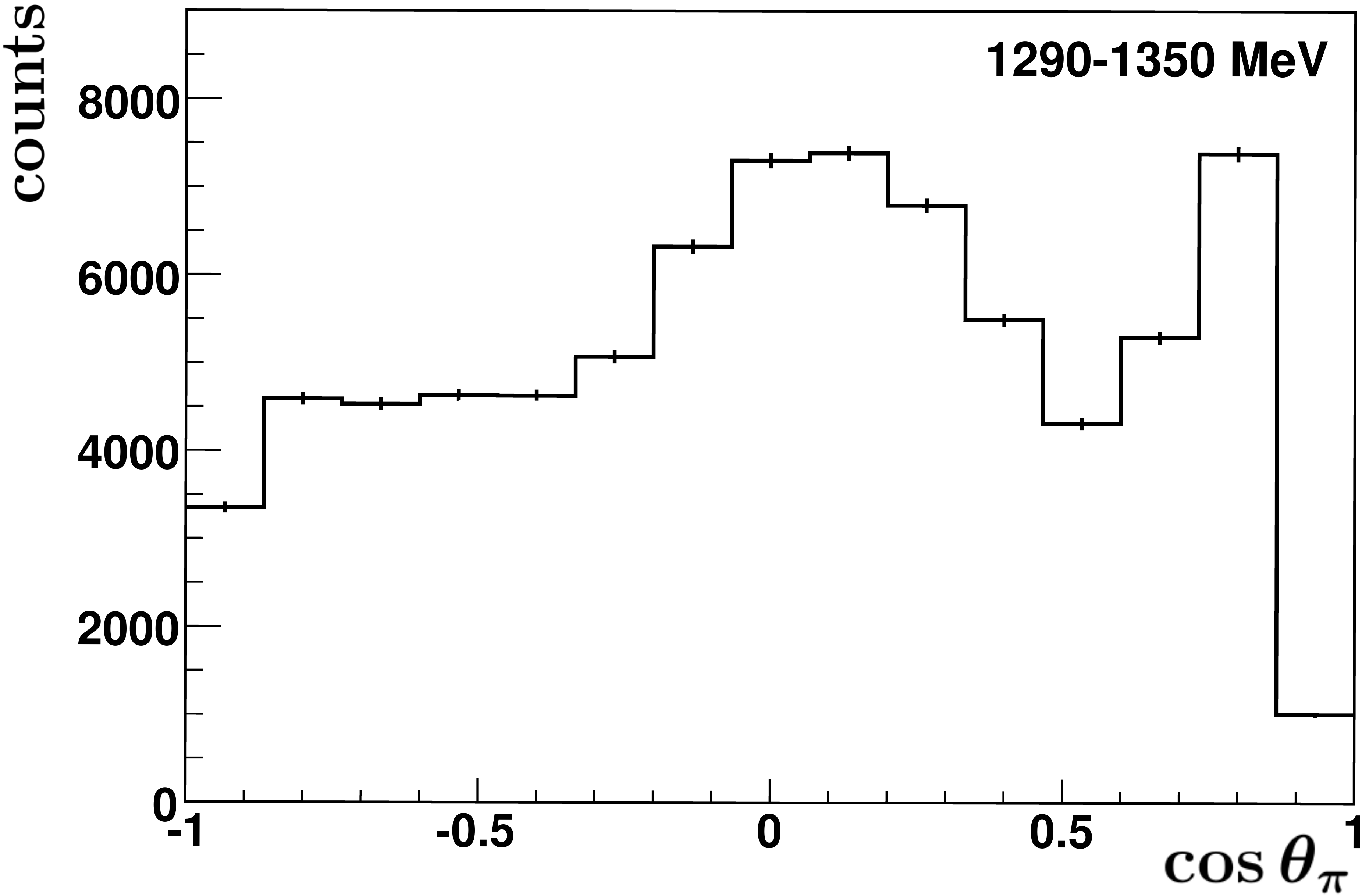}&\hspace*{-0.05cm}
\includegraphics[width=0.312\textwidth, angle=0]{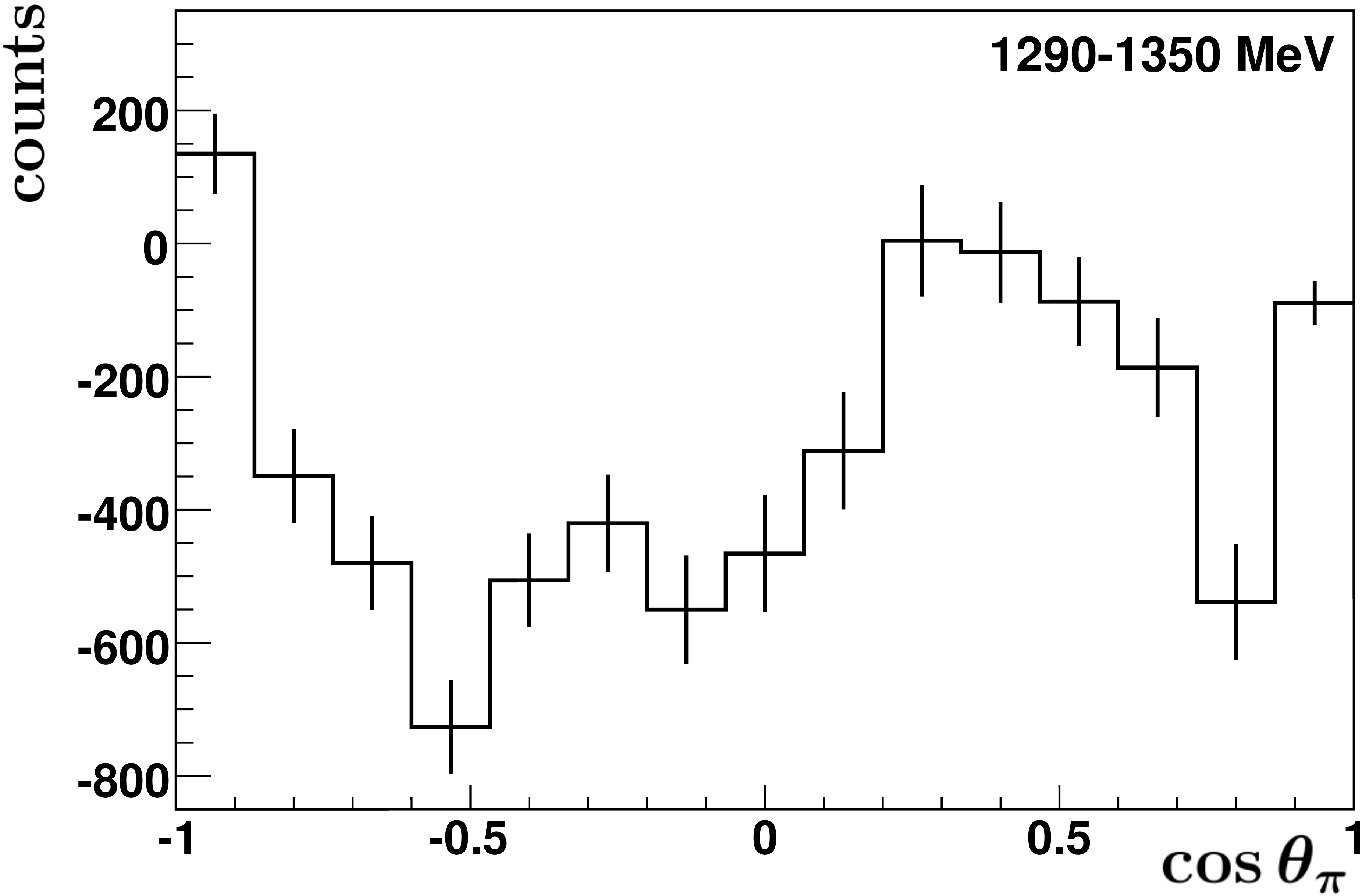}

\end{tabular}
\caption{Left: $N_{1/2}$ (dashed blue) and $N_{3/2}$ (solid black)
  separately; Middle: $N_{1/2}+N_{3/2}$; And right: $N_{1/2}-N_{3/2}$
  as function of $\cos{\theta_\pi}$.}
  \label{countrates2}
\end{figure*}

The observable~$E$ is a function of the photon energy and of $\cos\theta$, where $\theta$ denotes
the angle of the $\pi^0$ relative to the photon beam axis in the center-of-mass frame.
Fig.~\ref{countrates2} shows the count rates $N_{1/2}$ and $N_{3/2}$, the difference
$N_{1/2}-N_{3/2}$, and the sum $N_{1/2}+N_{3/2}$ for one specific energy bin, $E_\gamma = 1290 -
1350$~MeV.

For $\gamma p \to p\pi^0$, photoproduction off unpolarized carbon and oxygen nuclei does not depend on the helicity
of the photon beam. The contribution of the carbon and oxygen nuclei to the count rate difference $N_{1/2}-N_{3/2}$
vanishes. Yet, the carbon and oxygen nuclei do contribute to the sum $N_{1/2}+N_{3/2}$. Naively, one might expect
that all protons bound in the carbon and oxygen nuclei (32 out of the 74 nucleons in C$_4$H$_9$OH) contribute
to the sum $N_{1/2}+N_{3/2}$.
This is, however, not true. The kinematics of the reaction off bound nucleons differs from the reaction of free protons
due to the Fermi motion, and the contribution of the carbon and oxygen nuclei after all selection criteria are applied
is smaller than naively expected. 
For this reason, the {\it dilution factor} needs to be determined for each bin in
photon energy and $\cos\theta$. It depends significantly on the cuts applied in the data selection. 

\paragraph{The dilution factor: }
The dilution factor, $d$, for the reaction of interest is defined as the ratio of the number of events produced off
hydrogen nuclei ($N_{\rm H}$, free protons) to the number of events produced off all nucleons in the butanol
target ($N_{\rm C_4H_9OH}$) in the final data set:
\begin{equation}
\label{eqn_d1}
d(E_{\gamma},\theta)=\frac{N_{\rm H}(E_{\gamma},\theta)}{N_{\rm C_4H_9OH}(E_{\gamma},\theta)}\,.
\end{equation}
The number of events produced off hydrogen is given by the number of events produced off butanol
minus the number of events which are assigned to the production off the nucleons bound in carbon
and oxygen ($N^{But}_{\rm C}$). The latter is the number of $\pi^0p$ events which were produced off bound C/O-nucleons
and which passed all cuts of the analysis chain:
\begin{equation}
\label{eqn_d2}
N_H(E_{\gamma},\theta)=N_{\rm C_4H_9OH}(E_{\gamma},\theta)-s(E_{\gamma})\cdot N_{\rm C}(E_{\gamma},\theta)\,,
\end{equation}
where $s(E_{\gamma})$ is a scaling factor, which was determined using the additional data taken with the carbon foam
target ($N^{But}_{\rm C} = s\cdot N_{\rm C}$). Events produced off hydrogen and carbon/oxygen have different kinematic
distributions. This allowed us to separate the two reactions by performing additional measurements using
a carbon foam target within the polarized target cryostat filled with helium.
This assumes that the difference between
photoproduction off carbon and off oxygen does not play a significant
role. This is supported by the study in Ref.~\cite{MacCormick:1997ek},
which showed that events due to pion production off carbon and oxygen nuclei have similar kinematic distributions.

\begin{figure*}[ht]
  \includegraphics[width=0.32\textwidth, angle=0]{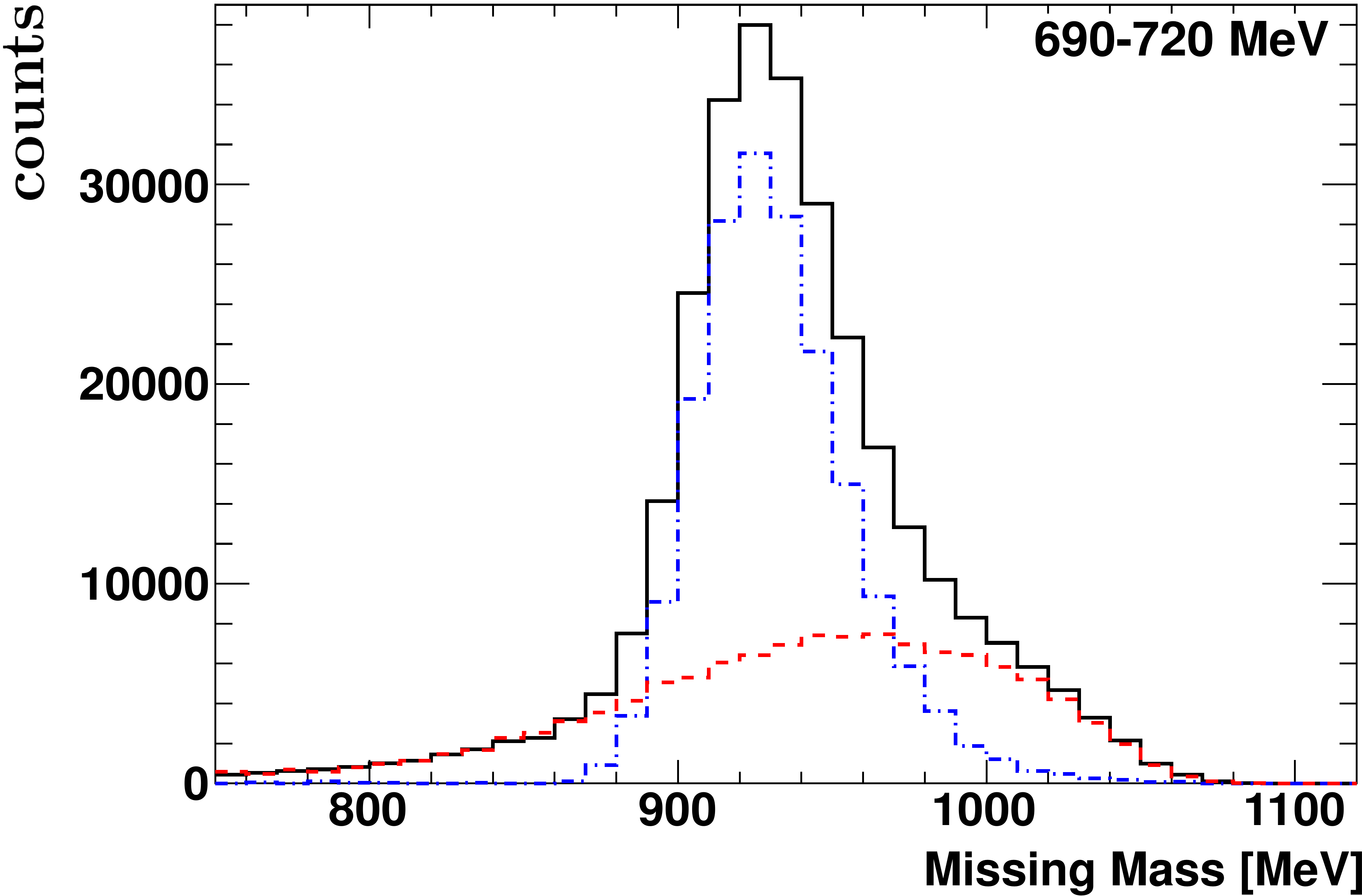} \hspace*{-0.cm}
  \includegraphics[width=0.32\textwidth, angle=0]{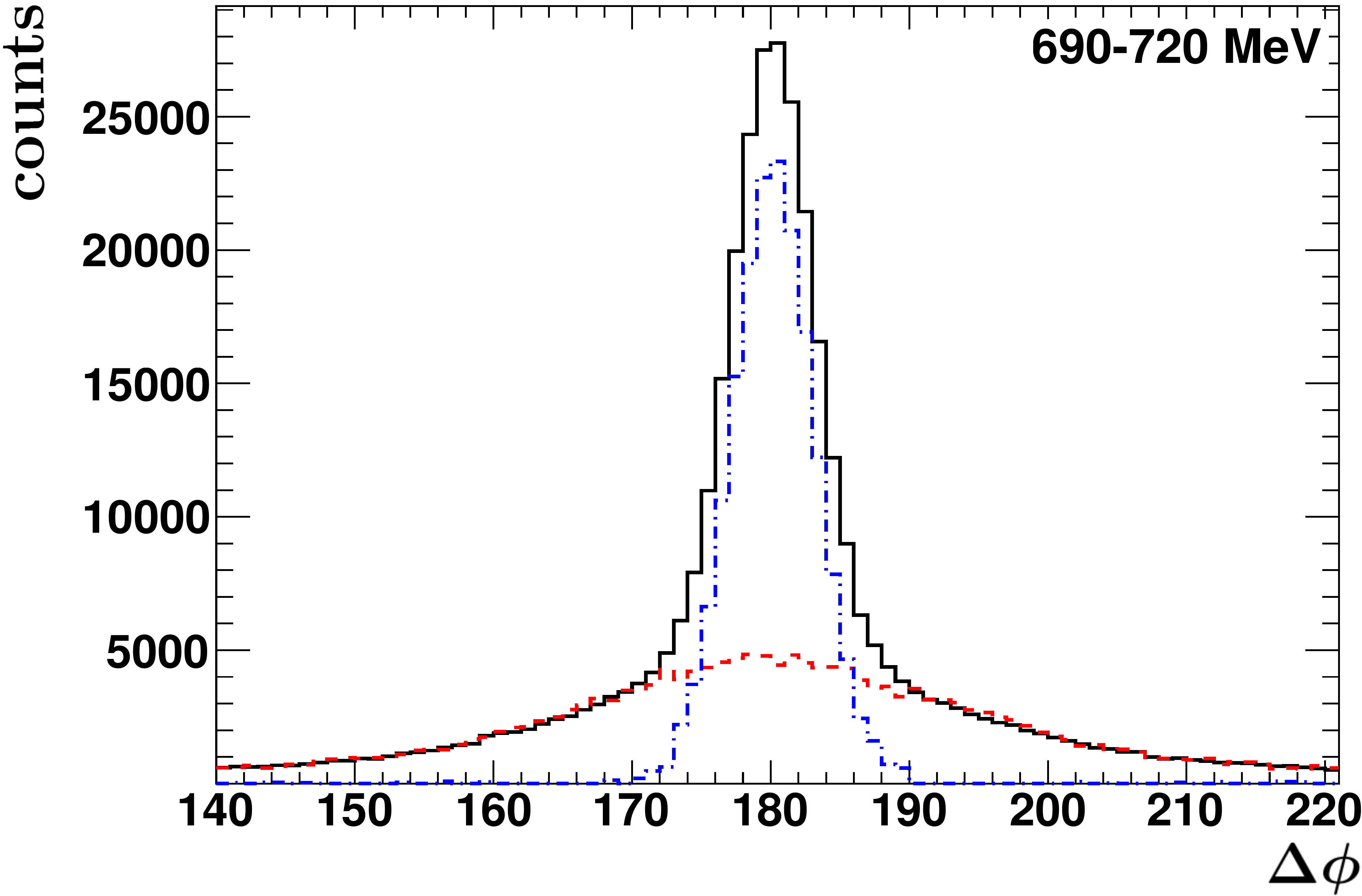} \hspace*{-0.cm}
  \includegraphics[width=0.32\textwidth, angle=0]{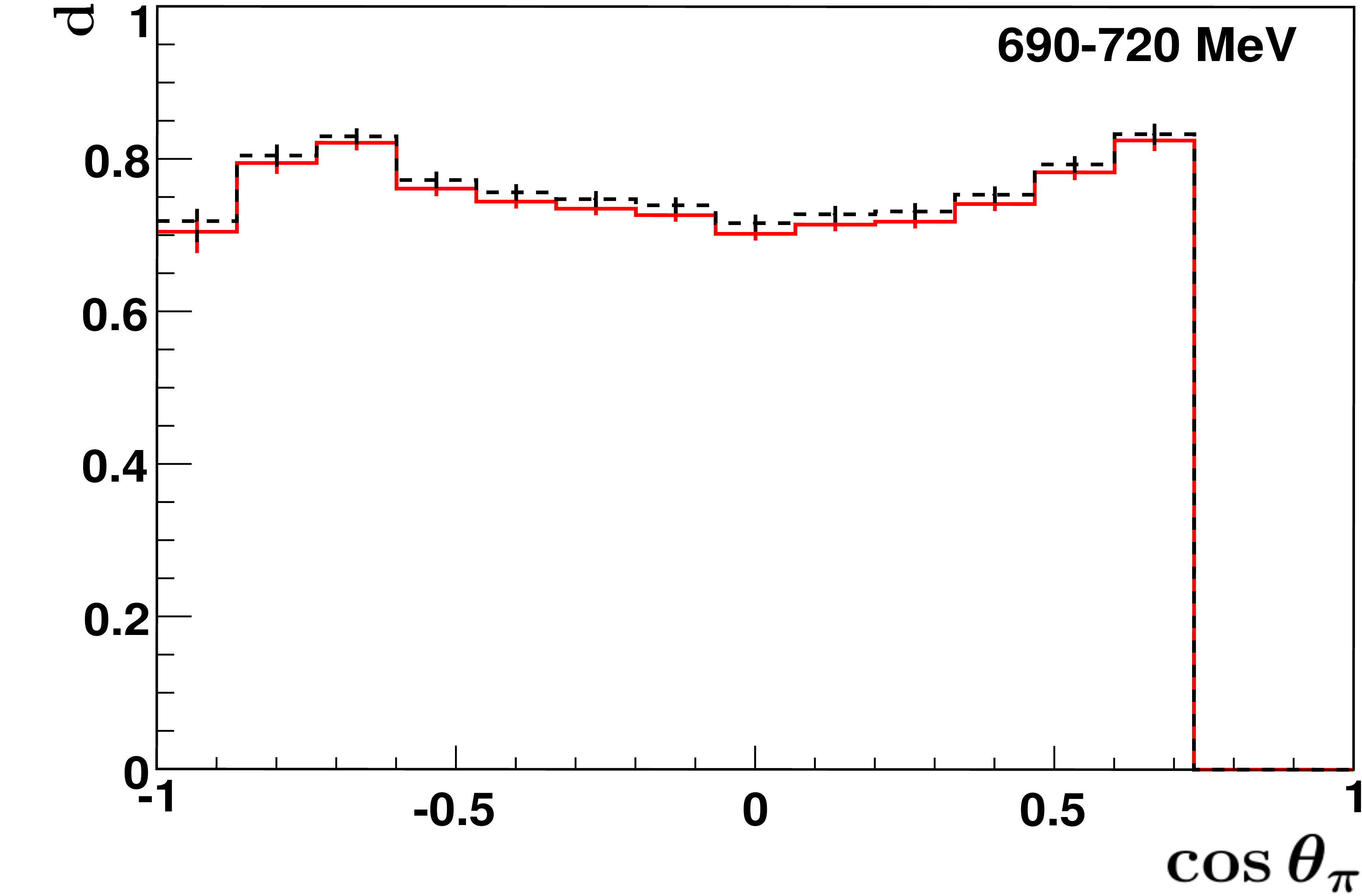}\\[+1.5ex]
  \includegraphics[width=0.32\textwidth, angle=0]{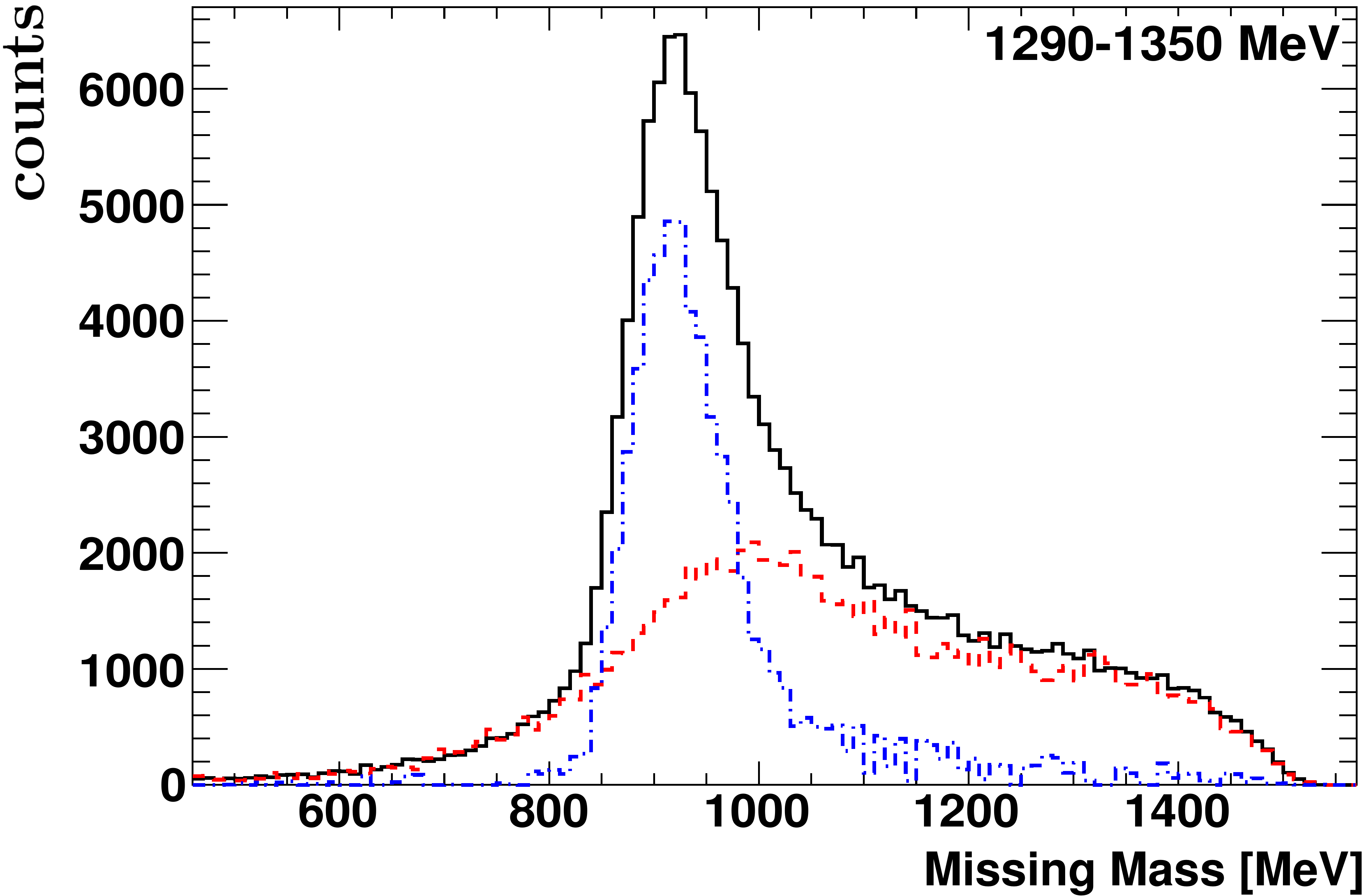} \hspace*{-0.cm}
  \includegraphics[width=0.32\textwidth, angle=0]{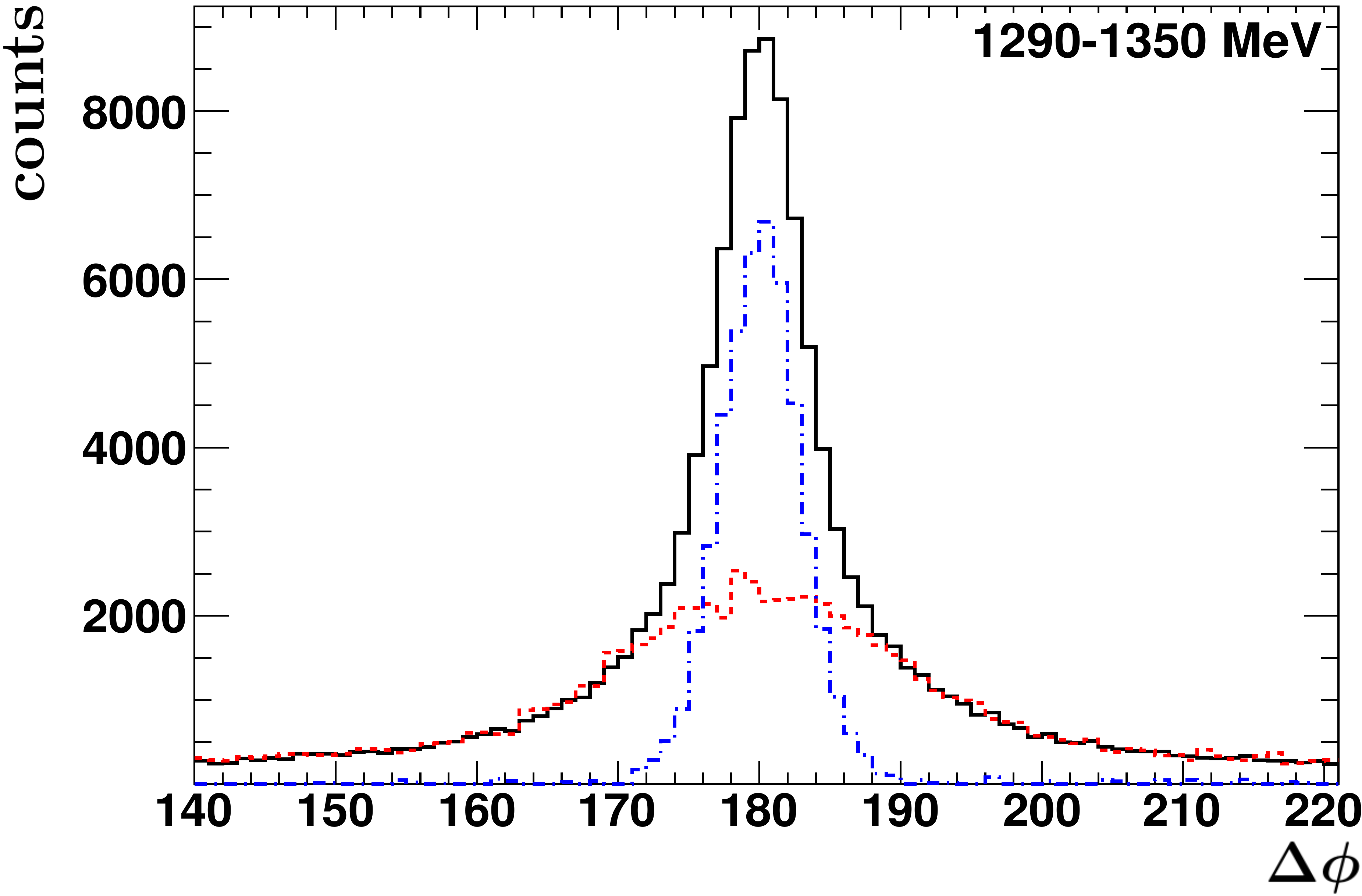} \hspace*{-0.cm}
  \includegraphics[width=0.32\textwidth, angle=0]{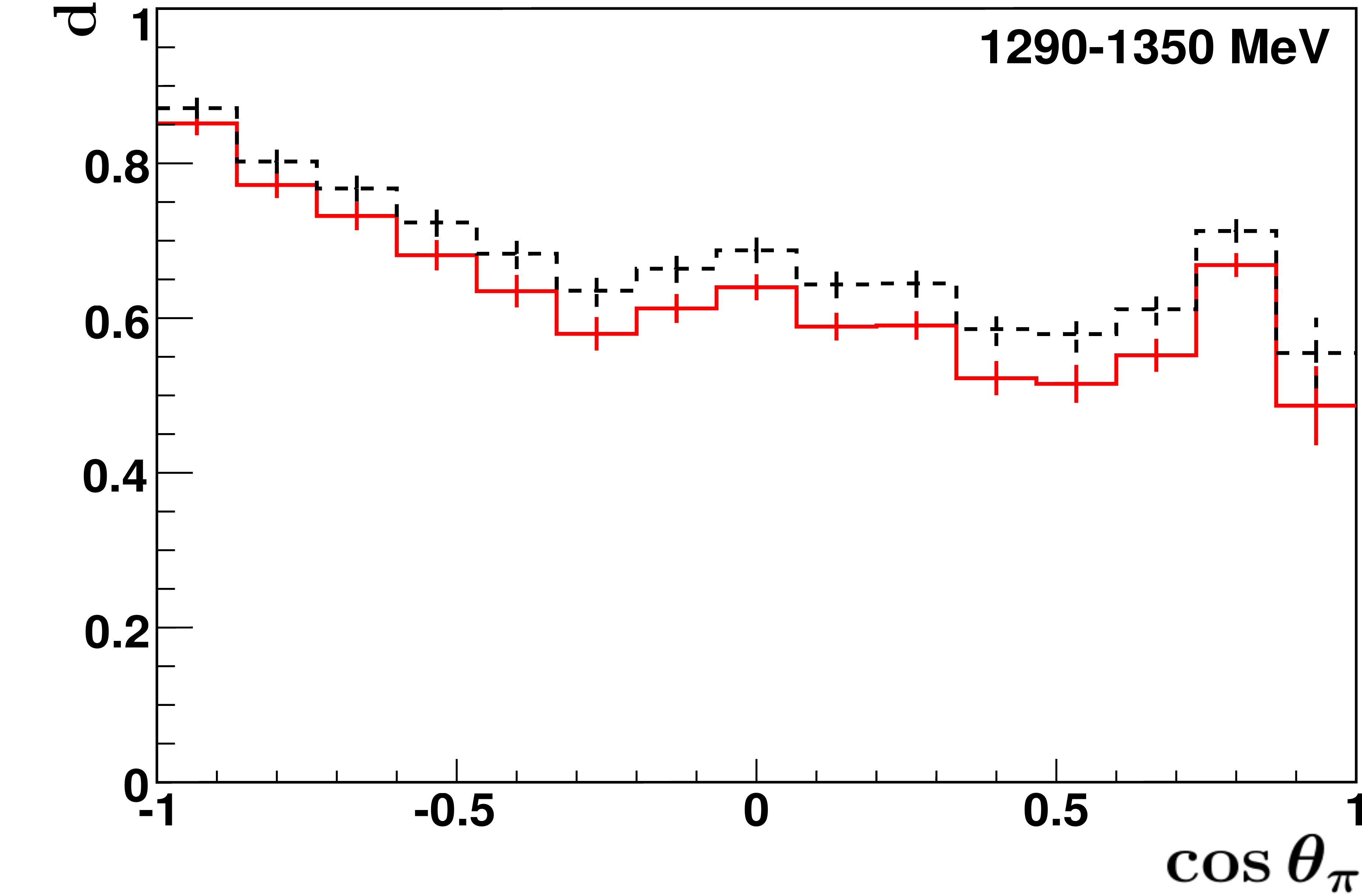}\\[+1.5ex]
  \includegraphics[width=0.32\textwidth, angle=0]{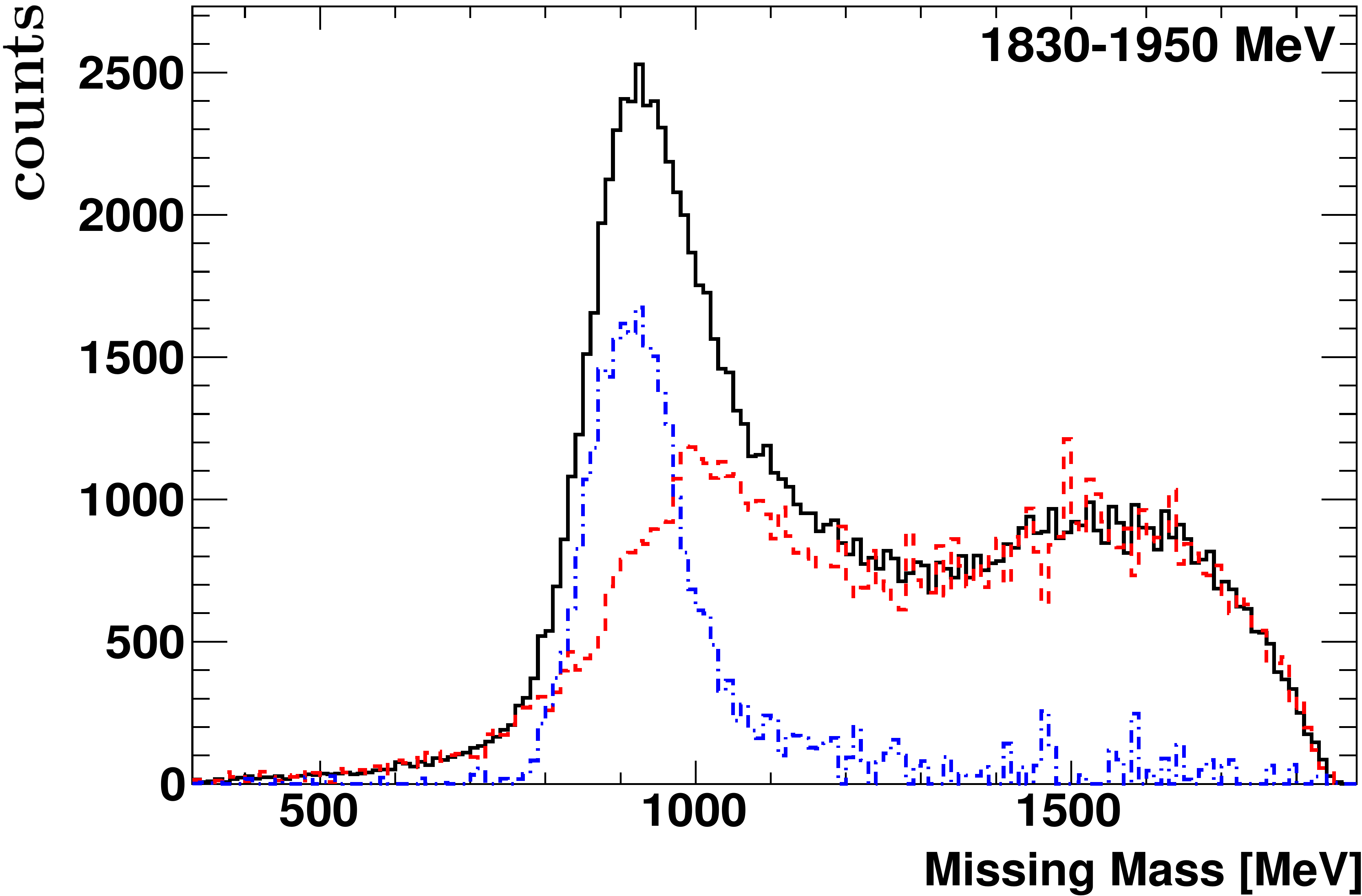} \hspace*{-0.cm}
  \includegraphics[width=0.32\textwidth, angle=0]{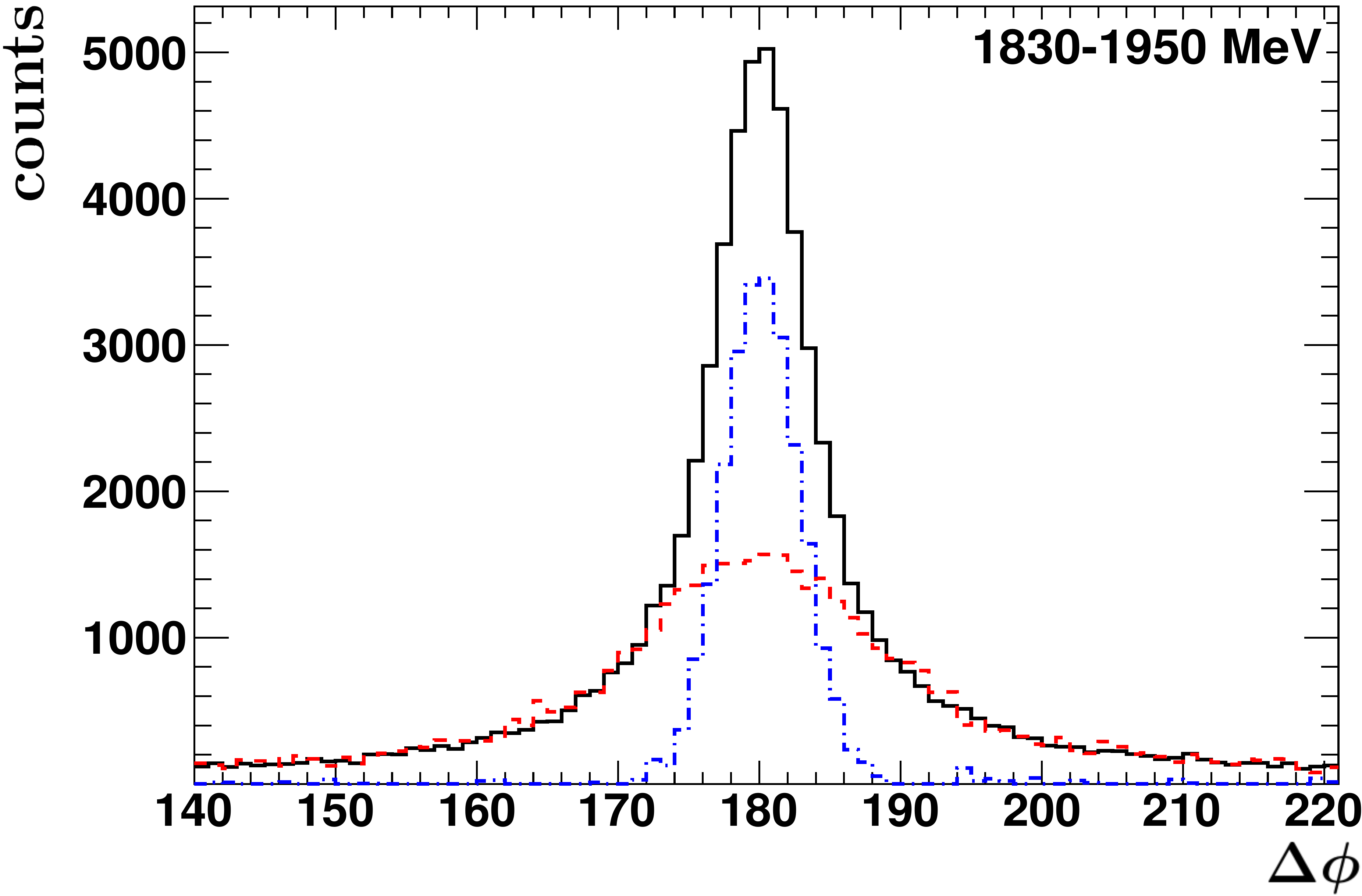} \hspace*{-0.cm}
  \includegraphics[width=0.32\textwidth, angle=0]{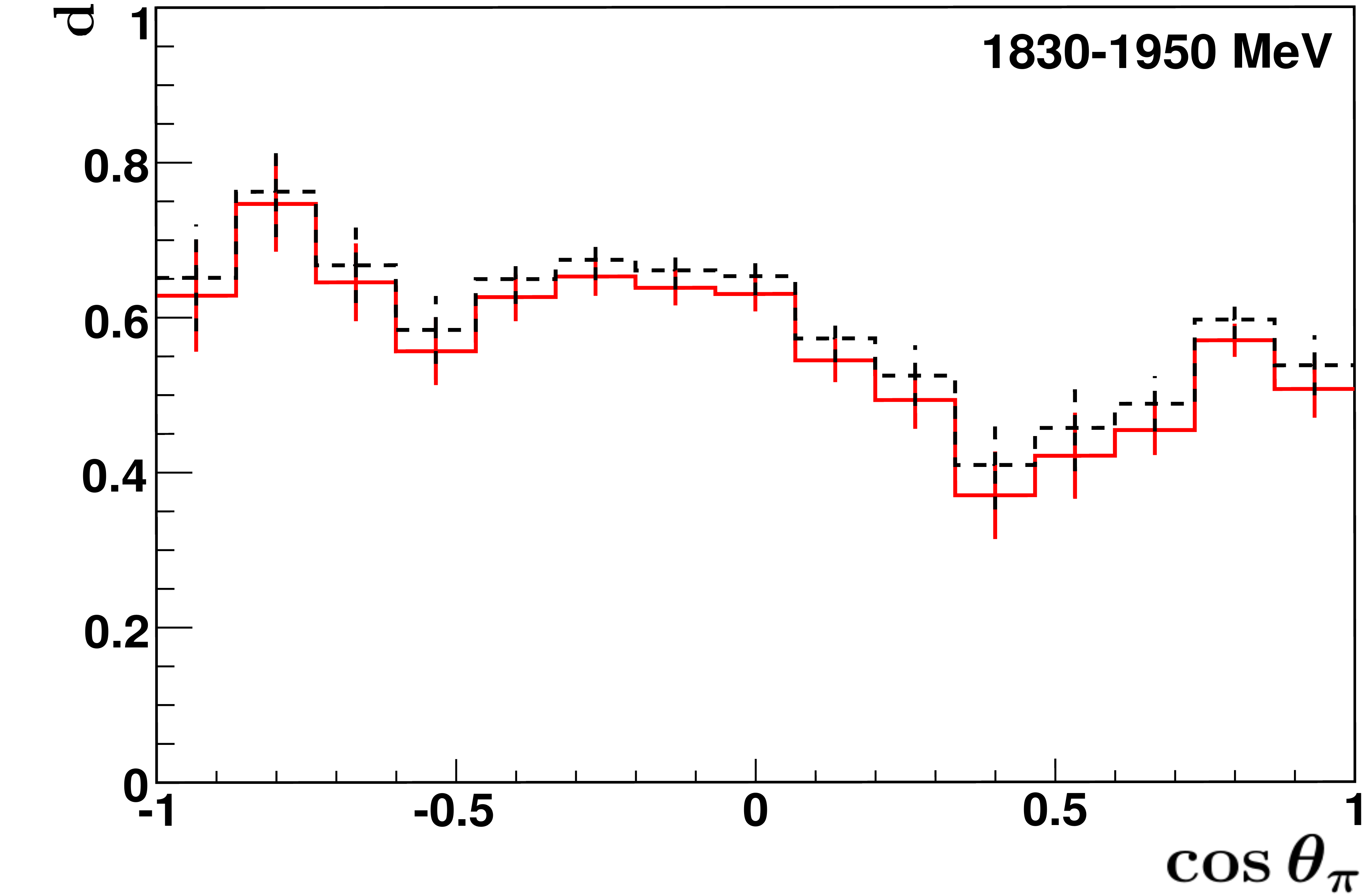}
  \caption{\label{dil}(Color online) Missing-mass spectra (first column),
  coplanarity spectrum (second column) for $E_{\gamma} = 690 - 720 $~MeV (1. row),
   $E_{\gamma}=1290 - 1350 $~MeV (2. row), and $E_{\gamma}=1830 - 1950 $~MeV (3. row).
   Black solid: butanol data, red dashed: scaled carbon data, blue dashed-dotted:
   $N_B-sN_C$ data. Last column: Dilution factor for both methods: missing mass in
   black dashed and coplanarity in red solid. Using the missing 
   mass method results in a slightly larger dilution factor. This difference might be explained 
   due to slightly different conditions in the butanol and carbon measurement (e.g. target positioning), 
   which have a larger impact on the missing mass compared to the coplanarity spectrum. The difference 
   in the dilution factor is considered in the systematic error.}
\end{figure*}

The scaling factor $s(E_{\gamma})$ was determined using two different methods. The first column of
Fig.~\ref{dil} shows missing-mass spectra for three ranges of photon energy integrated over the
solid angle: The solid histogram (black) represents the missing-mass distribution from the butanol
target (MM$_{\,\rm C_4H_9OH}$), and the dashed (red) histogram represents the missing mass
distribution from carbon (MM$_{\,\rm C}$) already scaled to the distribution from butanol. To
determine the scaling factor $s(E_{\gamma})$, the distribution from carbon was fitted to the
background events at the left side of the missing-mass spectrum (see Fig.~\ref{dil}). Subtracting
the scaled distribution measured with carbon from the distribution measured with butanol results in
the dashed-dotted (blue) histogram.

In the second method, the coplanarity spectrum is fitted in a similar way. Due to Fermi motion, the
coplanarity spectrum does not only show a Gaussian peak of small width at 180$^\circ$ as expected
for hydrogen, but also an additional rather broad distribution. The carbon distribution, following
this broad additional distribution, was fitted to the coplanarity spectrum for $100^\circ \leq
\Delta\phi \leq 160$ and $200^\circ \leq \Delta\phi \leq 260$ and the respective scaling factor was
again determined. Subtracting the scaled carbon from the butanol distribution results in the
dashed-dotted (blue) histogram with hardly any visible background.

After determining the scaling factors for the different energy bins, the dilution factor could be
derived as a function of photon energy and $\cos\theta$ using Eqns.~(\ref{eqn_d1}) and
(\ref{eqn_d2}). The dilution factor is shown as a dashed black histogram in the third column of
Figure~\ref{dil} for the first (MM) and as solid red histogram for the second (coplanarity) method.
It is nearly 0.8 at the lowest photon energy and decreases to 0.5 above 2~GeV.

The double-polarization observable $E$ was finally determined using the dilution factor from
the second method (coplanarity spectra). Compared to the missing-mass distribution, the number of
butanol and carbon events in the fit region are larger  and the coplanarity spectrum does hardly
change its shape with energy in contrast to the missing-mass spectrum.
Differences in the dilution factor from the two methods 
were taken into account as a systematic uncertainty.

\paragraph{The polarization: }
The last step to be considered in the determination of $E$ was the polarization of the
circularly-polarized photon beam and the longitudinally-polarized target (Eqn.~\ref{E_formula}).
Each event in the count-rate difference $N_{1/2}-N_{3/2}$ has been weighted by $\frac{1}{P_TP_z}$.
The target polarization $P_T$ was determined from NMR measurements before and after each
polarization phase and interpolating in between the two points using the measured relaxation time.
Thus, the polarization of the target was known at each point in time.

The electron beam polarization was determined using the M\o ller polarimeter (see
Section~\ref{moeller}). The polarization transferred from the electron to the photon is a
function of its energy and was determined using
Eqn.~(\ref{eqn_olsen}). It is denoted as $P_z$.

\begin{figure*}[pt]
  \centering
  \includegraphics[width=0.999\textwidth,height=0.945\textheight]{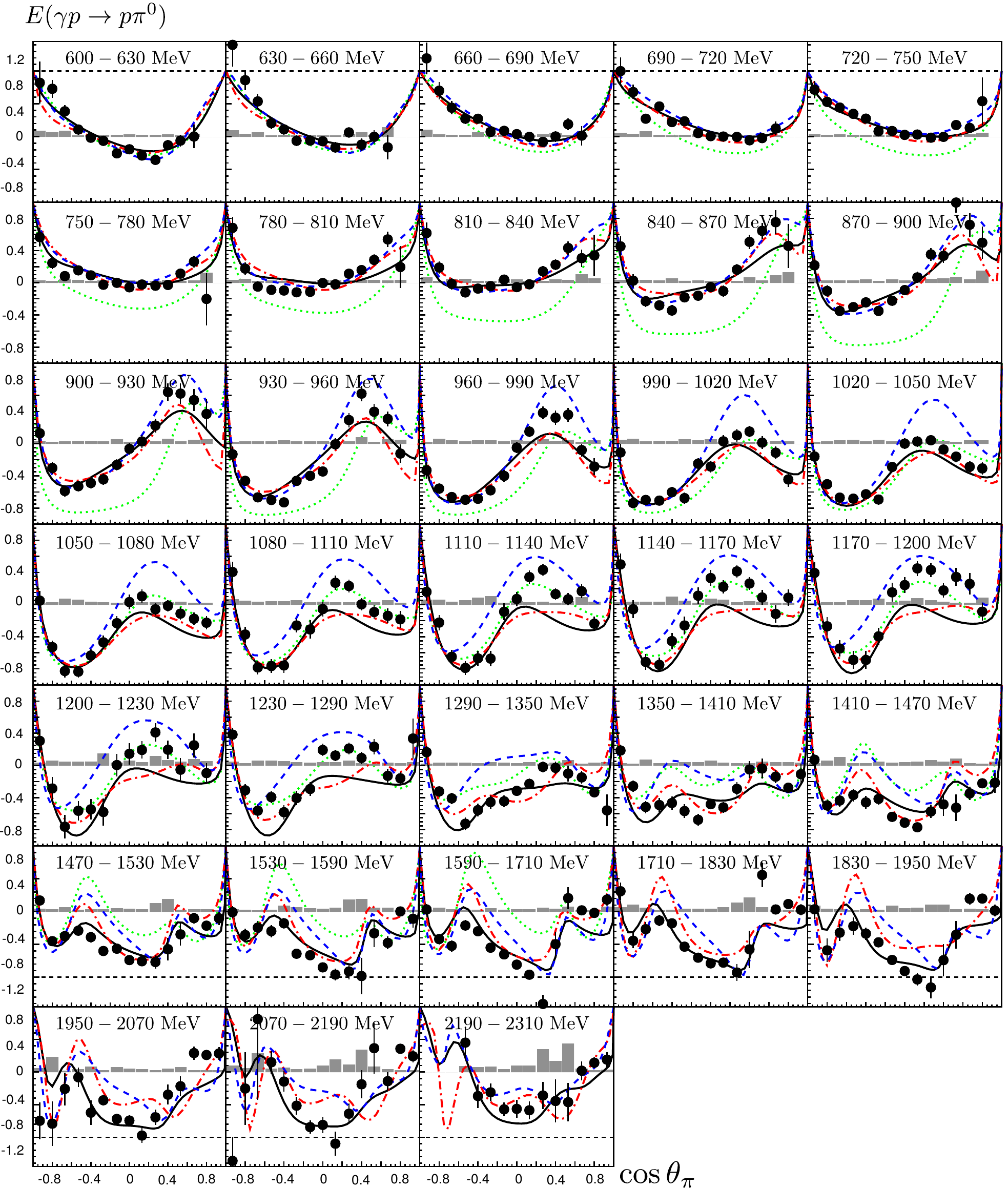}
  \caption{Double-polarization observable E for different beam energies from 600 to 2310~MeV. The data are compared with the PWA predictions BnGa 2011-02 (black solid line)~\cite{Anisovich:2011fc}, J\"uBo 2015-B~\cite{Ronchen:2015vfa} (blue dashed line, 2015-B~included already the $G$-data bins given in~\cite{Thiel:2012yj}), MAID 2007 (green dotted line)~\cite{Drechsel:2007if}, and the SAID prediction for solution CM12 (red dashed-dotted line)~\cite{Workman:2012jf}. 
The gray area represents the systematic error.
}
  \label{Efigs-pred}
\end{figure*}

\begin{figure*}[pt]
  \centering
\includegraphics[width=0.999\textwidth,height=0.945\textheight]{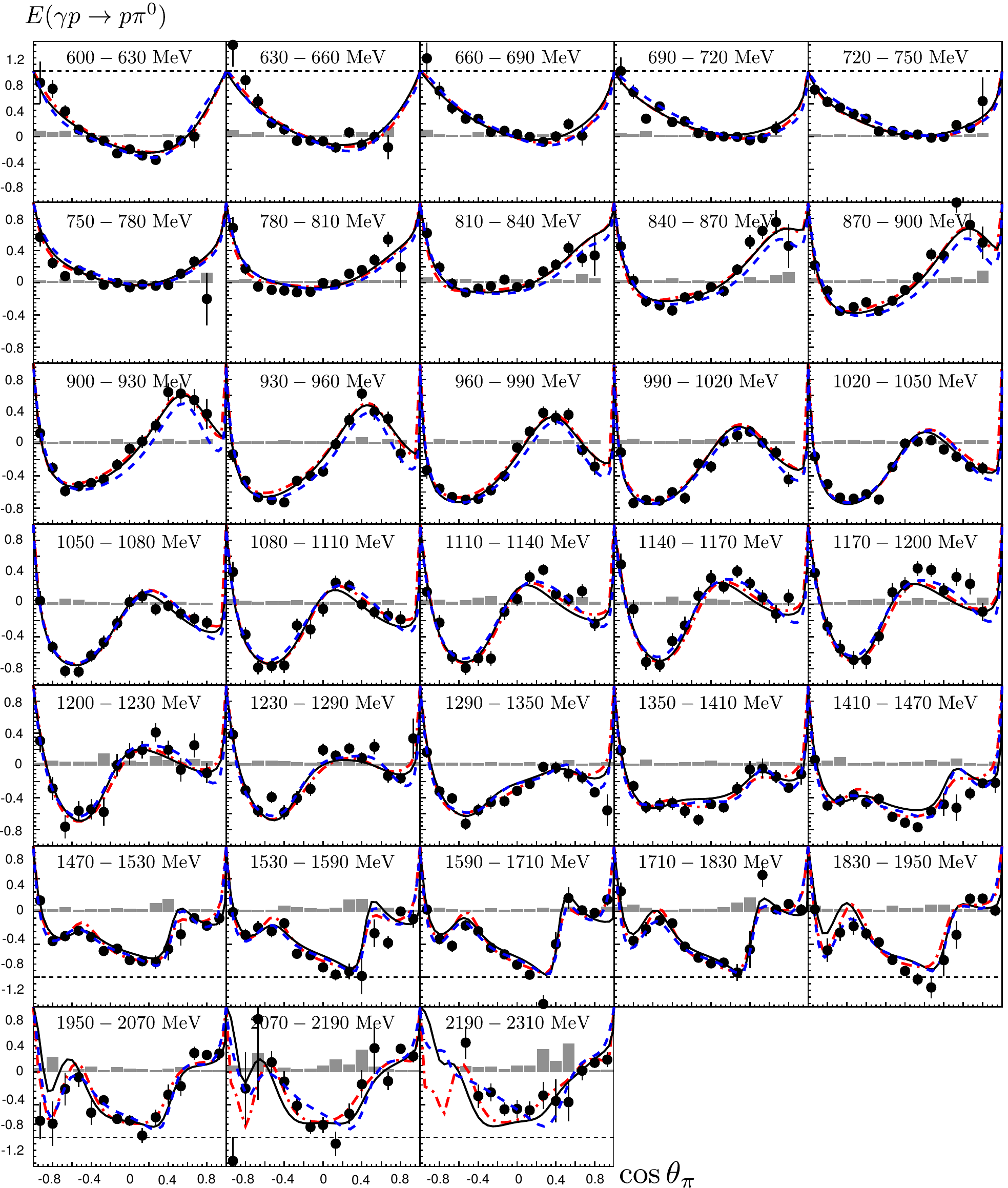}
  \caption{Double-polarization observable E for different beam energies from 600 to 2310~MeV. The data are compared to the PWA results from three fits including this data set: BnGa 2014-02 (black solid line)~\cite{Sokhoyan:2015fra,Beck:2016hcy},
    J\"uBo 2016-1 (blue dashed line){~\cite{Beck:2016hcy}}, and SAID 2015 (PD03) (red dashed-dotted line){~\cite{Beck:2016hcy}}. The gray area represents the systematic error. 
}
  \label{Efigs-fit}
\end{figure*}

\paragraph{The double-polarization observable \boldmath$E$: }
The results on $E$ using Eqn.~(\ref{E_formula}) are presented in Figs.~\ref{Efigs-pred} and
\ref{Efigs-fit} in comparison to different partial wave analysis solutions (predictions (Fig.~\ref{Efigs-pred}) and fits (Fig.~\ref{Efigs-fit})). 
For $E_\gamma <1230$~MeV, the data are shown in 30-MeV-wide bins. For
$1290<E_\gamma <1590$~MeV, 60-MeV-wide bins are used, three 120-MeV-wide bins cover the energy
range up to 2310~MeV. The solid-angle coverage is arranged in 15~bins in $\cos\theta$.
The total statistical uncertainty includes contributions from the statistical uncertainties in the
event numbers $N_{1/2}$ and $N_{3/2}$ and contributions from the statistical uncertainty in the
determination of the dilution factor. The systematic uncertainties, shown as gray bands 
in
Figs.~\ref{Efigs-pred} and \ref{Efigs-fit}, were obtained from a comparison of the results from
different analyses: On the one hand, two different methods to determine the dilution factor were
used as discussed above. 
On the other hand, different cut values 
were applied in the data selection. The latter makes sure that the results do not depend on the
applied cuts in the event selection. 
The systematic uncertainty in the target and beam polarization was
estimated to be 2\,\% and 3.3\,\%, respectively. These contributions were added in quadrature.


Angular momentum conservation guarantees that \\ $\sigma_{3/2}\,(\cos\theta=\pm 1)$ vanishes at forward and
backward angles, hence $E\,(\cos\theta=\pm 1)=1$. This requirement is compatible with the
experimental findings. As a function of energy, the $\cos\theta$ distributions change steadily. At
low photon energies ($<$900~MeV), we expect only a few low-spin resonances to contribute, and hence
no oscillatory behavior, in agreement with the experimental findings. At higher photon energies,
more structure in the angular distribution is expected and seen in the data.

\paragraph{Partial wave analyses methods: }
Different partial wave analyses have been performed by the SAID, MAID, J\"ulich-\-Bonn (J\"uBo) and
Bonn-Gatchina (BnGa) groups which can be compared to the data. We outline shortly the basic
ingredients used in the fits. A more detailed description of the partial wave analysis methods of
the four PWA groups and references to the data used in the fits are presented
in~\cite{Beck:2016hcy}.

SAID (GWU/INS)~\cite{Workman:2011vb}  uses a Chew-Mandelstam formulation of the scattering matrix,
which is parametrized in the form of a K-matrix \cite{Arndt:2006bf}. A fit to data on $\pi N$
elastic scattering, charge exchange and $\pi^-p\to \eta n$ determines a {\it hadronic rescattering
matrix} which encodes the hadronic channel coupling (or the effects of rescattering in the final
state), and returns masses, widths, and $\pi N$ branching ratios of 12 $N^*$ and 9 $\Delta^*$
resonances. The model was extended to include data on photoproduction (CM12)~\cite{Workman:2012jf}
and the photocouplings of 7 low-mass $N^*$ and 5 $\Delta^*$ resonances were determined.

The MAID group \cite{Drechsel:2007if} combines the 13 four-star resonances with
masses below 2~GeV known in 2006 \cite{Yao:2006px}
and a common background in a unitary formalism to fit
data on pion photo- and electroproduction from Bates/MIT, ELSA/Bonn,
MAMI/Mainz, and Jefferson Lab. Masses, widths and $\pi N$ branching ratios
of the resonances are taken from \cite{Yao:2006px}.

A dynamical coupled-channel approach is employed by the J\"ulich-Bonn (J\"uBo) group. The  model
guarantees unitarity and analyticity, and incorporates general S-matrix principles. In a first
step, data on $\pi N$ elastic scattering and on $\pi N\to \eta N, K\Lambda,$ and $K\Sigma$ were
fitted \cite{Ronchen:2012eg}; these data define the poles and residues of the 24 $N^*$ and
$\Delta^*$ resonances. In a second step, data on pion photoproduction were used to determine the
photocouplings of the contributing resonances \cite{Ronchen:2014cna}.

The Bonn-Gatchina group~\cite{Anisovich:2011fc} employs covariant amplitudes in a K-matrix formalism
to perform combined analyses of most known data on single and
double-meson production in photon- and pion-induced reactions; weak-decay modes (like $\gamma N$ in production
and decay) are treated in the form of production ($P$) or decay ($D$) vectors which do not contribute
to rescattering. Two equivalent classes of solutions compatible with all data included at that time
were presented in~\cite{Anisovich:2012ct}.

As discussed below, the data presented here have been included together with further new polarization data
into the different partial wave analyses. The respective results are summarized below. A detailed
discussion on the PWA-results is given in~\cite{Beck:2016hcy}.

\paragraph{Partial wave analyses - predictions: }
In Fig.~\ref{Efigs-pred} the results are compared to the predictions of different partial wave
analyses performed by the SAID (CM12), MAID, J\"ulich-Bonn (2015-B) and Bonn-Gatchina (BnGa
2011-02) groups. The comparison shows a wide spread of the predicted results even in the low-energy
region and thus, underlines the need for new polarization data. In the low-energy region (600 to
780~MeV), only the BnGa, J\"uBo and SAID predictions are compatible with the new data. In
describing the cross section, MAID underestimates the contributions with helicity $A_{1/2}$ and
overestimates those with helicity $A_{3/2}$.
In the 800-950~MeV range, the J\"uBo, BnGa and SAID predictions agree fairly well with the data.
From 1.0 to 1.4~GeV, the general trend of the data is predicted by all PWAs, even though
quantitatively, the agreement is fair, at most. Above this energy, most PWA predictions (except
BnGa) deteriorate further. While all predictions show
three local minima, the J\"uBo, SAID and MAID  predictions have a pronounced maximum at
$\cos\theta\approx -0.45$. Likely, it originates from the interference of $\Delta(1950)7/2^+$
production and background amplitudes; it is too strong in these three analyses since the
contributions of other resonances and their interference with $\Delta(1950)7/2^+$ are not included
in the predictions. In the MAID online version, the enhancement in the prediction disappears when
the $N(1520)3/2^-$ helicity couplings are reduced by a factor of 4. The qualitative agreement
between the data and the BnGa prediction is surprisingly good in the high-mass region.

\paragraph{Partial wave analyses - new fits:} Figure~\ref{Efigs-fit} shows the data on $E$ again
but now with fits of the PWA groups. The fits take into account not only the new data on $E$
presented here, but also other recently published polarization data: new data on $\gamma p\to
\pi^0p$ were reported from Mainz, a measurement of the polarization transfer from a polarized
photon beam to a recoiling nucleon, $C_x$, \cite{Sikora:2013vfa} and low-energy data on
$d\sigma/d\Omega$ and $\Sigma$ \cite{Hornidge:2012ca}; the beam asymmetry $\Sigma$ was reported
from JLab data on $\gamma p\to \pi^0p$ and $\gamma p\to \pi^+n$ \cite{Dugger:2013crn}; the
variables $P,C_x,C_z$ were measured for one energy relevant here, $E_\gamma =1845$~MeV, at JLab
\cite{Luo:2011uy}; data on $G$ \cite{Thiel:2016chx} and on $T,P,H$
\cite{Hartmann:2014mya,Hartmann:2015kpa} were taken at Bonn. 
\begin{figure}[h!]
\epsfig{file=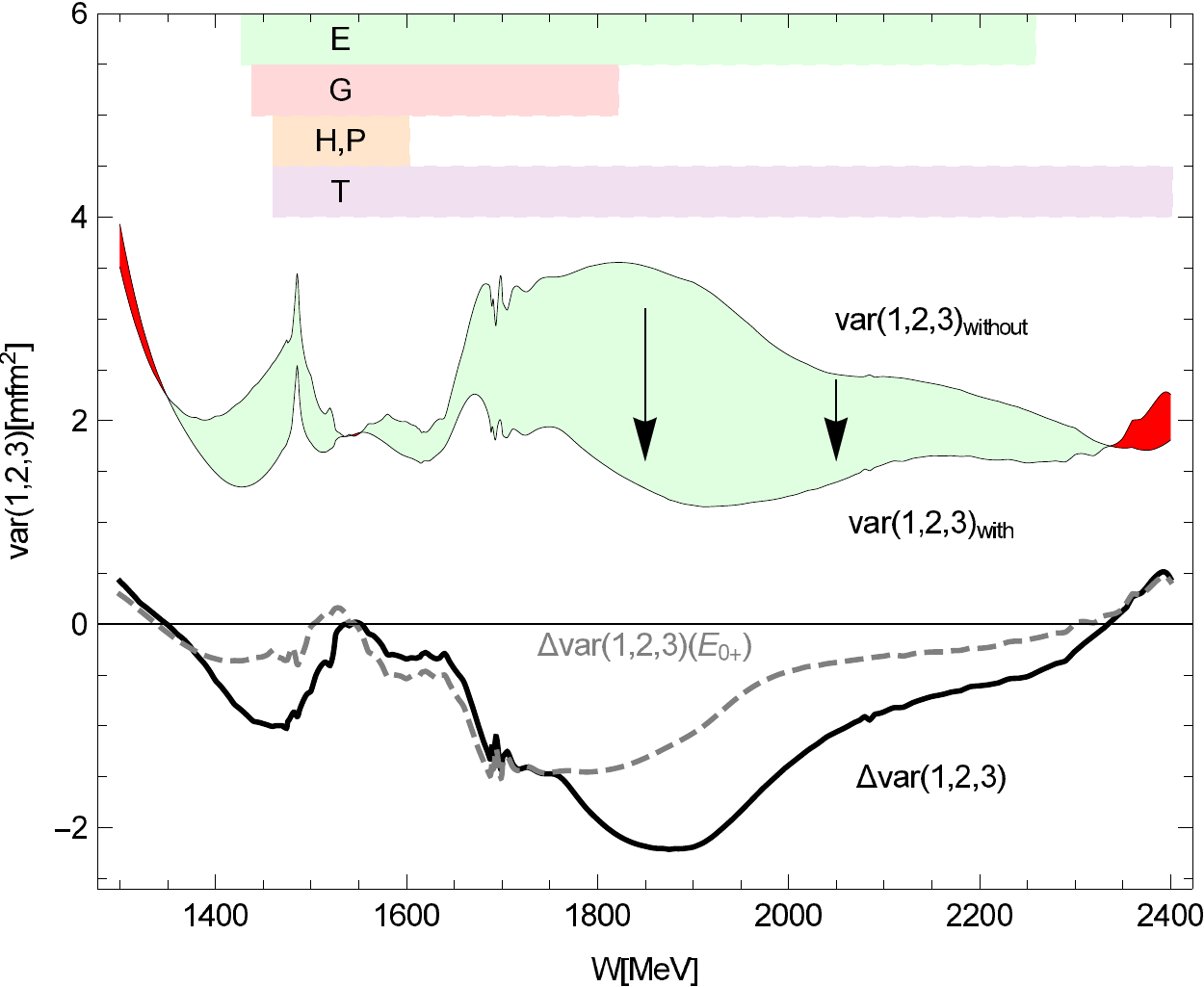,width=0.46\textwidth}
\caption{\label{allthree} The
variance of all the three PWAs (J\"uBo, SAID, BnGa) summed over all $\gamma p\to \pi^0 p$
multipoles up to $L=4$. The range covered by the new double-polarization observables are indicated
by shaded areas. Over the largest part of the energy range, the new data have enforced an
improvement of the overall consistency. The improvement is displayed as light green area and,
separately as difference of the variance. The contribution to the improvement from the $E_{0+}$
wave is shown as the dashed curve. Ranges with an overall deterioration are marked in red (figure
taken from~\cite{Beck:2016hcy}). The spike slightly below $W=1.5$~GeV reflects discrepancies in the
description of the $\eta p$ threshold effect the different PWAs. A wider peak below $W=1.7$~GeV might stem
from slightly different $N(1680)5/2^+$ properties used in the three PWAs. Both become less
pronounced when the new data are included in the fits. }
\end{figure}
All four PWAs are capable of
reproducing the data quantitatively, but small differences remain. A detailed comparison of the PWA
results of the different PWA groups based on the the data presented here as well as on the data
mentioned above is given in~\cite{Beck:2016hcy}. Here we only summarize the main result. In the
limit of a complete database with a complete angular coverage and large statistics, the amplitudes
of the different PWAs should converge to the same physical solution. Since the new data represent a
significant improvement of the existing database and represent an important step toward a complete
experiment, one would expect a convergence of the solutions of the different PWAs toward the same
solution even though the amplitudes do not yet reach the point of being really identically.
Fig.~\ref{allthree}, taken from~\cite{Beck:2016hcy} compares the different amplitudes in terms of
multipole amplitudes (for details see~\cite{Beck:2016hcy}). For this purpose, the variance between
two models, 1 and 2, was calculated as the sum over the squared differences of the
 16 (complex) $\gamma p\to \pi^0 p$ multipoles ${\cal M}$ up to $L=4$:
\begin{equation}
\hspace{-.0mm}
\mathrm{var}(1,2) = \frac12 \sum_{i=1}^{16}({\cal M}_1(i)
- {\cal M}_2(i))({\cal M}_1^*(i) - {\cal M}_2^*(i))\,.
\end{equation}
The mean of the variance $$\mathrm{var}(1,2,3)=\frac{1}{3}\cdot \displaystyle\sum_{i,j=1,i\neq
j}^{3}\mathrm{var}(i,j)$$ of the three PWAs (J\"uBo, SAID and BnGa), before and after the new data
were included in the fit, is shown in Fig.~\ref{allthree}. The overall spread of the three partial
wave analyses is reduced considerably due to the impact of the new polarization observables of
Refs.~\cite{Hartmann:2015kpa,Thiel:2016chx} and this work. As visible in Fig.~\ref{allthree}, a
significant fraction of the improvement stems from the $E_{0+}$ multipole exciting the $J^P=1/2^-$
wave (and thus the resonances $N(1535)$, $\Delta(1620)$, $N(1650)$, $N(1895)$, and $\Delta(1900)$).

\begin{figure}[b!]
\resizebox{0.485\textwidth}{!}{%
\includegraphics{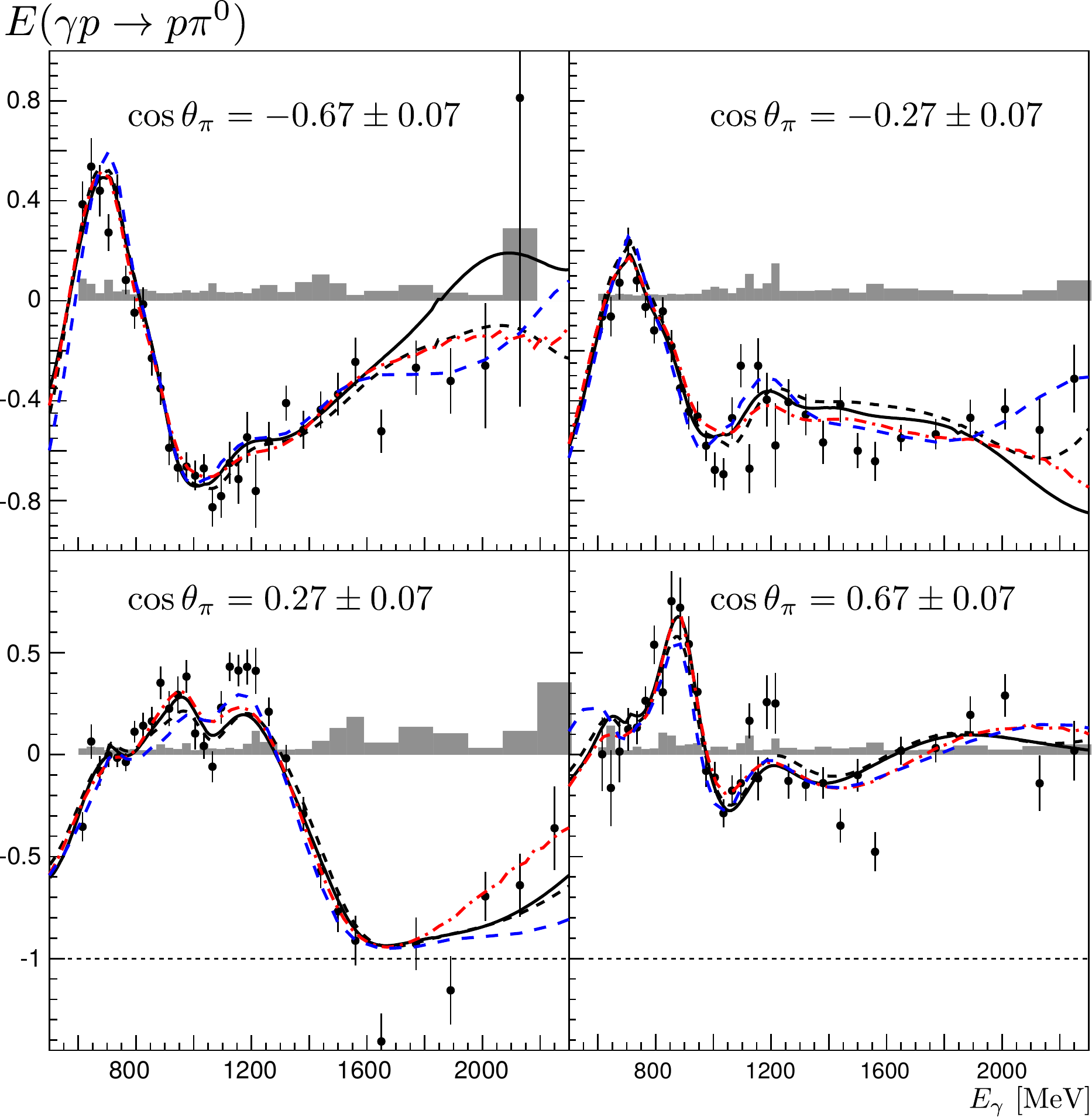}}
\caption{\label{Fig_E_energy_dependence}       
The energy dependence of E for four different angles, compared to different solutions
of the different PWAs: BnGa2014-02 (black solid), BnGa2014-01 (black dashed)~\cite{Sokhoyan:2015fra,Beck:2016hcy},
J\"uBo 2016-1 (blue dashed)~\cite{Beck:2016hcy}, SAID 2015 (PD03) (red dashed-dotted)~\cite{Beck:2016hcy}. All solutions S
shown are based on fits, which included the data presented here. The gray area represents the systematic error.
}
\end{figure}

Figure~\ref{Fig_E_energy_dependence} displays the energy dependence of the double polarisation observable 
$E$ for four different angular
bins in comparison to the final solutions of the different PWAs. At high photon energies, 
significant differences between the different PWAs are visible. Obviously additional data,
extending the energy range of the polarization observables discussed here, improving their
statistics, and the measurement of additional polarization observables are needed to further
constrain the PWA amplitudes to one common solution.

\paragraph{Spin-dependent cross sections: }
From the double-polari\-zation observable~$E$, the spin-dependent total cross sections $\sigma_{1/2}$
and $\sigma_{3/2}$ can be calculated via
\begin{eqnarray}
 \sigma_{1/2}&=&\sigma_0(1+E)\,,\\
 \sigma_{3/2}&=&\sigma_0(1-E)\,.
\end{eqnarray}
For the unpolarized cross section $\sigma_0$, the BnGa2014-02 fit to the data was used. 
\begin{figure*}[pt]
\begin{tabular}{cc}
\includegraphics[width=0.48\textwidth]{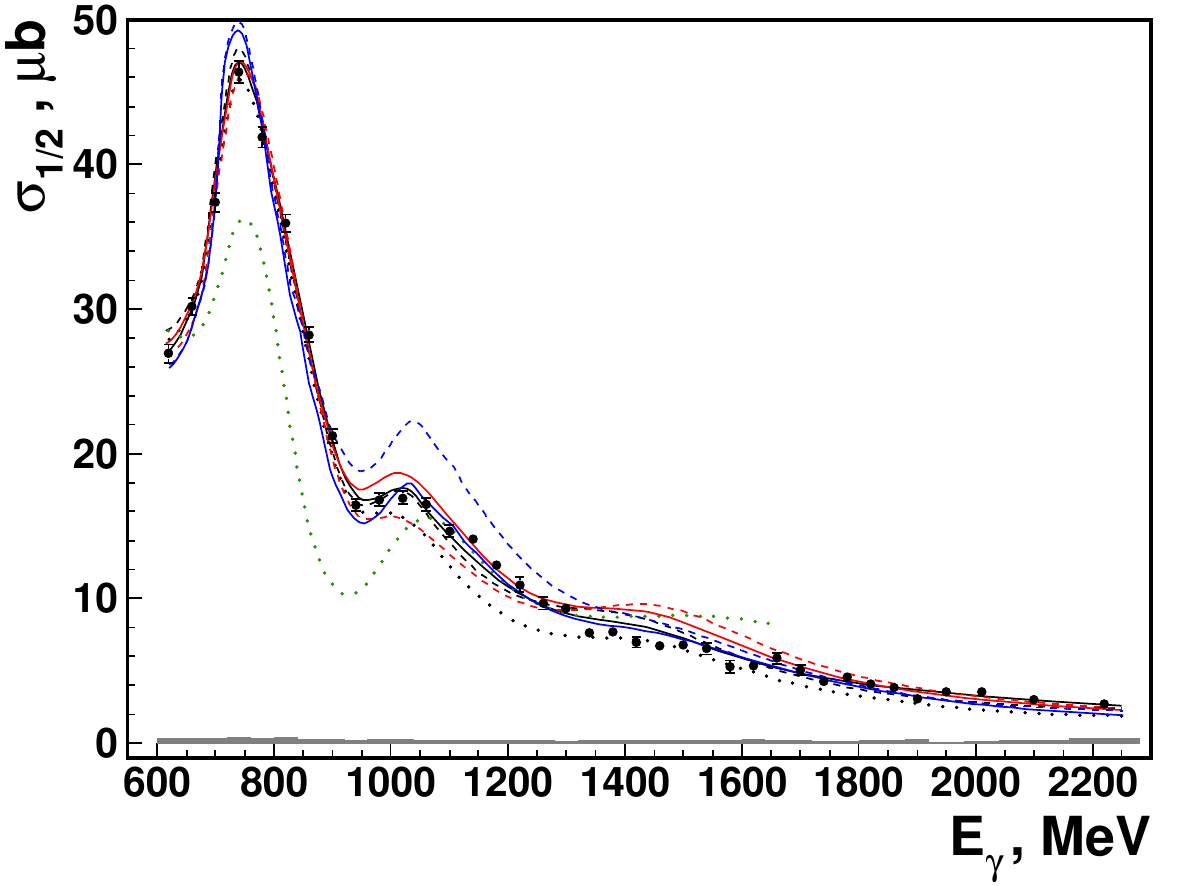}&
\includegraphics[width=0.48\textwidth]{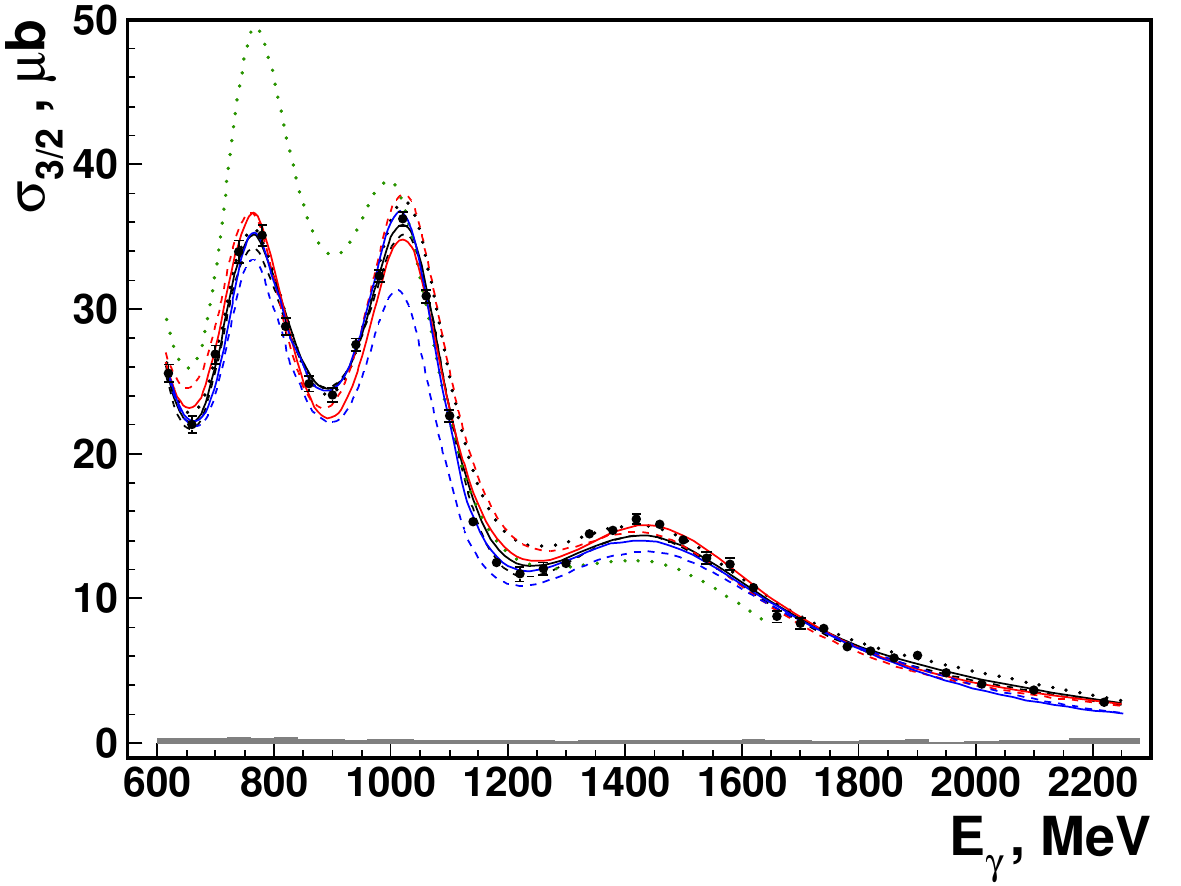}
\end{tabular}
 \caption{$\sigma_{1/2}$ and $\sigma_{3/2}$ and PWA results from different solutions including the data
 presented here: BnGa 2014-02 (black solid)\cite{Sokhoyan:2015fra,Beck:2016hcy}, 2014-01 (black dashed), SAID (PD03)~\cite{Beck:2016hcy}
 (red solid, curve calculated used solution PD03 for $E$ and CM12 for the cross section), J\"uBo 2016-01~\cite{Beck:2016hcy} (blue solid), and not including this data:
 BnGa 2011-02 (black dotted), SAID CM12 (red dashed),  J\"uBo 2013-01 (blue dashed), MAID 2007 (green dotted). The gray area represents the systematic error.
\vspace{8mm}}
\label{Fig_Sigma12_32}
 \begin{tabular}{ccc}
\includegraphics[width=0.32\textwidth]{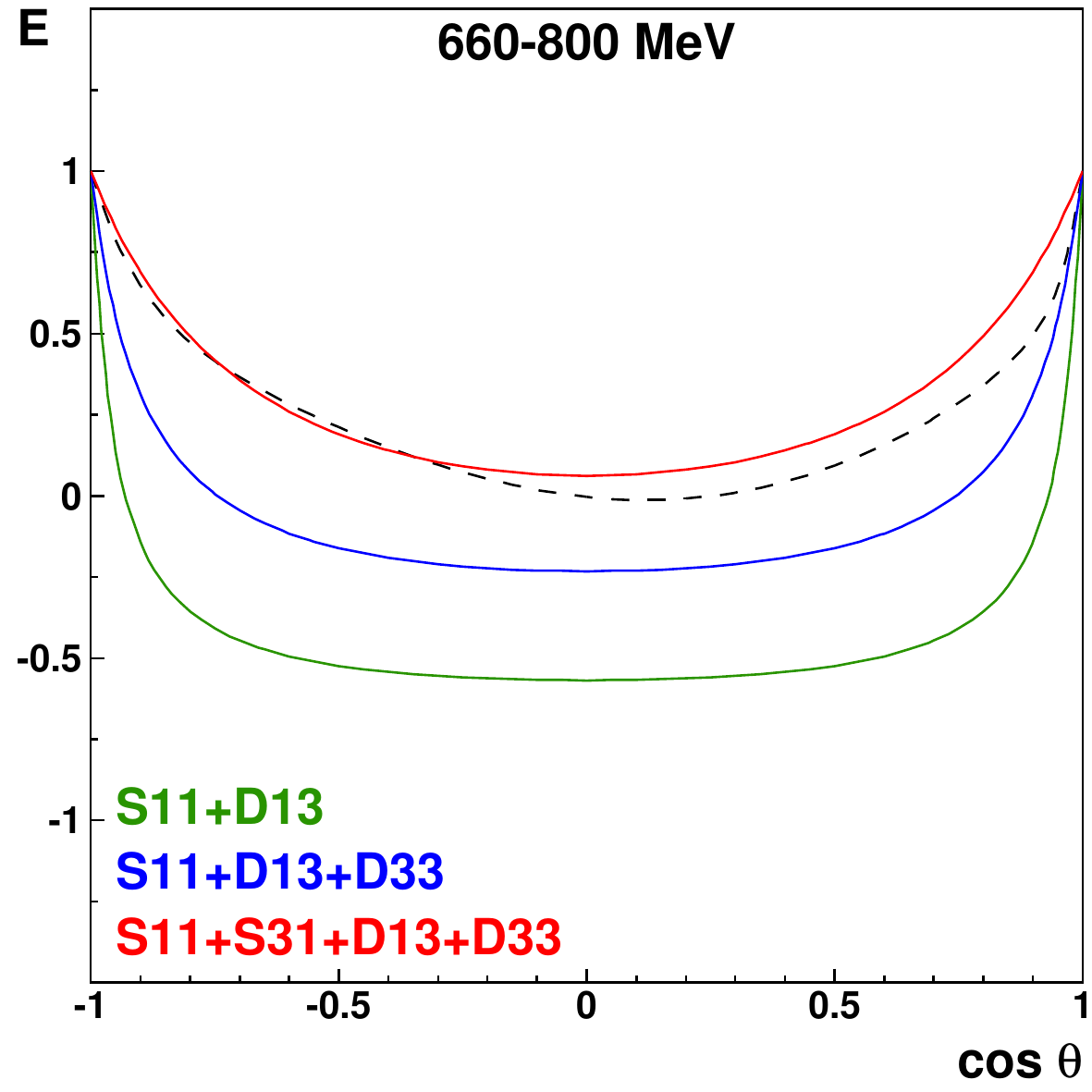} &
\includegraphics[width=0.32\textwidth]{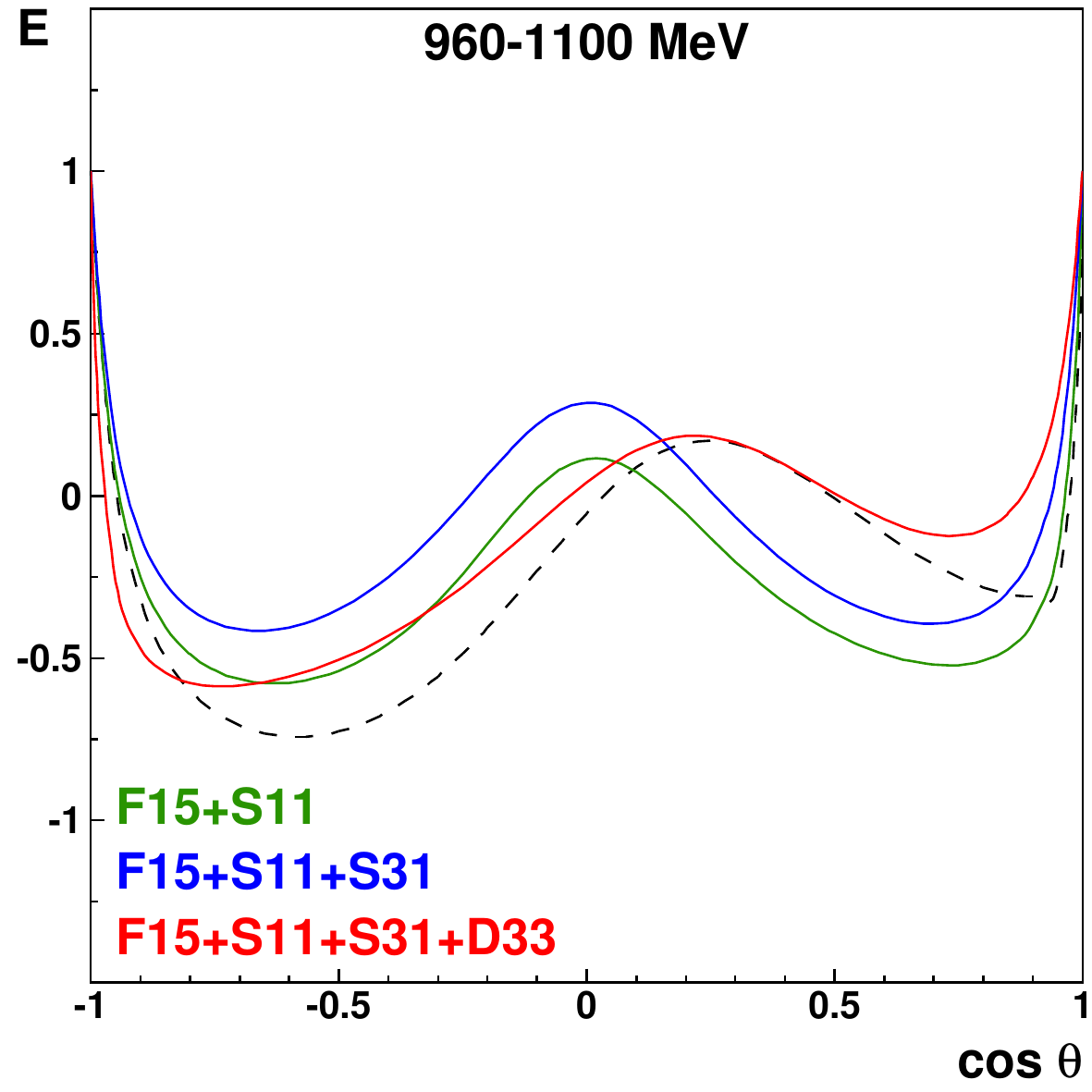} &
\includegraphics[width=0.32\textwidth]{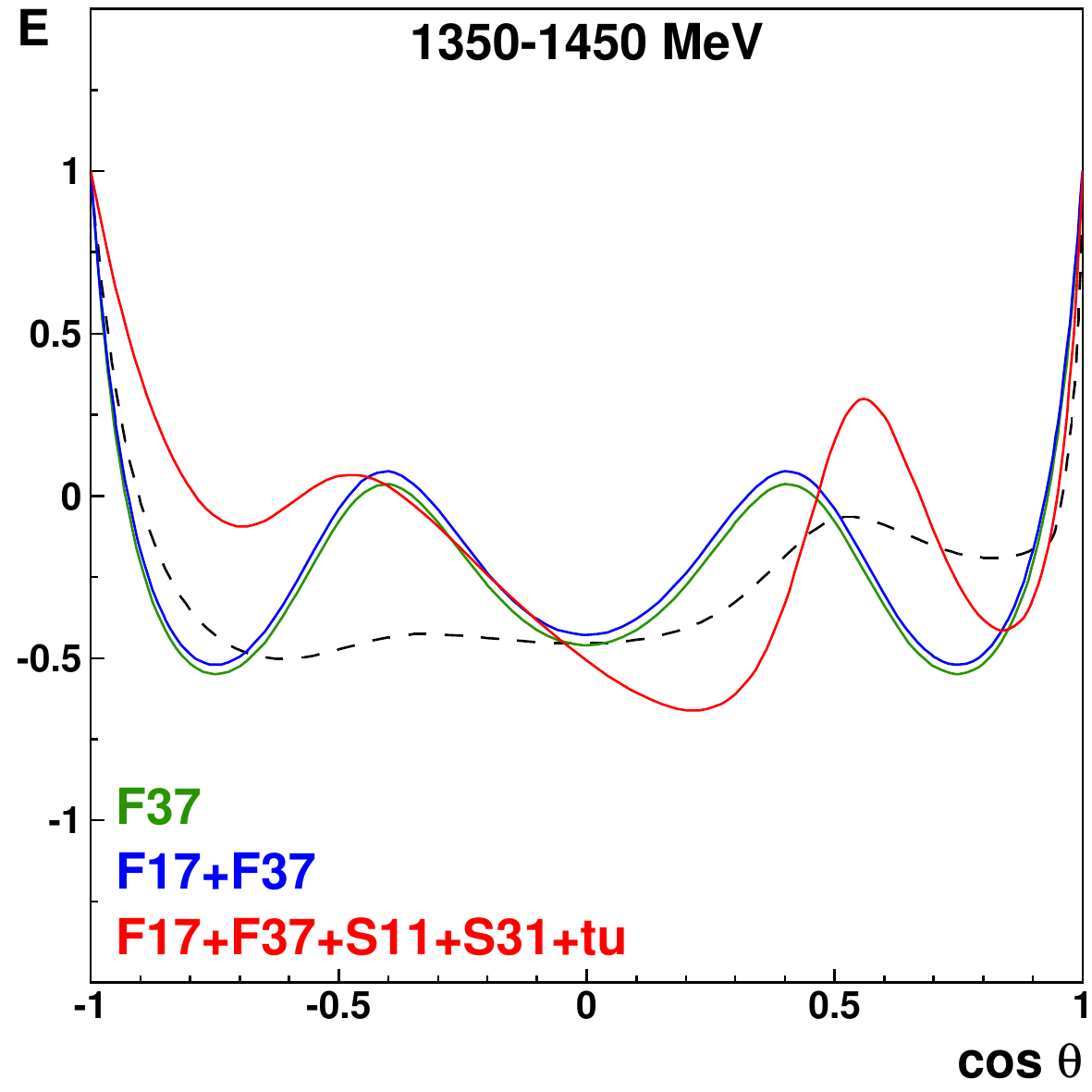} \\
 \end{tabular}
 \caption{Black dashed: BnGa fit to the data (BG2014-02~\cite{Sokhoyan:2015fra,Beck:2016hcy}). Different colors: single partial wave contributions within the description of the BnGa-PWA as indicated in the figures. For further details, see text.
}
 \label{E_pwa}
\end{figure*}
The
calculation of the total cross section requires the integration over the full solid angle. In the
angular range, where no data points exist, the double-polarization observable~$E$ was extrapolated
to the unmeasured angular range using the BnGa fit. The resulting total cross sections for
$\sigma_{1/2}$ and $\sigma_{3/2}$ are shown in Fig.~\ref{Fig_Sigma12_32}, in comparison to the
different PWA predictions.

The two distributions are very different. In the second resonance region, covering the photon
energy range from 660 to 900~MeV, the $\sigma_{1/2}$ peak is significantly stronger than the peak
in $\sigma_{3/2}$. This is due to the sizable contribution of the $N(1535)1/2^-$ resonance which
contributes to $\sigma_{1/2}$ only. Some early partial wave analyses, MAID and SAID-SN11,
underestimated the $N(1535)1/2^-$ contribution to $\gamma p\to N\pi$, and assigned this intensity
to the $N(1520)3/2^-$ resonance. The third resonance region covers the energy range from 900 to
1200~MeV. In this region, the MAID and SAID-SN11 and, to a smaller extent, the SAID-CM12 and BnGa
2011-02 partial wave analyses overestimate the $\sigma_{3/2}$ contribution, at the expense of
$\sigma_{1/2}$, while J\"uBo underestimated the $\sigma_{3/2}$ contribution.

The description of the data by the most recent fits of the different PWA groups (which included the
data presented here) improved quite significantly. The fourth resonance region,
1200\,$<E_\gamma<$\,1700~MeV, is dominated by $\sigma_{3/2}$ but none of the fits is fully
satisfactory. Likely, the statistical significance of the high-mass data in Fig.~\ref{Efigs-fit}
is not large enough to constrain the fit.

\paragraph{The dominant partial wave contributions:}
It may be illuminating to compare the results on $E$ with simple
models. In Fig.~\ref{E_pwa}, we display the result of the latest BnGa
fit with the predictions based
on a few partial waves.
The observed angular distribution of $E$ for \mbox{660\,$<E_\gamma<$ 800~MeV} exhibits a single minimum
almost reaching \mbox{$E\approx 0$} just above $\cos\theta=0$. If there was only a contribution from
$J^P=1/2^-$ ($S_{11}$), $E=1$ would hold for all values of $\cos\theta$. This is obviously not
true. Adding the amplitude for $N(1520)3/2^-$ ($D_{13}$) (with the helicity couplings as determined
in the fit) leads to a much deeper minimum than observed. It requires contributions from
$\Delta(1620)1/2^-$ ($S_{31}$) and $\Delta(1700)3/2^-$ ($D_{33}$) to arrive close to the expected
curve. The remaining asymmetry requires additional contributions from odd waves, e.g. from
$\Delta(1600)3/2^+$ ($P_{33}$).

In the 960\,$<E_\gamma<$\,1100~MeV range, the angular distribution of $E$ is characterized by two
minima which originate from the interference of $J^P=1/2^-$ ($N(1535)1/2^-$, $\Delta(1620)1/2^-$)
and $J^P=5/2^+$ ($F_{15}$) ($N(1680)5/2^+$). The for\-ward-backward asymmetry due to the
interference of even and odd waves becomes more pronounced adding $\Delta(1700)3/2^-$.

The photon energy interval from  1350 to 1450~MeV exhibits an angular distribution of $E$ with
three local minima, or two intermediate maxima. Such a pattern can be reproduced by the $J^P=7/2^+$
partial wave ($N(1990)7/2^+$ ($F_{17}$), $\Delta(1950)7/2^+$ ($F_{37}$)). However in the data, the
local minimum at $\cos\theta\approx-0.45$ is considerably lower than the minimum at
$\cos\theta\approx0.9$. Adding $J^P=1/2^-$ waves ($N(1535)1/2^-$, $\Delta(1620)1/2^-$) and $t$- and
$u$-channel contributions, the data represented by the final BnGa curve (dashed black) are still not
yet reasonably well reproduced: many partial waves are required (and do contribute in the BnGa
fit).


\section{\label{Summary}\boldmath Summary}

We have reported a measurement of 467 data points on the helicity asymmetry
$E=(\sigma_{1/2}-\sigma_{3/2})/(\sigma_{1/2}+\sigma_{3/2})$. The data cover the $E_\gamma$ energy
range from 600 to 1230~MeV in 30~MeV wide bins, 1230 - 1590~MeV in 60~MeV wide bins, and the range
from 1590 - 2310~MeV in 120~MeV wide bins. The bin widths are chosen in view of the available
statistics and of the dependence of the angular distributions on the photon energy. The data are
presented in 15 slices in $\cos\theta$; in most energy bins, the solid angle coverage is almost complete.

The data presented here, and other new polarization data measured at Bonn, JLab, and Mainz
were included in the PWAs of the BnGa, J\"uBo, and SAID groups. The new data have a significant
impact on the fit. While the predictions exhibit large discrepancies among themselves and with the
data, there is good agreement between data and all new fits. Smaller discrepancies will need further
studies, both experimentally and in new partial wave analyses.

\subsection*{Acknowledgements}
We thank the technical staff of ELSA and the participating institutions for their invaluable
contributions to the success of the experiment. We acknowledge support from the \textit{Deutsche
Forschungsgemeinschaft (SFB/TR16, SFB/TR110)}, the \textit{U.S. Department of energy, Office of Science, Office of Nuclear Physics under Awards No. DE-FG02-92ER40735}, the \textit{Russian Foundation for Basic Research}, and the  \textit{Schweizerischer Nationalfonds (200020-156983, 132799, 121781, 117601)}. 
We thank the J\"uBo, SAID and the MAID group for providing their PWA-results.


\end{document}